\documentclass [aps,pra,amssymb,amsmath,onecolumn, showpacs]{revtex4-1}
\usepackage[latin9]{inputenc}
\usepackage{amsmath}
\usepackage{amssymb}
\usepackage{graphicx}
\usepackage{verbatim}
\usepackage{bm}
\usepackage{color}

\usepackage{graphicx,epsfig,amsfonts,amssymb}
\usepackage{bm}
\usepackage{times}
\usepackage{lipsum}
\usepackage{verbatim}

\newcommand{\nn}{\nonumber}

\newcommand{\sgn}{\operatorname{sgn}}

\newcommand{\ra}{\rangle}
\newcommand{\rar}{\rightarrow}

\newcommand{\be}{\begin{eqnarray}}
\newcommand{\ee}{\end{eqnarray}}

\newcommand{\bs}{\begin{equation}\begin{split}}
\newcommand{\es}{\end{split}\end{equation}}

\begin{document}
\title{A large class of solvable multistate Landau-Zener models and quantum integrability}
\author{Vladimir Y. Chernyak$^{a,b}$}
\email{chernyak@chem.wayne.edu}


\author{Nikolai A. Sinitsyn$^{c}$}
\email{nsinitsyn@lanl.gov}

\author{Chen Sun$^{c,d}$}
\email{chen.sun.whu@gmail.com}
\affiliation{$^a$ Department of Chemistry, Wayne State University, 5101 Cass Ave, Detroit, Michigan 48202, USA}
\affiliation{$^b$ Department of Mathematics, Wayne State University, 656 W. Kirby, Detroit, Michigan 48202, USA}
\affiliation{$^c$Theoretical Division, Los Alamos National Laboratory, Los Alamos, NM 87545, USA}
\affiliation{$^d$Department of Physics, Texas A\&M University, College Station, TX 77843, USA}

\begin{abstract}
The concept of quantum integrability has been introduced recently for quantum systems with  explicitly time-dependent  Hamiltonians  \cite{commute}. Within the multistate Landau-Zener (MLZ) theory, however, there has been a successful alternative approach to identify and solve complex time-dependent models  \cite{quest}. Here we compare both methods by applying them to a new class of exactly solvable MLZ models.  This class contains systems with an arbitrary number $N\ge4$ of interacting states and shows a quickly growing with $N$ number of exact  adiabatic energy crossing points, which appear  at different moments of time. At each $N$, transition probabilities in these systems can be found analytically and exactly but
complexity and variety of solutions in this class also  grow with $N$ quickly.
We illustrate how   common features of solvable MLZ systems appear from quantum integrability and develop an approach to further classification of solvable MLZ problems.
\end{abstract}

\maketitle



\section{Introduction}

In quantum mechanics, the concept of integrability is controversial \cite{integ-q,yuz-q}. In classical physics, integrability means the equality of the number of invariants of motion to the number of degrees of freedom. So, there is no controversy: for a finite classical integrable system one can derive its trajectory of motion analytically given initial conditions.  Interestingly, all quantum systems  have similar properties: for any $N\times N$ Hamiltonian matrix there are $N$ independent matrices that commute with it. Also, by finding $N$ eigenstates and eigenvalues of a  time-independent Hamiltonian one can write the solution of the evolution equation for the state vector at any time given the initial conditions.  These facts, however, do not make all quantum mechanical problems easy to understand.


The most accepted definition of quantum integrability means presence of a nontrivial symmetry, i.e. Yang-Baxter relations, leading, e.g., to the validity of  Bethe's ansatz  in a class of 1D  models. To extend it to
a broader class of finite size systems, authors of \cite{yuz-q} proposed to call a quantum system integrable if its Hamiltonian depends on a continuous spectral parameter $u$ so that there are also nontrivial operators that depend on $u$ polynomially and commute with the Hamiltonian at all values of this parameter. Such models do show characteristics that are usually attributed to solvable models. For example, their spectra have exact crossing points of energy levels at some values of $u$, and statistics of gaps between their energy levels can be Poissonian.

However, so defined integrable finite size quantum Hamiltonians do not generally have explicit solutions for  eigenstates. For example, apart from the thermodynamic limit $N\rar \infty$, Bethe ansatz usually leads to eigenstates that depend on parameters that satisfy nonlinear algebraic equations. Numerical studies of such equations  can be as complex as direct diagonalization of the matrix Hamiltonian. So, can the notion of quantum integrability be more useful when dealing with finite size quantum problems?


The multistate Landau-Zener (MLZ) theory \cite{majorana,be,four-LZ,mlz-all} (see also \cite{mlz-misc} for some recent applications of this theory) has recently introduced a different view on application of quantum integrability. MLZ theory provides a new approach to obtain broad families of parameter dependent models with numerous exact energy level crossings, and with the possibility to obtain completely analytical description of the evolution matrix in terms of commonly known functions.
This theory  describes explicitly time-dependent dynamics according to the nonstationary Schr\"odinger equation of the form
\begin{equation}
i\frac{d}{dt}\psi=\hat{H}(t)\psi,\quad  {H}(t) = {A} +{B}t,
\label{mlz}
\end{equation}
where $H(t)$ is the matrix representation of $\hat{H}(t)$;  ${A}$ and ${B}$ are constant Hermitian $N\times N$ matrices (we set $\hbar=1$). One can always choose the {\it diabatic basis} in which the matrix $B$ is diagonal, and if any pair of its elements are degenerate then the corresponding off-diagonal elements of the matrix ${A}$ can be set to zero by a time-independent change of the basis, that is
\begin{align}
&B^{ab}= \delta^{ab}\beta_a, \quad  A^{ab}=0\,\,{\rm if} \,\, \beta_{a}=\beta_{b},
\label{diab3}
\end{align}
where we mark elements of $N\times N$ matrices by upper indexes for the reasons that will be clear later.
Constant parameters $\beta_a$  are called the {\it slopes of diabatic levels}, diagonal elements of the Hamiltonian in the diabatic basis,
$B^{aa}t+A^{aa}$, are called {\it diabatic energies}, and  nonzero off-diagonal elements of the matrix ${A}$ in the diabatic basis are called the {\it coupling constants}. 
A MLZ model is called {\it solvable} if one can find the probability $N\times N$ matrix ${P}$, with elements $P^{ab}=|S^{ab}|^2$, where $S^{ab}$ is the amplitude of the diabatic state  $a$ at $t  \rightarrow +\infty$, given that at $t \rightarrow -\infty$ the system was in the  diabatic state $b$.

There are two different approaches to identify solvable models of the form (\ref{mlz}). The earlier one was based on the discovery of
{\it integrability conditions} (ICs) \cite{six-LZ,quest} in MLZ theory, which are similar to invariants of motion in classical mechanics. ICs impose certain constraints on MLZ model parameters. If these constraints are satisfied, the transition probability matrix elements can be written explicitly in terms of elementary functions  \cite{quest}.  ICs, as they were defined in \cite{quest}, remain unproved  mathematically but there have been numerous rigorous tests of their predictions without any counterexample. Numerical evidence for validity of ICs is overwhelming \cite{quest,six-LZ,four-LZ,DTCM}.

More recently,  Ref.~\cite{commute} showed that explicitly time-dependent models, including MLZ systems, can be solved if
one can identify a family with several nontrivial Hamiltonians $\hat{H}_j({\bm \tau})$, that depend on parameters $\tau^j$, $j=0,\ldots,M$, where $M\ge 1$ and $\tau_{0}\equiv t$ is real time. These Hamiltonians must  satisfy conditions
\be
[\hat{H}_i,\hat{H}_j]=0, \quad i,j=0,\ldots, M,
\label{cond1}
\ee
\be
\partial \hat{H}_i/\partial \tau^j =  \partial \hat{H}_j/\partial \tau^i,
\label{cond2}
\ee
where we identify $\hat{H}_0(\tau^{0})$ with $\hat{H}(t)$ in the original model, which in our case is a MLZ model of the form (\ref{mlz}).
One can then consider the original equation (\ref{mlz}) as a component of the multiple-time Shr\"odinger equation for the state vector $ \Psi(\bm{\tau}) \equiv \Psi(\tau^0,\ldots, \tau^M)$:
\begin{equation}
\label{system1}
 i \frac{\partial}{\partial \tau^j} \Psi(\bm{\tau}) = \hat{H}_{j} (\bm{\tau}) \Psi(\bm{\tau}), \; \phantom{\sum} j = 0, 1, \ldots, M.
\end{equation}
Note that we reserve lower matrix indexes to distinguish among different Hamiltonians and corresponding time variables.
Conditions (\ref{cond1})-(\ref{cond2}) guarantee that the system (\ref{system1})
has a single-valued solution in the form of an ordered exponential along a path ${\cal P}$ that starts at a reference time-point $\bm{\tau}_{in}$ and ends at  ${\bm{\tau}}$:
\begin{eqnarray}
 \label{time-ord}
\Psi(\bm{\tau}) = T \exp\left(-i \int_{{\cal P}} \hat{H}_{j} d\tau^{j}\right) \Psi(\bm{\tau}_{in}).
\end{eqnarray}
Here, we introduced a convention to sum over repeated lower and upper multi-time indexes.  The original scattering MLZ problem (\ref{mlz}) is recovered if we set initial and final integration points at, respectively, $\tau^0_{in}=- \infty$ and  $\tau^0_{fin}=+ \infty$, at constant values of other  parameters $\tau^j$. Explicit solution of this problem becomes possible because, apart from the initial and final points, the path  ${\cal P}$ in (\ref{time-ord}) can be chosen arbitrarily without affecting the end state \cite{commute}. In particular, one can choose it to be always in the region with $|{\bm \tau}| \rar \infty$, where WKB approximation becomes exact.  Ref.~\cite{commute} has shown that some of the previously conjectured solutions of MLZ systems can be rigorously derived using this approach.


Many questions about MLZ integrability remain open. There is still no derivation of the previously known ICs from the
integrability structure (\ref{cond1})-(\ref{cond2}). The latter has been very effective in proving validity of already derived solutions rigorously. However, so far it has not been used to find truly new solvable MLZ models.

 Moreover, apart from satisfying ICs, known solvable models   have many  other unexplained properties.
 First,
 ICs require appearance of a number of exact energy level crossings, which are generally hard to find. In \cite{quest}, simple perturbative tests for exactness of level crossings in MLZ systems were very successful in search for such models. This is surprising and very fortunate because perturbatively derived relations among parameters should generally be only necessary but not sufficient for appearance of an exact crossing point.
Second, the strategy used in \cite{quest} to find new solvable models was to continuously deform  parameters of matrices ${A}$ and ${B}$ in (\ref{mlz}) so that ICs remained  satisfied, thus creating a continuous family of solvable models. A curious observation of \cite{quest} was that transition probability matrices in such generated families of solvable models were independent of deformation parameters.
 Namely, if we create the skew-symmetric matrix $\hat{\gamma}$ with elements
\be
\gamma^{ab}=\frac{|A^{ab}|^2}{\beta_{a}-\beta_{b}},
\label{skew}
\ee
then in solvable models that were connected by continuous deformations of parameters $A^{ab}$ and $\beta^a$, the matrix $\hat{\gamma}$ and the transition probability matrices remained the same \cite{quest}. In addition, matrices $\hat{\gamma}$  usually have additional unexplained symmetries, e.g., the same values $\gamma^{ab}$  appear multiple times, possibly with different signs, in different places of $\hat{\gamma}$ \cite{six-LZ,four-LZ,quest}.
All these observations, if generally true and understood, can be quite handy because they strongly simplify equations that determine parameters of a solvable model.

Our article has two goals. First, we compare the two currently available approaches  by  deriving a new large family of solvable MLZ models  with arbitrary number $N$ of  interacting states and some unifying property. Hence, the first part of our article can be considered as a review of the methods based on ICs \cite{quest} and quantum integrability \cite{commute}  with a novel solvable MLZ class as an example and application. Based on this comparison, we claim that  previously known ICs must follow from Eqs.~(\ref{cond1})-(\ref{cond2}).
We turn then to our second goal of using this quantum integrability to classify solvable MLZ models and explain numerous previous observations about them including existence of ICs themselves. We show that this goal is achievable within classes of equations (\ref{system1}) with specific types of  singularities.

The structure of our article is as follows. In  section~\ref{main-sec}, we present a new solvable MLZ model, namely, the class of arbitrarily large $N$-state MLZ Hamiltonians  with  parameter constraints that make  these models solvable.
In sections~\ref{test5-sec} and  \ref{testN-sec}, we explore transition probabilities and corresponding phase diagrams in five and six state sectors of our model. We also  provide numerical checks for theoretical predictions.
In section~\ref{derive-sec}, we derive  our model with previously used ICs \cite{quest}.
In section~\ref{bowtie-sec}, we show how our model can be generated and solved starting from Eqs.~(\ref{cond1})-(\ref{cond2}). We also provide the proof of the existence of the  exact energy crossing points.
In section~\ref{dist-sec}, we briefly discuss common properties found in our model and other solvable MLZ systems.
Section~\ref{mtlz-sec} has the goal to provide a unifying approach to a broader class of systems.
 It develops generalization of our model to a much broader class that we named the Multi-Time Landau-Zener (MTLZ) problem. We show how numerous facts about solvable MLZ systems can be
proved rigorously within this large class, and suggest a direction for further classification of such models.
We summarize our findings in the conclusion.

\section{The model of two states interacting with a set of levels}
\label{main-sec}
We start with describing the new solvable MLZ model.
We searched for it starting with
the most general Hamiltonian that describes arbitrary interactions of two diabatic levels with  $N-2$ other states:
\begin{equation}
 H(t)=\left( \begin{array}{cccccc}
b t & 0 & g_{13} & g_{14} & \ldots & g_{1N}\\
0 & -b t &  g_{23} &  g_{24} & \ldots & g_{2N}\\
g_{13} & g_{23} &  b_3 t+e_3  & 0 &\ldots &0\\
g_{14}  & g_{24} & 0  & b_4 t+e_4  &\ldots &0\\
\vdots & \vdots & \vdots & \vdots & \ddots  &\vdots\\
g_{1N} & g_{2N} & 0 & 0 & \ldots  & b_N t+e_N
\end{array}\right),
\label{hal}
\end{equation}
where $g_{ij}$, $b_i$, and $e_i$ are constant parameters.

For simplicity, we used the time translation freedom to set diabatic energy crossing of first two levels at $t=0$ and also used the gauge freedom \cite{be} to make slopes of these levels different only by sign. We will assume that $b>0$. All couplings $g_{ij}$ between levels $i$ and $j$ are assumed to be real. There is no direct coupling  between levels 1 and 2, as well as between any two levels with indexes higher than 2. We also assume that there are no parallel levels, namely, $b_i\ne b_j$ $\forall$ $i\ne j$.

We will claim  that the MLZ model \eqref{hal} is solvable  if its parameters satisfy following conditions:
\begin{align}
\label{b-con}
&b<|b_i|,\\
\label{e-con}
&e_i=\lambda_i e\sqrt{\frac{b_i^2}{b^2}-1},\quad e\ge 0, \\
\label{g-con1}
&\sum_{i=3}^N \frac{g_{1i}^2}{b_i-b}=0,\\
\label{g-con-2}
& g_{2i}=\tau_i g_{1i}\sqrt\frac{b_i+b}{b_i-b},\\
\label{sign-con}
& \lambda_i\tau_i \sigma_i=\lambda_j\tau_j \sigma_j,
\end{align}
where $i,j=3,4,\ldots, N$ and $\lambda_i$, $\tau_i$ and $\sigma_{i}$ are signs of, respectively, $e_i$, $g_{2i}/g_{1i}$, and $b_i$, i.e.,
\begin{align}
\lambda_i=\pm 1,\quad \tau_i=\pm 1,\quad \sigma_{i}=\sgn(b_i).
\end{align}
\begin{figure}[!htb]
\scalebox{0.5}[0.5]{\includegraphics{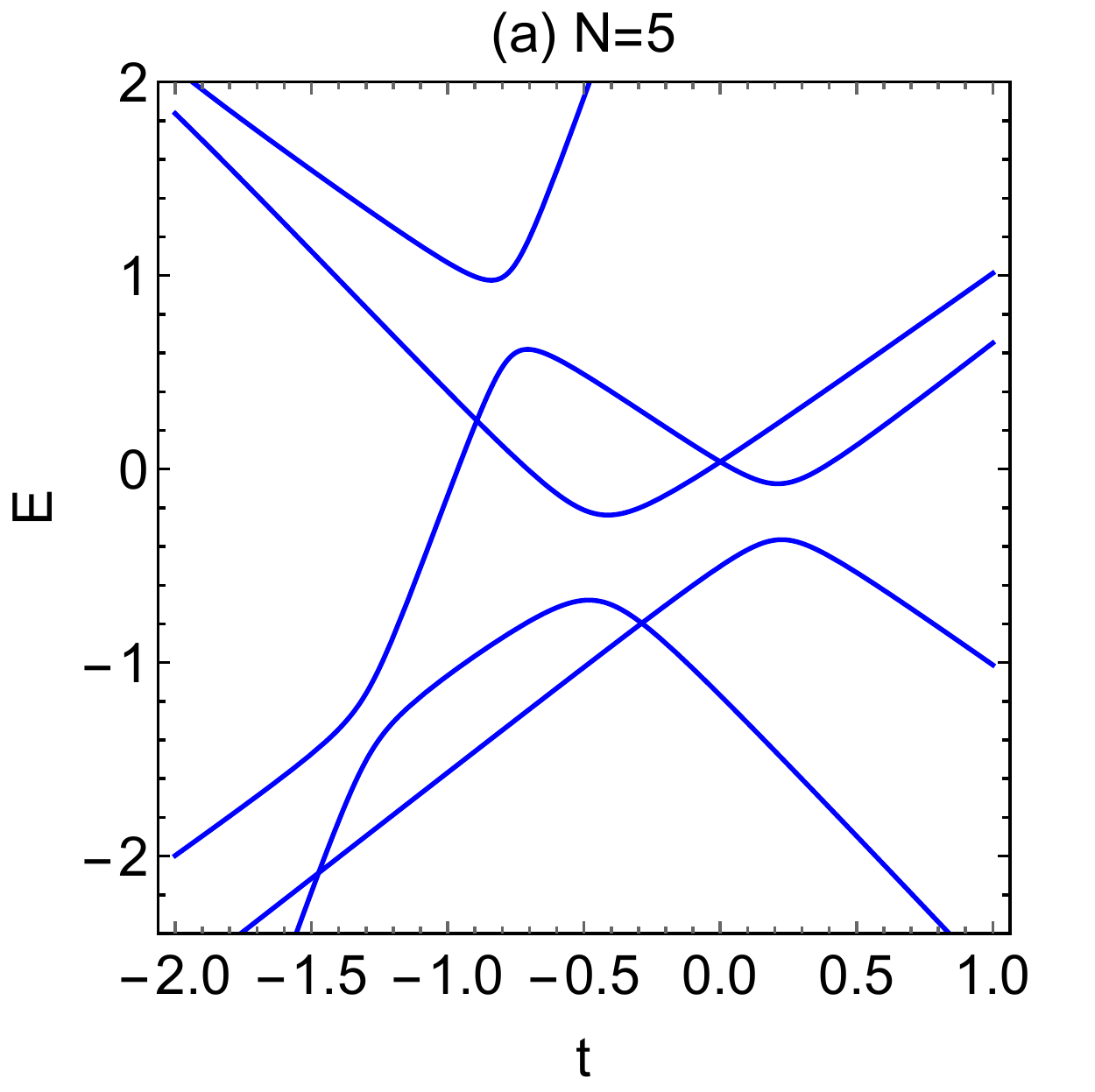}}
\scalebox{0.5}[0.5]{\includegraphics{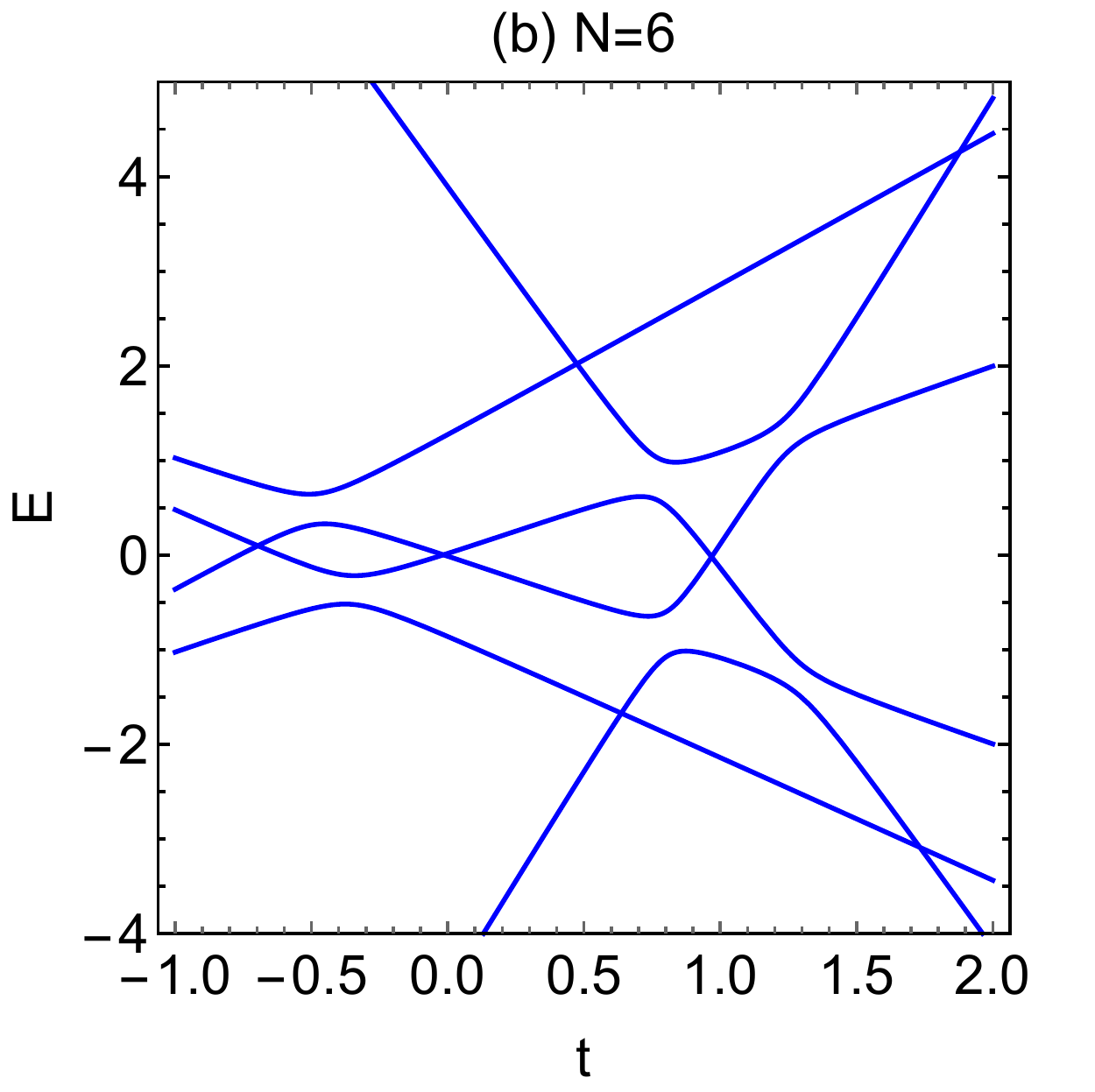}}\\
\scalebox{0.5}[0.5]{\includegraphics{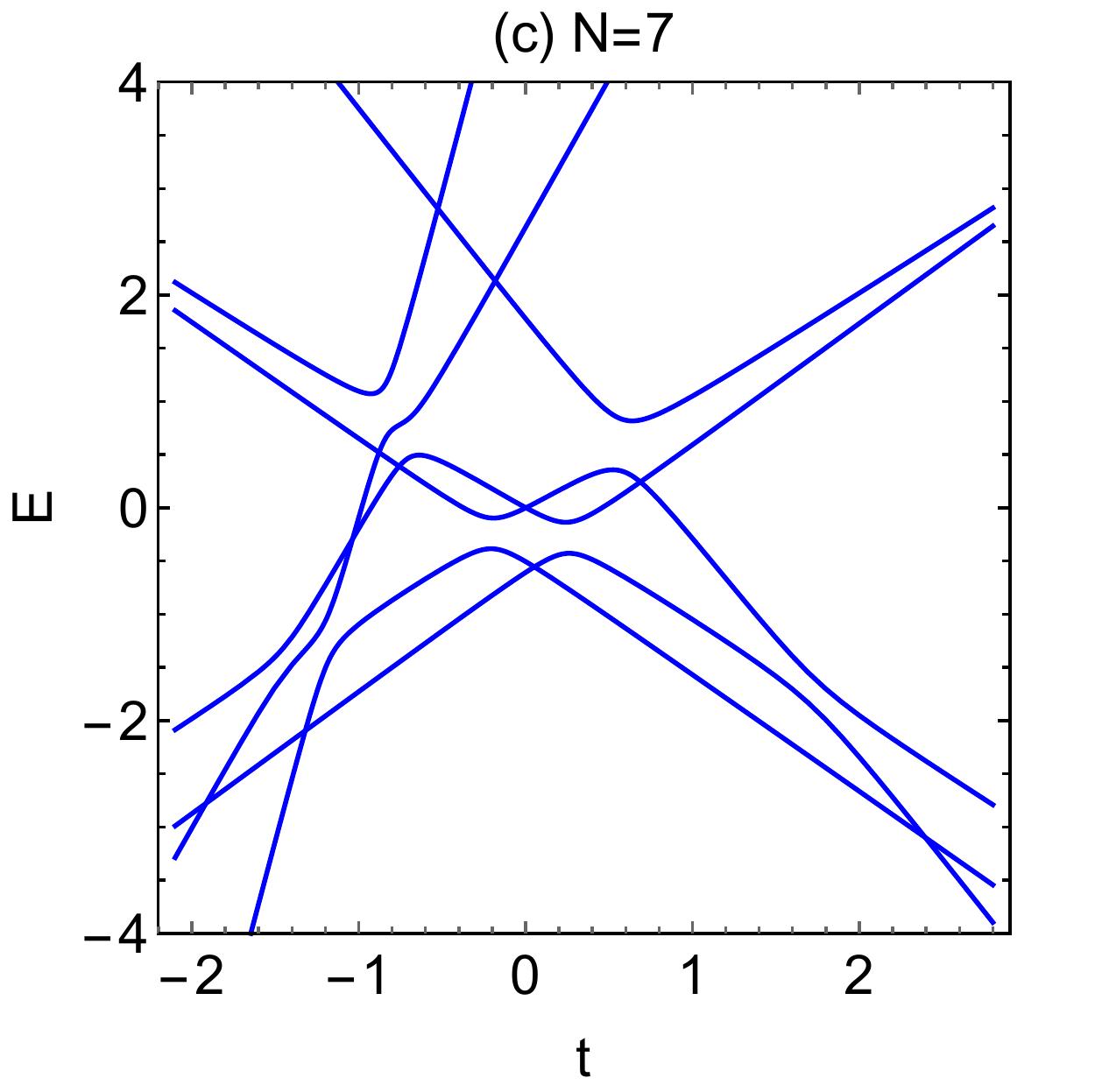}}
\scalebox{0.5}[0.5]{\includegraphics{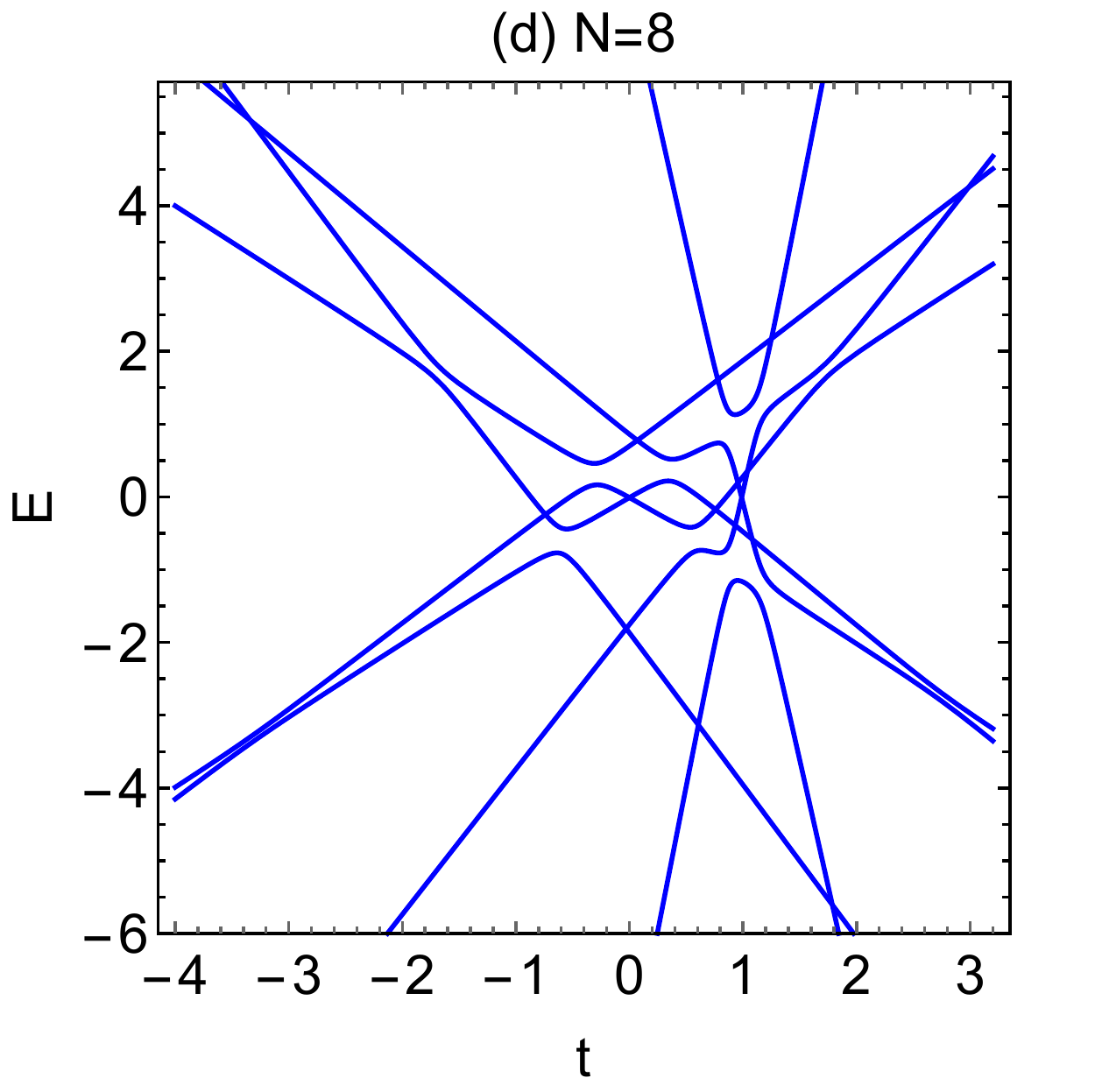}}\\
\scalebox{0.5}[0.5]{\includegraphics{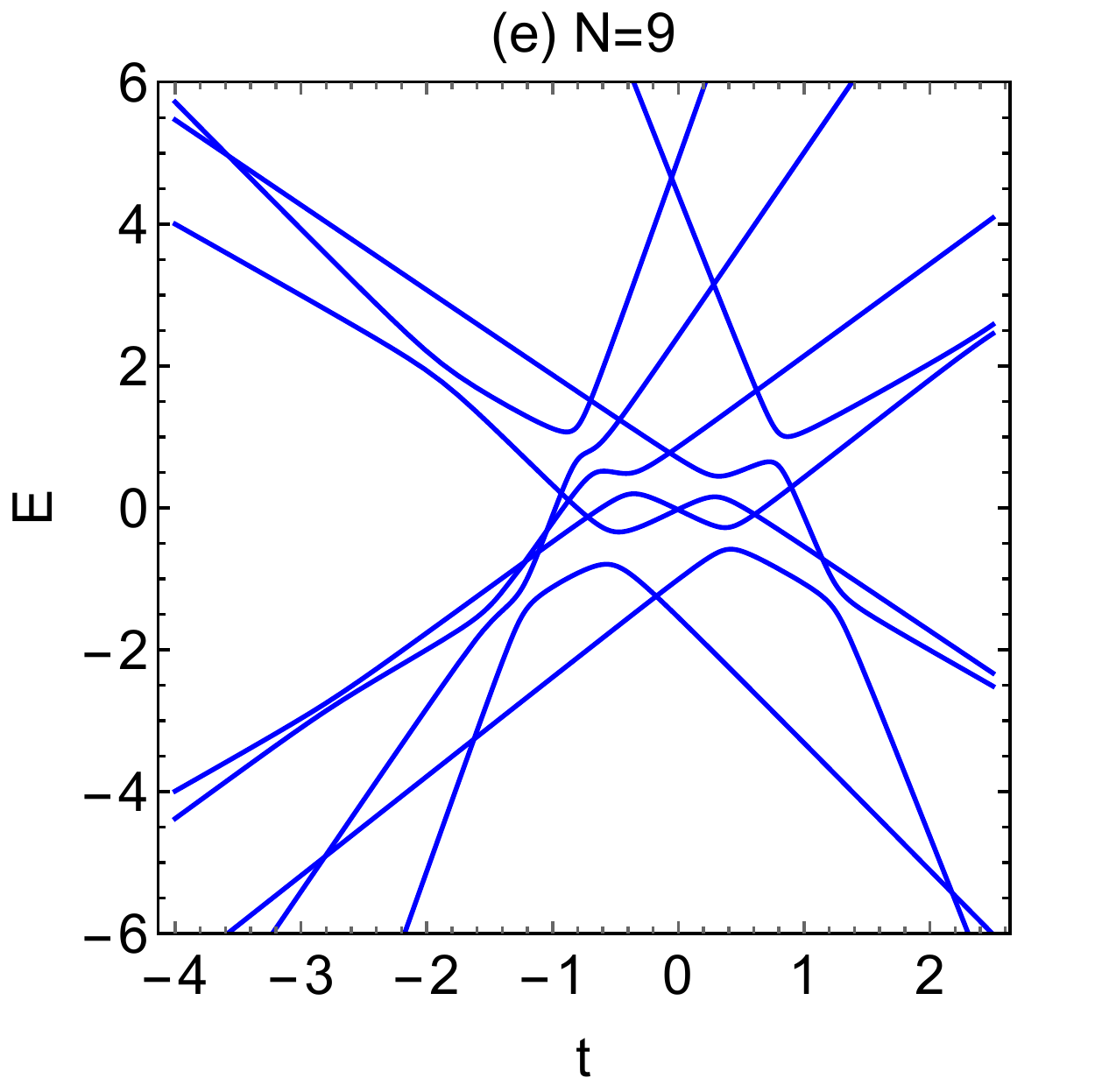}}
\scalebox{0.48}[0.48]{\includegraphics{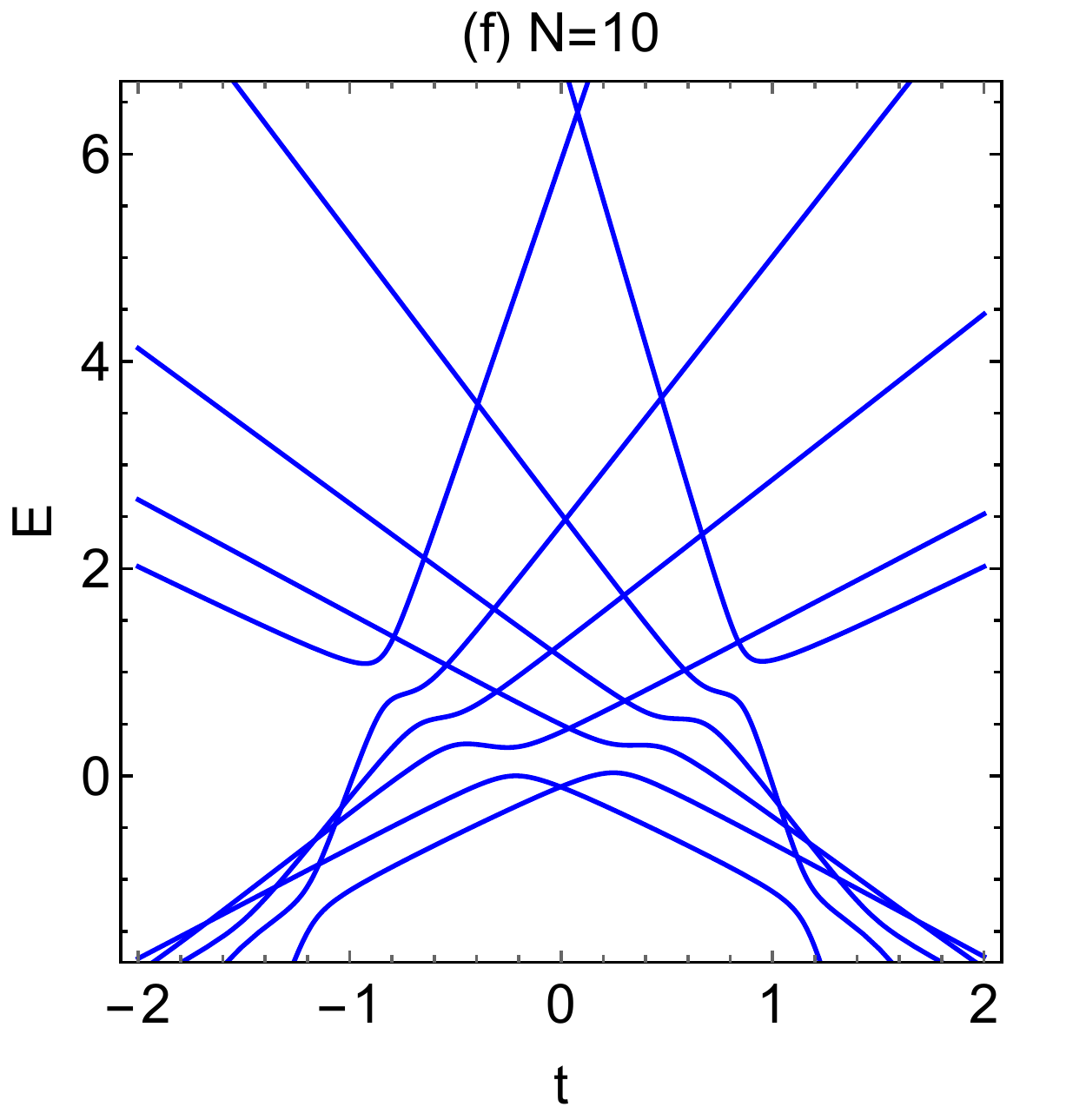}}
\caption{Adiabatic energies as functions of time for models (\ref{hal}) with constraints (\ref{b-con})-(\ref{sign-con}) and $N=5,6,7,8,9,10$. ICs predict, respectively, 4, 7, 11, 16, 21, and 29 exact energy level crossings that are found in figures (a)-(f). The choices for $\lambda_i$ are: (a) $\lambda_3=-\lambda_4=-\lambda_5=1$; (b) $-\lambda_3=\lambda_4=-\lambda_5=\lambda_6=1$; (c) $\lambda_3=\lambda_4=-\lambda_5=-\lambda_6=\lambda_7=1$; (d) $-\lambda_3=-\lambda_4=\lambda_5=\lambda_6=-\lambda_7=\lambda_8=1$; (e) $\lambda_3=\lambda_4=\lambda_5=\lambda_6=-\lambda_7=\lambda_8=-\lambda_9=1$; (f) $\lambda_3=\lambda_4=\lambda_5=\lambda_6=\lambda_7=\lambda_8=\lambda_9=\lambda_{10}=1$. 
Other  parameters are chosen randomly. 
}\label{exact_crossing}
\end{figure}

By ``solvable"  we mean here that the Hamiltonian satisfies ICs in MLZ theory \cite{quest} and belongs to a family of operators satisfying conditions (\ref{cond1})-(\ref{cond2}).
Let us now explicitly mention some of the properties of constraints (\ref{b-con})-(\ref{sign-con}) (for their derivations see Section \ref{derive-sec}):

1) Equation~\eqref{b-con} says that all slopes $b_i$, $i=3,\ldots, N$, should be larger than $b$ in absolute magnitude.

2) Constant diagonal elements of the Hamiltonian are fully determined by other parameters up to the rescaling factor $e$.

3) According to (\ref{g-con1}), all $b_i$ cannot have the same sign.

4) According to (\ref{g-con-2}), every $g_{2i}$ can be expressed in terms of the corresponding $g_{1i}$ and level slopes. Consider expressions for pairwise transition probabilities:
\begin{align}\label{p12}
&p_{1i}=e^{-\frac{2\pi g_{1i}^2}{|b-b_i|}},\quad q_{1i}=1-p_{1i},\nn\\
&p_{2i}=e^{-\frac{2\pi g_{2i}^2}{|b+b_i|}},\quad q_{2i}=1-p_{1i}, \quad i=3, \ldots, N.
\end{align}
Using the constraint between $g_{1i}$ and $g_{2i}$, we find
\be
p_{1i}=p_{2i}, \quad q_{1i}=q_{2i}.
\label{p-eq}
\ee
So, although couplings of levels 1 and 2 are different, characteristic sets of pairwise transition probabilities for them are equal.

5) The multiplication of signs $\lambda_i\tau_i \sigma_i$ for all $i=3,4,\ldots,N$ should be the same: either 1 or $-1$.

6)  In the sector of this model with $N$ interacting states, there are totally $2N-3$ independent continuous parameters:  $e$, $N-2$ couplings $g_{1i}$, and $N-2$ independent
level slopes.
Also, our model depends  on discrete sign parameters $\lambda_i$ and $\tau_i$ that, as we will show, describe phases with different behavior of transition probability matrices.


7) Constraints on couplings depend only on combinations
\be
\frac{g_{1i}^2}{b_i- b}, \quad \frac{g_{2i}^2}{b_i+b}.
\label{comb-1}
\ee
Hence, if we start from one case that satisfies (\ref{e-con})-(\ref{sign-con}) and change some $b_i$ continuously ($i\ge 3$),   then if we adjust $g_{1i}$ and $g_{2i}$ to keep (\ref{comb-1}) constant, i.e. conserving the matrix (\ref{skew}), we will find that such a deformed model is also solvable and has the same numerical values of transition probabilities.

Spectrum of this MLZ model as function of $t$ shows quickly growing with $N$ number of exact energy crossing points. The number of such points is the same as the number of zero couplings in the Hamiltonian (\ref{hal}), which is $1+(N-2)(N-3)/2$.
Figure~\ref{exact_crossing} shows examples of numerically calculated spectra in cases with $N=5,6,7,8,9,10$, which correspond to, respectively, $4,7,11,16,22,29$ exact adiabatic energy level crossing points that  appear generally at different values of $t$. These crossings do not seem  following from any known discrete symmetry, such as Kramers degeneracy. Moreover, the fact that $N$ can be an arbitrary integer (larger than 3), means that our models cannot   be generally represented as a direct product of independent smaller  MLZ systems or obtained by populating such systems by noninteracting bosons and fermions, as it was discussed in \cite{multiparticle}. In this sense, our model has  features of quantum integrable systems that were discussed in
\cite{yuzbashyan-LZ}.


Importantly, MLZ integrability means not only presence of numerous exact energy level crossings but also that we can describe dynamics of a system analytically, i.e., we can derive matrices of transition probabilities in
any sector of the model. In the next two sections, we provide  examples of derivation of transition probabilities in  the model (\ref{hal}). We postpone the proof of validity of integrability conditions in the model (\ref{hal}) for arbitrary $N$ to section~\ref{derive-sec}.

\section{Transition probabilities in the 5-state sector}
\label{test5-sec}

Consider the case with $N=5$ and  the Hamiltonian
\begin{equation}
 H=\left( \begin{array}{ccccc}
b t & 0 & g_{13} & g_{14} &  g_{15}\\
0 & -b t &  g_{23} &  g_{24} &  g_{25}\\
g_{13} & g_{23} &  b_3 t+e_3  & 0 &0\\
g_{14}  & g_{24} & 0  & b_4 t+e_4   &0\\
g_{15} & g_{25} & 0 & 0   & b_5 t+e_5
\end{array}\right).
\label{}
\end{equation}
Due to the constraint on sum of $g_{1i}^2$, at least one of the three slopes $b_i$ should be positive, and at least one should be negative. So, we will set $b_3>b_4>0>b_5$.

\begin{figure}[!htb]
\scalebox{0.1}[0.1]{\includegraphics{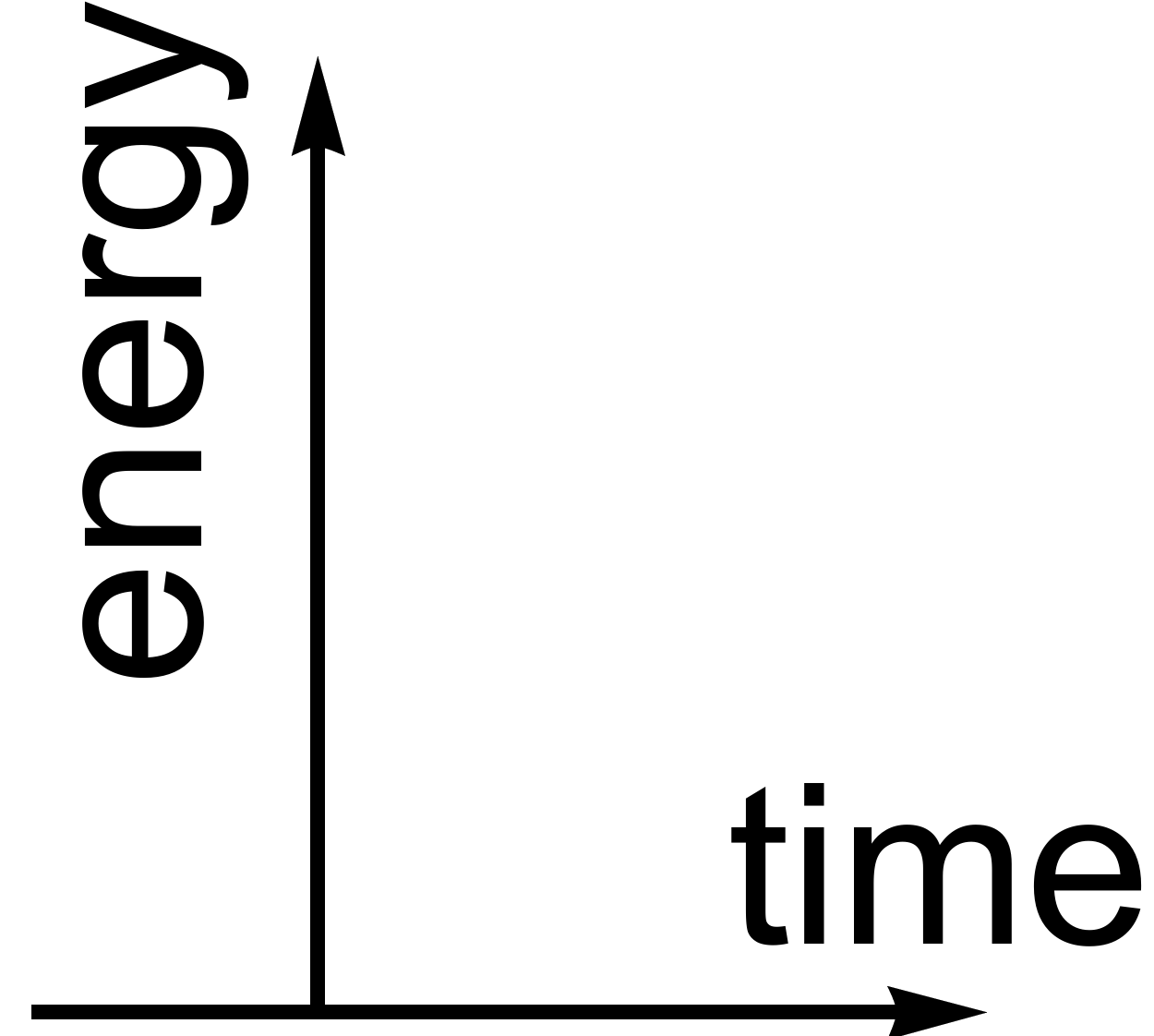}}~
\scalebox{0.4}[0.4]{\includegraphics{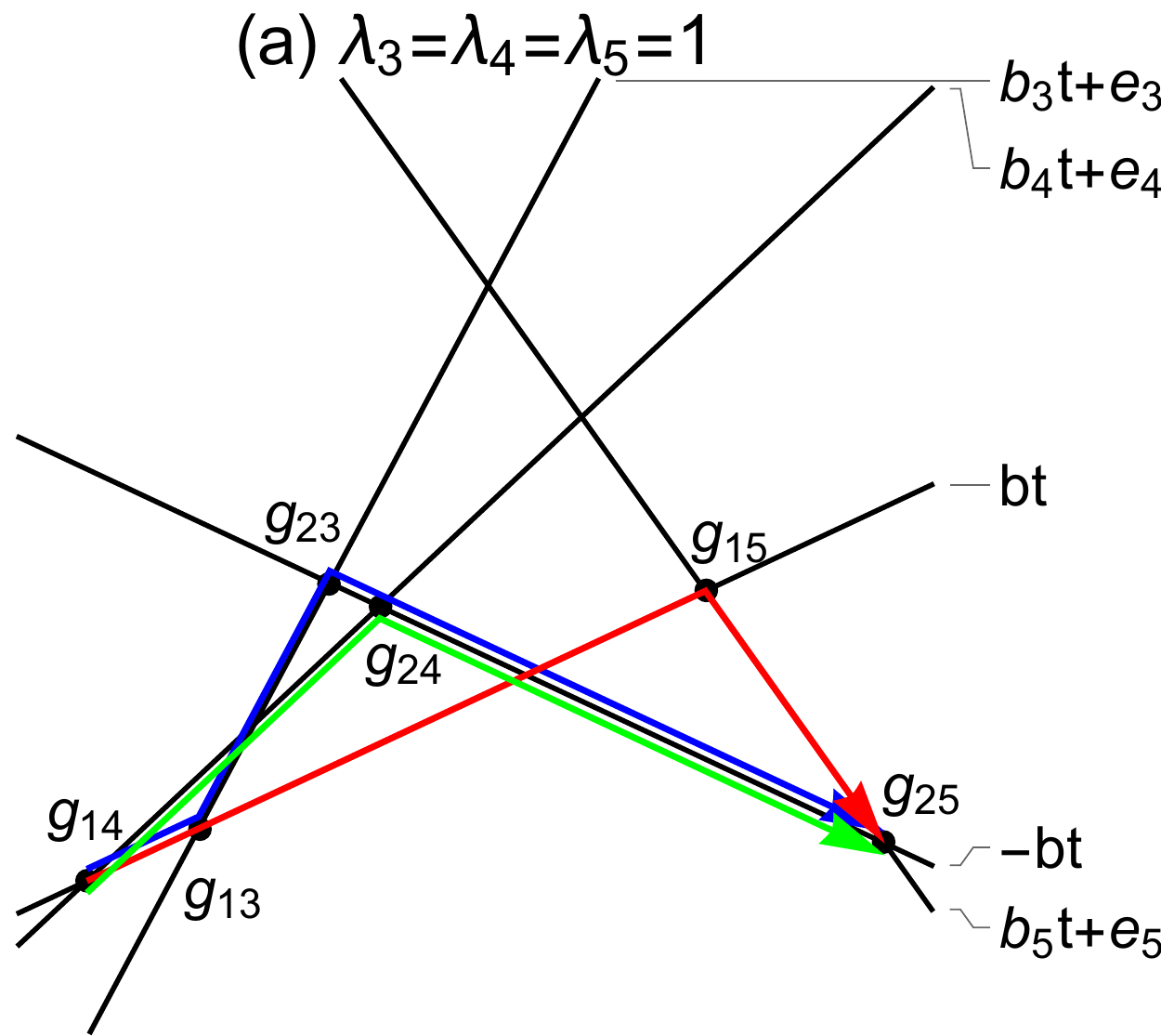}}~ ~ ~
\scalebox{0.4}[0.4]{\includegraphics{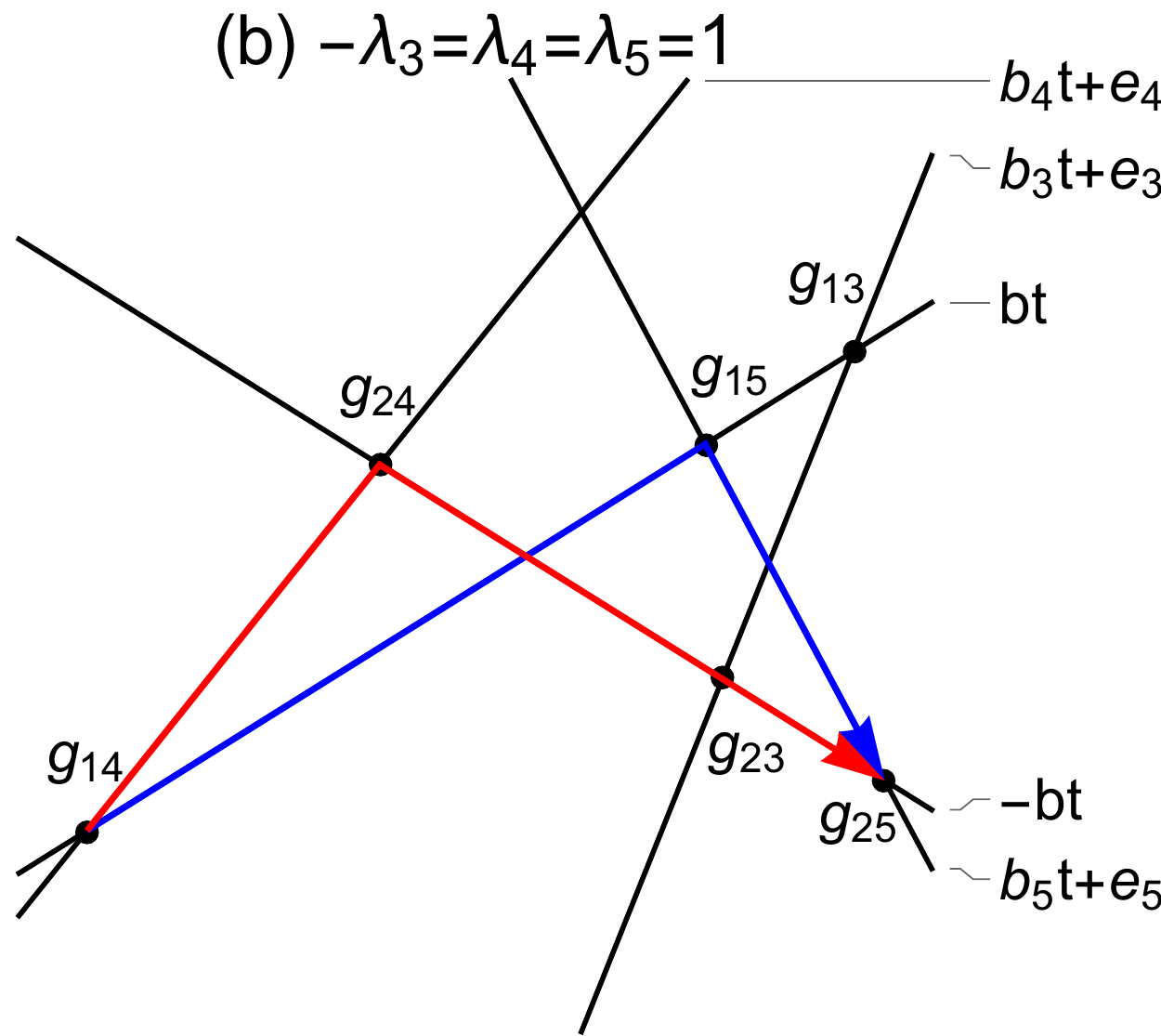}}~ ~ ~
\scalebox{0.4}[0.4]{\includegraphics{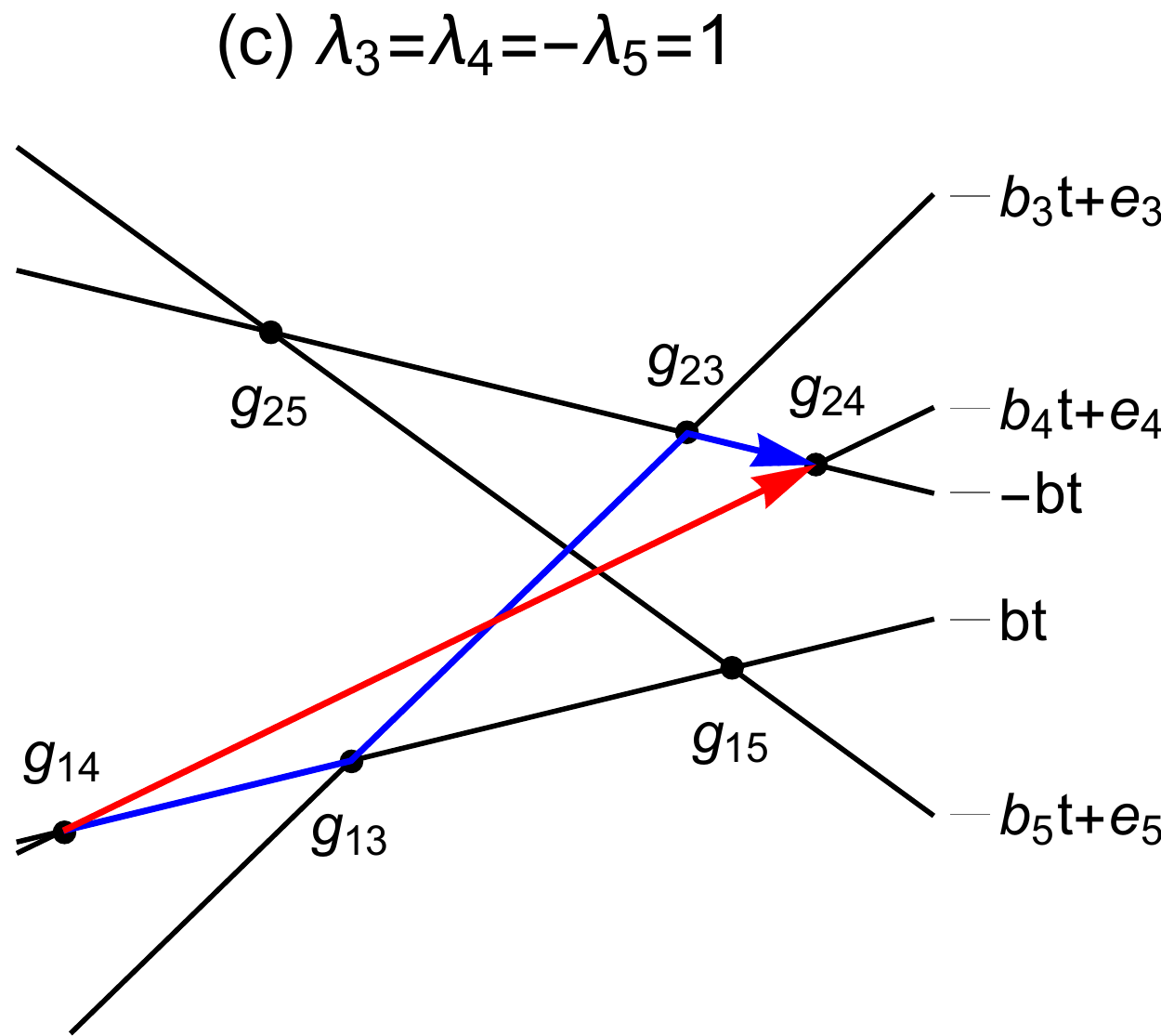}}
\caption{Diabatic level diagrams for three different phases with $N=5$:  (a) Phase 1 with $\lambda_3=\lambda_4=\lambda_5=1$; (b)  Phase 2 with  $-\lambda_3=\lambda_4=\lambda_5=1$; (c)  Phase 3 with  $\lambda_3=\lambda_4=-\lambda_5=1$. Other parameters are: $e=1$, $\rho=1$, $b=1$, $b_3=4$, $b_4=2$, and $b_5=-3$. (The units for all parameters are defined such that they are dimensionless. We will adopt this convention in all later figures.)}
\label{5-state_diagram}
\end{figure}

According to \cite{six-LZ,quest}, if ICs are satisfied then solution of the model is given by the semiclassical ansatz. To construct it,  one should first
draw the diabatic level diagram that shows time dependence of diabatic energy levels as functions of time, and mark nonzero pairwise couplings at corresponding level intersections.
Figure~\ref{5-state_diagram} shows topologically different examples of such diagrams for $N=5$. To obtain the specific transition probability from level $i$ at $t\rar -\infty$ to level $j$ at $t\rar +\infty$, one  should then find
all semiclassical trajectories that connect these states propagating only forward in time. For example, Fig.~\ref{5-state_diagram}(a) shows three such trajectories connecting levels with slopes $b_4$ and $b_5$. They are marked by blue, green, and red arrows.

One should then prescribe an amplitude to each trajectory.
If the trajectory goes through a crossing of two levels $m$ and $n$ with direct coupling $g_{mn}$ and does not change the level after the crossing then the amplitude gains the factor $\sqrt{p_{nm}}$, where $p_{nm}=e^{-2\pi g_{nm}^2/|\beta_{n}-\beta_{m}|}$. If the level changes, then the trajectory gains the factor $i\sgn(g_{mn})\sqrt{1-p_{nm}}$. 
The total amplitude of this trajectory is the product of all such factors that it gains from all crossing points through which it passes.
The final transition probability is the absolute value squared of the sum of amplitudes of all such trajectories connecting states $i$ and $j$. Elementary examples of such calculations can be found in \cite{six-LZ, four-LZ}.

Let us denote
\begin{align}\label{}
\rho\equiv \lambda_i\tau_i \sigma_i.
\end{align}
The time moments of diabatic level crossings with nonzero couplings are $t_{1i}=-e_i/(b_i-b)$ and  $t_{2i}=-e_i/(b_i+b)$.
Using expression (\ref{e-con}) for $e_i$, we find:
\begin{align}\label{times-1}
&t_{1i}=-\rho\tau_i\frac{ e}{b}\sqrt{\frac{b_i+b}{b_i-b}},\quad t_{2i}=-\rho\tau_i\frac{ e}{b}\sqrt{\frac{b_i-b}{b_i+b}}.
\end{align}
Note that $\rho$  enters as a common factor in (\ref{times-1}), so its choice does not affect transition probabilities because changing this sign only changes order of factors contributing to trajectory amplitudes.
So, below  we will set $\rho=1$.

In  Table \ref{table1}, we show all possible cases of signs of $\lambda_i$'s or equivalently $\tau_i$'s that can potentially lead to different behavior of transition probabilities. For convenience, we also provide the
corresponding orders of the time moments of diabatic level crossings that contribute to trajectory amplitudes. The  order of some of these time moments cannot be uniquely specified. For example, determining the order of $t_{13}$ and $t_{25}$ in Case 2  depends on the relative value of $\sqrt{(b_3+b)/(b_3-b)}$ and $\sqrt{(b_5-b)/(b_5+b)}$, which cannot be determined without knowing the relation between $|b_3|$ and $|b_5|$. However, this ambiguity does not influence trajectory amplitudes because all such undetermined cases are between $t_{1i}$ and $t_{2j}$ with $i\ne j$, i.e., such crossing points 
are not connected directly by any semiclassical trajectory. 
\begin{table}
\caption{\label{table1} Order of time moments of diabatic level crossings at different signs of $\lambda_i$ at $N=5$, $b_3>b_4>0>b_5$, and $\rho=\lambda_i\tau_i\sigma_i=1$}
\begin{ruledtabular}
\begin{tabular}{ccccc}
Cases &$(\lambda_3,\lambda_4,\lambda_5)$ & $(\tau_3,\tau_4,\tau_5)$ & Order of time moments\\
\hline
1&$(1,1,1)$ &$(1,1,-1)$ & $t_{25}>t_{15}>0>t_{24}>t_{23}>t_{13}>t_{14}$\\
2&$(-1,1,1)$ &$(-1,1,-1)$ & $t_{13},t_{25}>t_{15},t_{23}>0>t_{24}>t_{14}$\\
3&$(1,-1,1)$ &$(1,-1,-1)$ & $t_{14}, t_{25}>t_{15},t_{24}>0>t_{23}>t_{13}$\\
4&$(1,1,-1)$ &$(1,1,1)$ & $0>t_{24}>t_{23}>-e/b>t_{13}>t_{14}$, and $0>t_{15}>-e/b>t_{25}$\\
5&$(-1,-1,1)$ &$(-1,-1,-1)$ & $t_{14}>t_{13}>e/b>t_{23}>t_{24}>0$, and $t_{25}>e/b>t_{15}>0$\\
6&$(-1,1,-1)$ &$(-1,1,1)$ & $t_{13}>t_{23}>0>t_{15},t_{24}>t_{14}, t_{25}$\\
7&$(1,-1,-1)$ &$(1,-1,1)$ & $t_{14}>t_{24}>0>t_{15},t_{23}>t_{13},t_{25}$\\
8&$(-1,-1,-1)$ &$(-1,-1,1)$ & $t_{14}>t_{13}>t_{23}>t_{24}>0>t_{15}>t_{25}$\\
\end{tabular}
\end{ruledtabular}
\end{table}

\begin{figure}[!htb]
\scalebox{0.5}[0.5]{\includegraphics{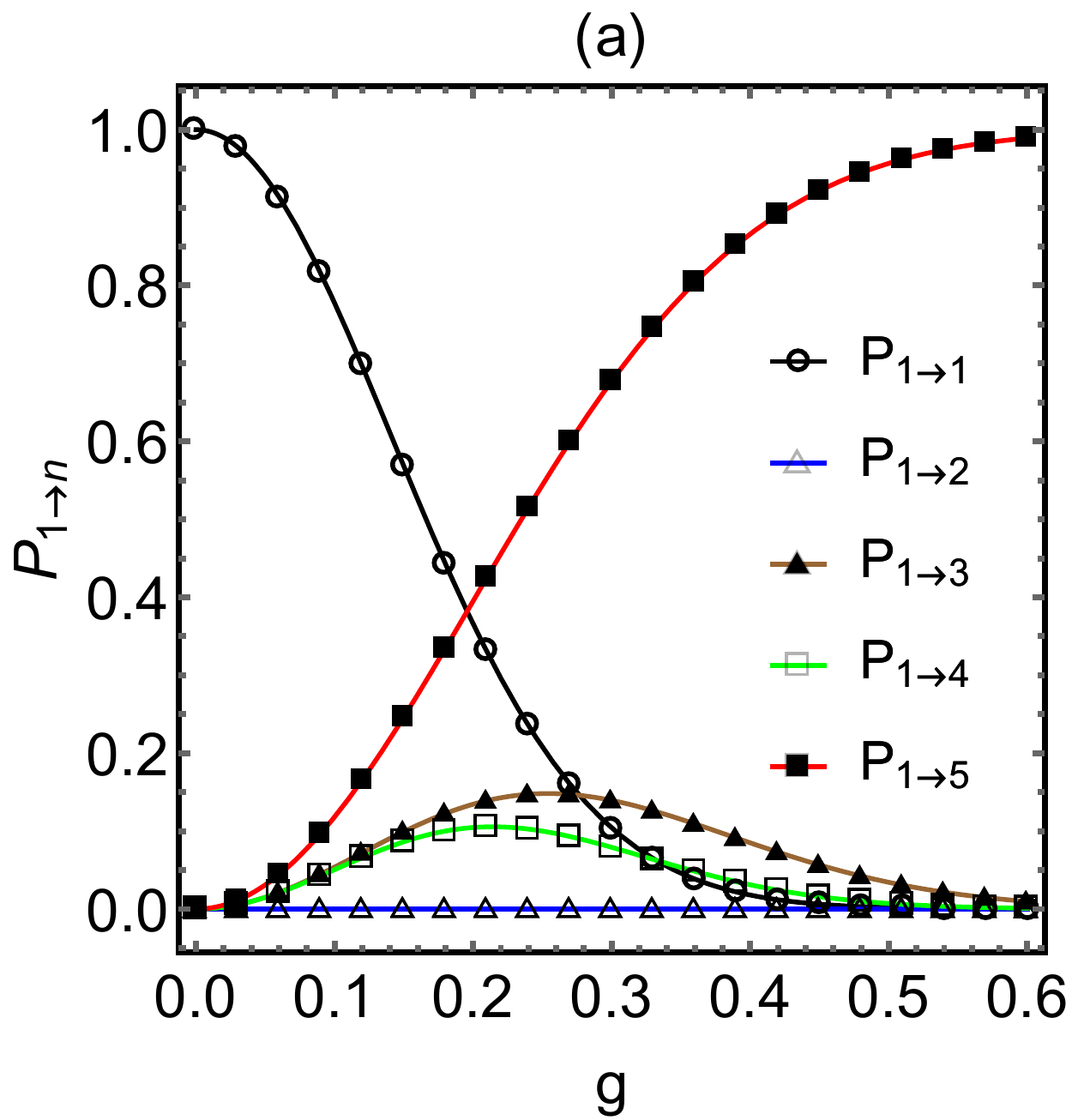}}~ ~ ~
\scalebox{0.5}[0.5]{\includegraphics{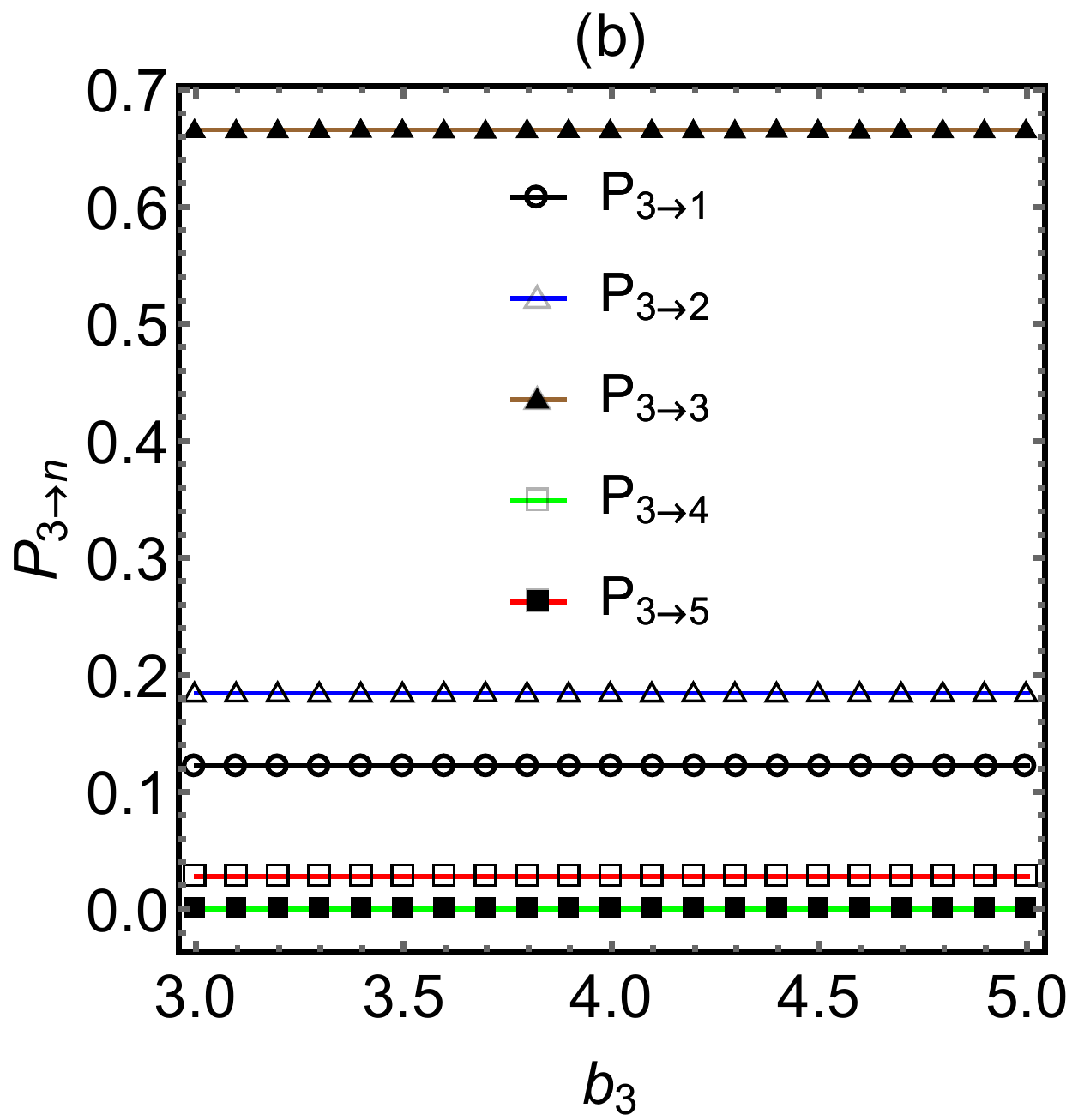}}
\caption{Transition probabilities in a 5-state model. Solid curves are predictions of Eq.~\eqref{P_case_1} and discrete points are results of numerically calculated transition probabilities for evolutions from $t=-500$ to $t=500$, with a time step $dt=0.005$. (a) Transition probabilities from level 1 to all diabatic states as functions of coupling $g$. Parameters are: $e=1$, $\rho=1$, $b=1$, $b_3=4$, $b_4=2$, $b_5=-2.5$, and $\lambda_3=\lambda_4=\lambda_5=1$; $g_{13}=g \sqrt{b_3/b - 1 }$, $g_{14}=g \sqrt{ b_4/b - 1 }$, $g_{15}=\sqrt 2 g \sqrt{1 - b_5/b}$;  $e_i$ and $g_{2i}$ are determined by  constraints \eqref{e-con} and \eqref{g-con-2}, respectively. (b) Transition probabilities from level 3 to all diabatic states as functions of $b_3$. Couplings $g_{13}$ and $g_{23}$ are chosen such that  $g_{13}^2/(b_3- b)$ and $g_{23}^2/(b_3+b)$ are constants, $g=0.18$, and all other parameters are the same as in (a). The agreement between theory and numerics is excellent.
}\label{5-state_numerics}
\end{figure}

 Although there are eight possibilities listed in Table \ref{table1}, we found that these cases group into only three different phases that correspond to different transition probability matrices.  This is because Cases $n$ and $9-n$ (1 and 8, 2 and 7, etc.) have opposite choices of signs of $\lambda_i$'s, and their orders of time moments are opposite to each other. Figure~\ref{5-state_diagram} shows the diabatic level diagrams for the three different phases. They have different  patterns or path interference, so it is expected that the corresponding transition probability matrices are also different.
To write these matrices explicitly, let us define
\begin{align}\label{piqi}
&p_{i}=e^{-\frac{2\pi g_{1i}^2}{|b-b_i|}},\quad q_{i}=1-p_{i}, \quad i=3,4, \ldots, N,
\end{align}
and note that using the constraint (\ref{g-con1}) for $N=5$ we have $p_5=p_3p_4$. We find then:

\underline{\bf Phase 1} corresponds to Cases 1 and 8 in Table~\ref{table1}. Calculations based on the semiclassical ansatz lead to the transition probability matrix:
\begin{align}\label{}
& P_{Case~1}=\hat P_{Case~8}=\left( \begin{array}{ccccc}
p_{3}^2p_{4}^2 & 0 &  p_{3}p_{4}q_{3} & p_3^2p_4q_4 &  q_5\\
0 & p_{3}^2p_{4}^2  & q_3 &  p_3 q_4 &  p_3p_4q_5\\
p_{3}p_{4}q_{3} & q_3  &  p_3^2  & p_3q_3q_4 &0\\
p_3^2p_4q_4  &  p_3 q_4 & p_3q_3q_4 &  (p_4+q_3q_4)^2  &0\\
q_5 & p_3p_4q_5 & 0 & 0   & p_3^2p_4^2
\end{array}\right).
\label{P_case_1}
\end{align}

\underline{\bf Phase 2} corresponds to Cases 2, 3, 6 and 7:

\begin{align}
\label{P_case_2}
& P_{Case~2}=\hat P_{Case~7}=\left( \begin{array}{ccccc}
           (p_3 p_4-q_3 q_4)^2 & p_4q_3^2 &  p_3q_3 &   p_4q_4 &  p_3q_5\\
           p_4q_3^2 & p_3^2p_4^2 &  p_3p_4q_3 & q_4 &  p_3p_4q_5 \\
            p_3q_3 &  p_3p_4q_3 & p_3^2 & 0 &  q_3q_5  \\
           p_4q_4 & q_4  & 0 & p_4^2 & 0\\
           p_3q_5 &  p_3p_4q_5 &  q_3q_5  & 0 & p_3^2p_4^2
         \end{array}\right).
\end{align}

\begin{align}
\label{case2-2}
& P_{Case~3}=\hat P_{Case~6}=\left( \begin{array}{ccccc}
          (p_3p_4-q_3q_4)^2 & p_3 q_4^2 &  p_3q_3 &  p_4q_4 &  p_4q_5 \\
           p_3 q_4^2 & p_3^2p_4^2 & q_3 &  p_3p_4q_4 &   p_3p_4q_5 \\
           p_3q_3  & q_3 & p_3^2 & 0 & 0 \\
           p_4q_4 &  p_3p_4q_4 & 0 & p_4^2 &  q_4q_5\\
           p_4q_5  &  p_3p_4q_5  & 0 &  q_4q_5 & p_3^2p_4^2
         \end{array}
\right).
\end{align}
The difference between (\ref{P_case_2}) and (\ref{case2-2}) is merely in renaming indexes of some of the levels.

\underline{\bf Phase 3} corresponds to Cases 4 and 5:
\begin{align}\label{}
& P_{Case~4}=\hat P_{Case~5}=\left( \begin{array}{ccccc}
             p_3^2p_4^2 & q_5^2 &  p_3p_4q_3 &  p_3^2 p_4q_4 & p_3p_4q_5 \\
             q_5^2 & p_3^2p_4^2 &   p_3p_4q_3 &  p_3^2 p_4q_4 &   p_3p_4q_5 \\
              p_3p_4q_3 &  p_3p_4q_3 & p_3 & p_3q_3q_4 &  q_3q_5 \\
             p_3^2 p_4q_4   &   p_3^2 p_4q_4   &  p_3q_3q_4 & (p_4+q_3q_4)^2 &  p_3q_4q_5 \\
             p_3p_4q_5 &  p_3p_4q_5 &  q_3q_5  &   p_3q_4q_5 & p_3^2p_4^2 \\
           \end{array}\right).
\label{P_case_4}
\end{align}
 Figure~\ref{5-state_numerics}
shows comparison between analytical predictions of  Eq.~\eqref{P_case_1} and results of numerical simulations for Phase 1 with $\lambda_3=\lambda_4=\lambda_5=1$. Agreement with  numerics is excellent.

Here we observe a common feature of transition probability matrices (\ref{P_case_1})-(\ref{P_case_4}): all of them are symmetric, i.e., $P^{ab}=P^{ba}$. This fact has an explanation. The semiclassical ansatz predicts that transition probabilities are
independent of parameter $e$. Formally, the case $e=0$ is not described by ICs but, since level slopes are all different, transition probabilities behave continuously upon variation of all parameters. So, setting $e=0$ does not change predictions (\ref{P_case_1})-(\ref{P_case_4}). In this case, our model describes situation when all diabatic levels cross in one point. Moreover, diabatic states split in two groups with zero direct couplings within states of the same group. The symmetry of the transition probability matrices in such models has been proved rigorously in \cite{cross}. Since it is valid at $e=0$, by continuity it must be valid for $e\ne 0$ if the semiclassical ansatz is valid.

\section{Models with  $N>5$}
\label{testN-sec}

\begin{figure}[!htb]
\scalebox{0.43}[0.43]{\includegraphics{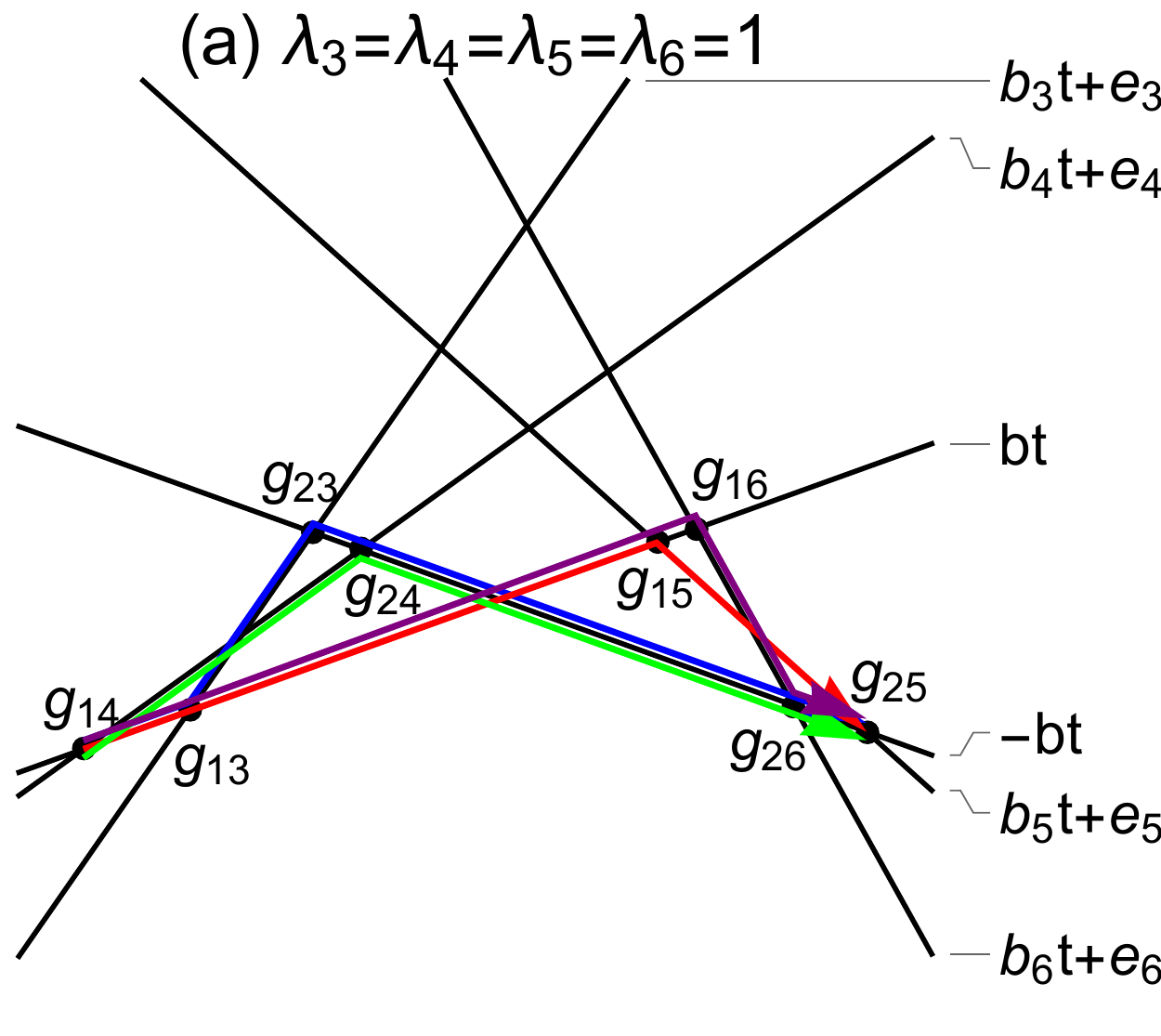}}~ ~ ~
\scalebox{0.43}[0.43]{\includegraphics{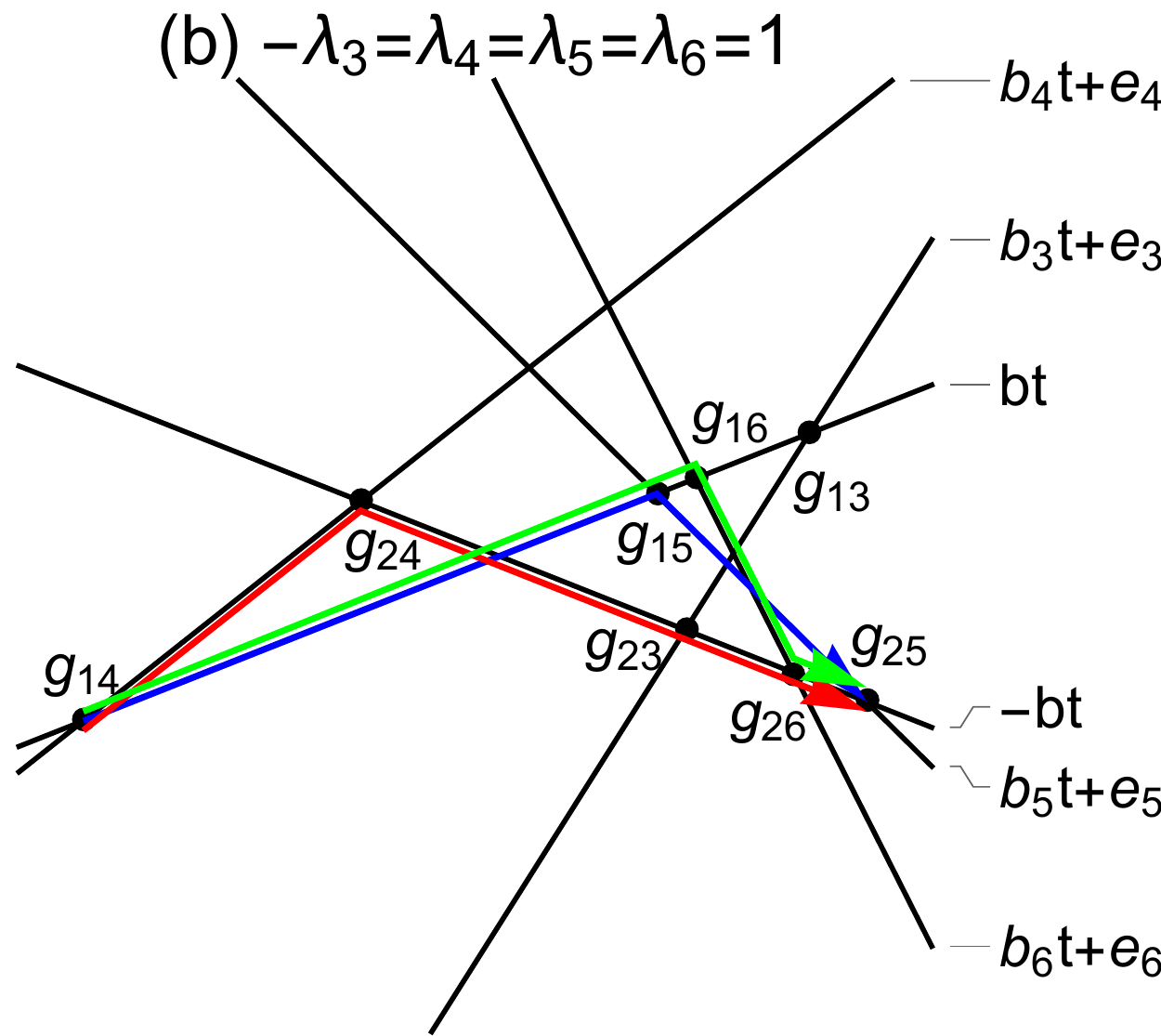}}~ ~ ~
\scalebox{0.43}[0.43]{\includegraphics{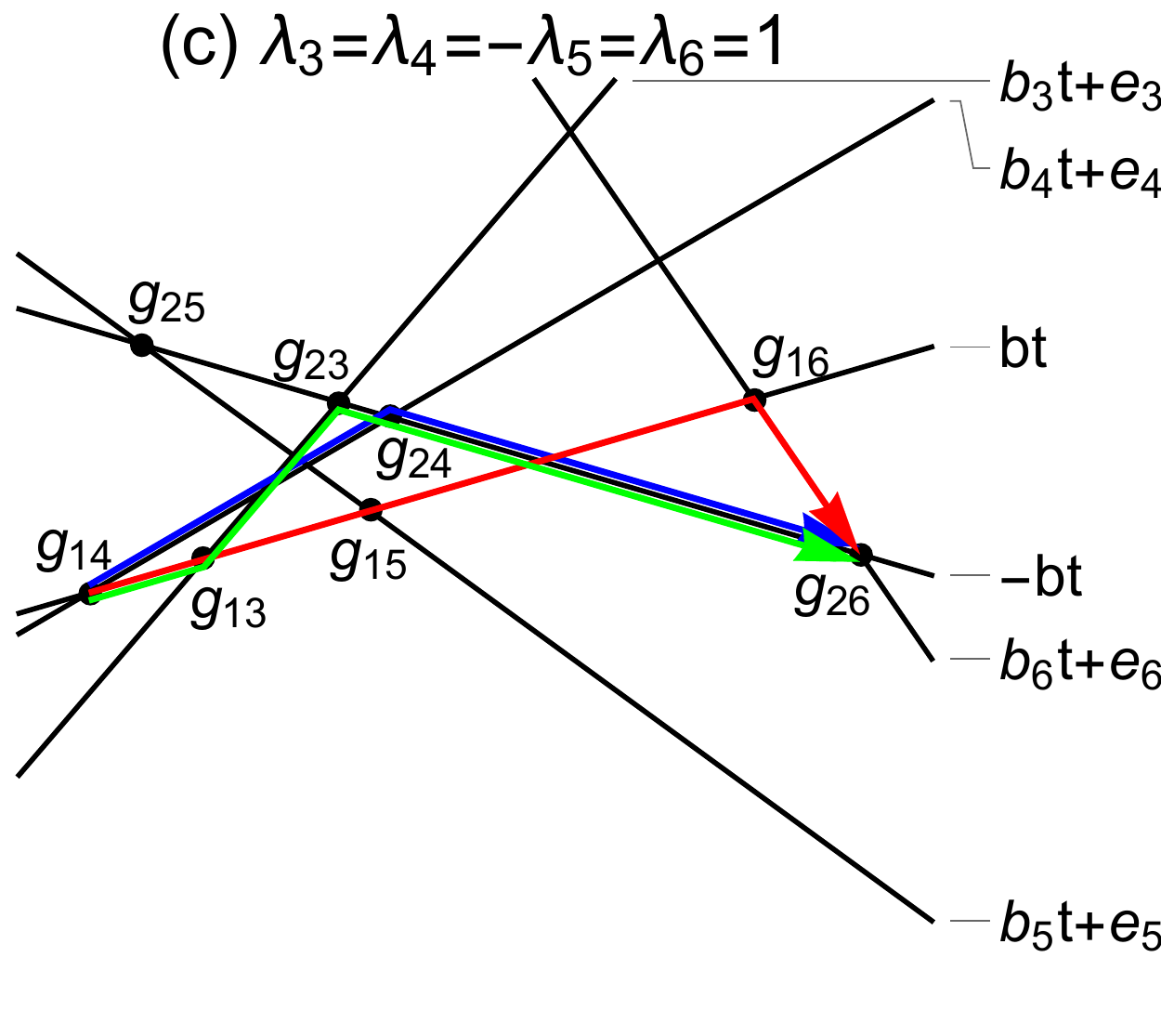}}~ ~ ~\\ ~ \\
\scalebox{0.1}[0.1]{\includegraphics{E-t_axes}}~
\scalebox{0.43}[0.43]{\includegraphics{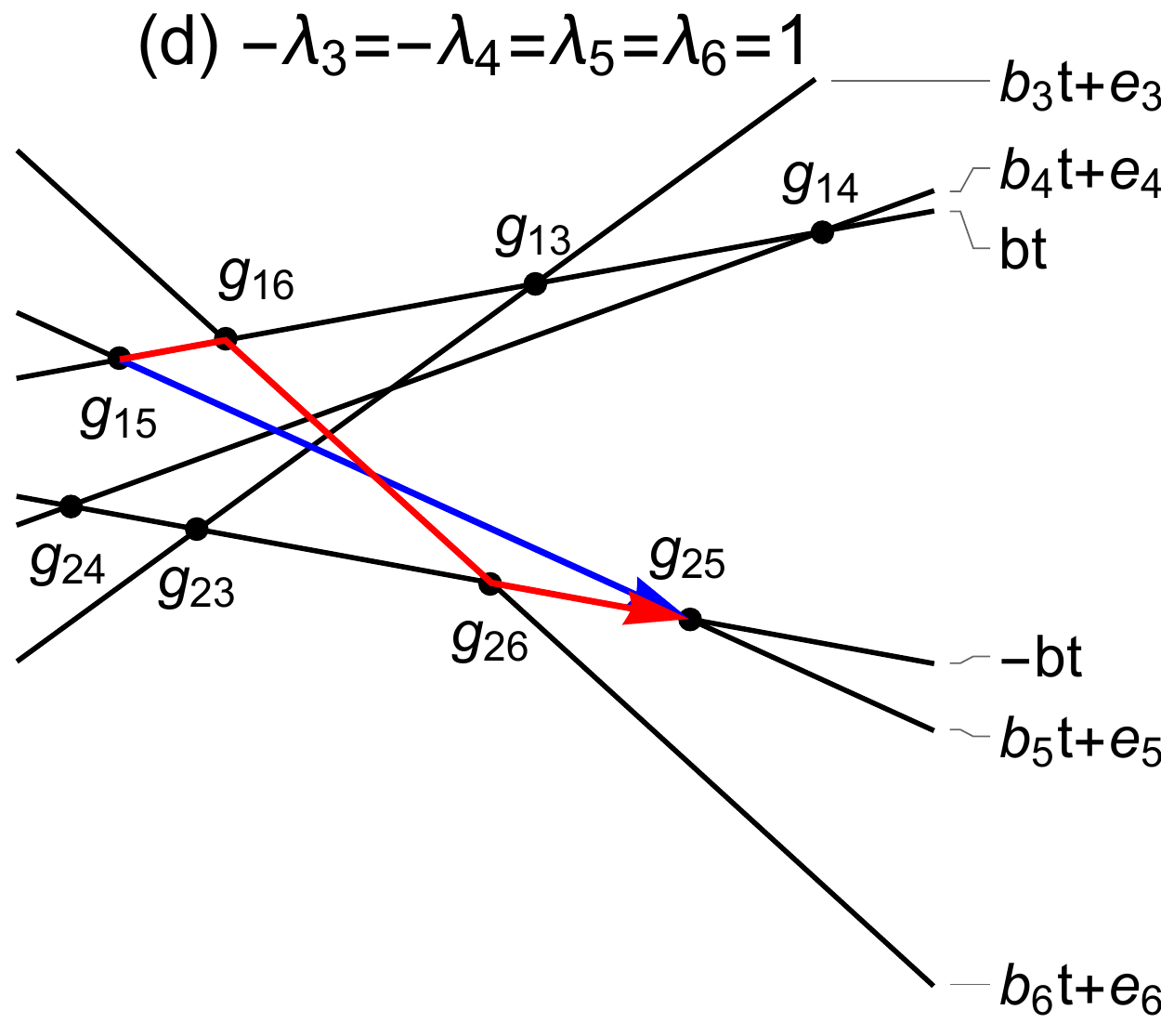}}~ ~ ~
\scalebox{0.43}[0.43]{\includegraphics{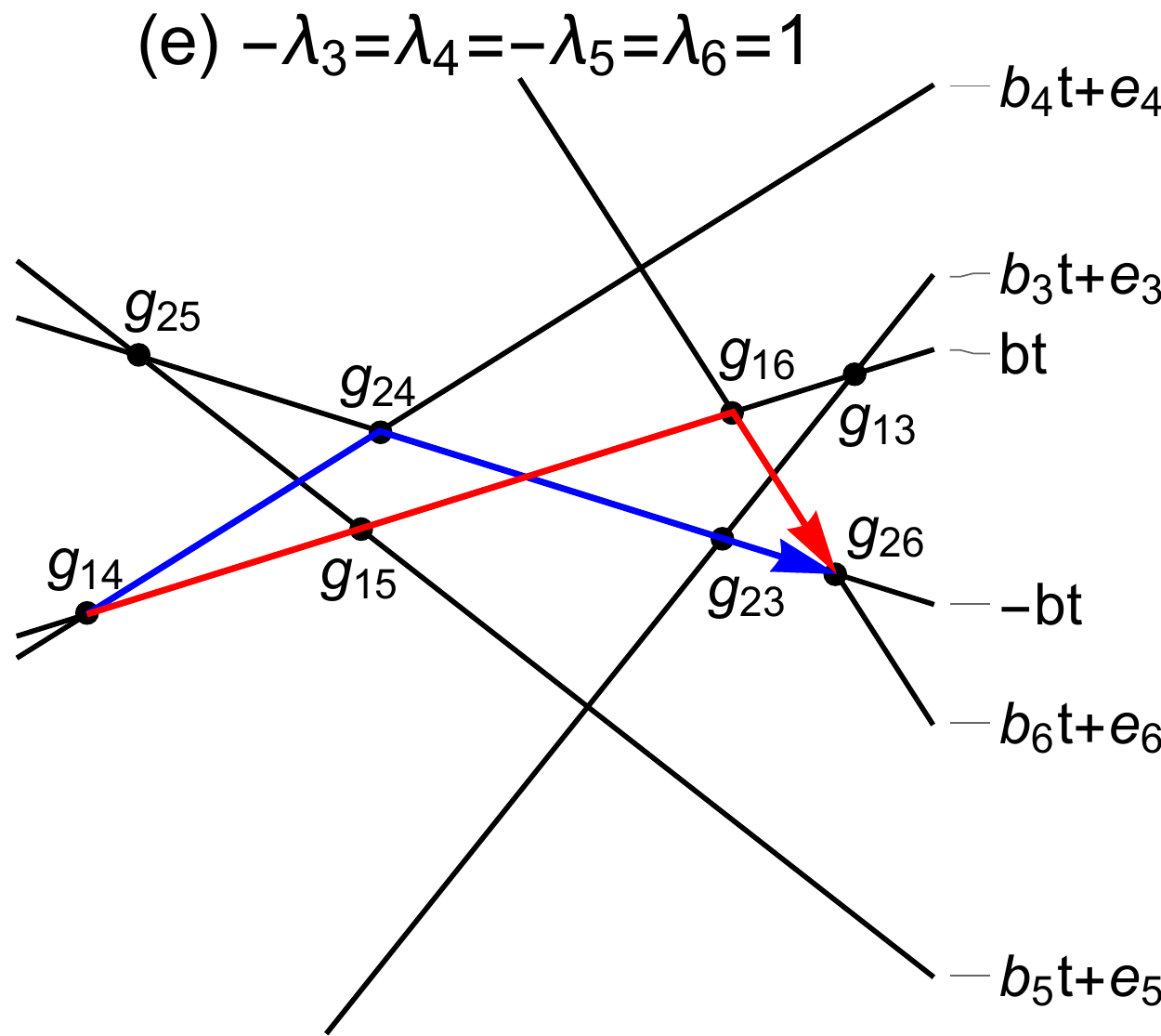}}
\caption{Diabatic level diagrams of 6-state models for different phases: (a) $\lambda_3=\lambda_4=\lambda_5=\lambda_6=1$; (b)  $-\lambda_3=\lambda_4=\lambda_5=\lambda_6=1$; (c) $\lambda_3=\lambda_4=-\lambda_5=\lambda_6=1$;  (d) $-\lambda_3=-\lambda_4=\lambda_5=\lambda_6=1$; (e) $-\lambda_3=\lambda_4=-\lambda_5=\lambda_6=1$. Other parameters are: $e=1$, $\rho=1$, $b=1$, $b_3=4$, $b_4=2$, $b_5=-2.5$, and $b_5=-5$. 
}\label{6-state_diagram}
\end{figure}
\begin{figure}[!htb]
\scalebox{0.5}[0.5]{\includegraphics{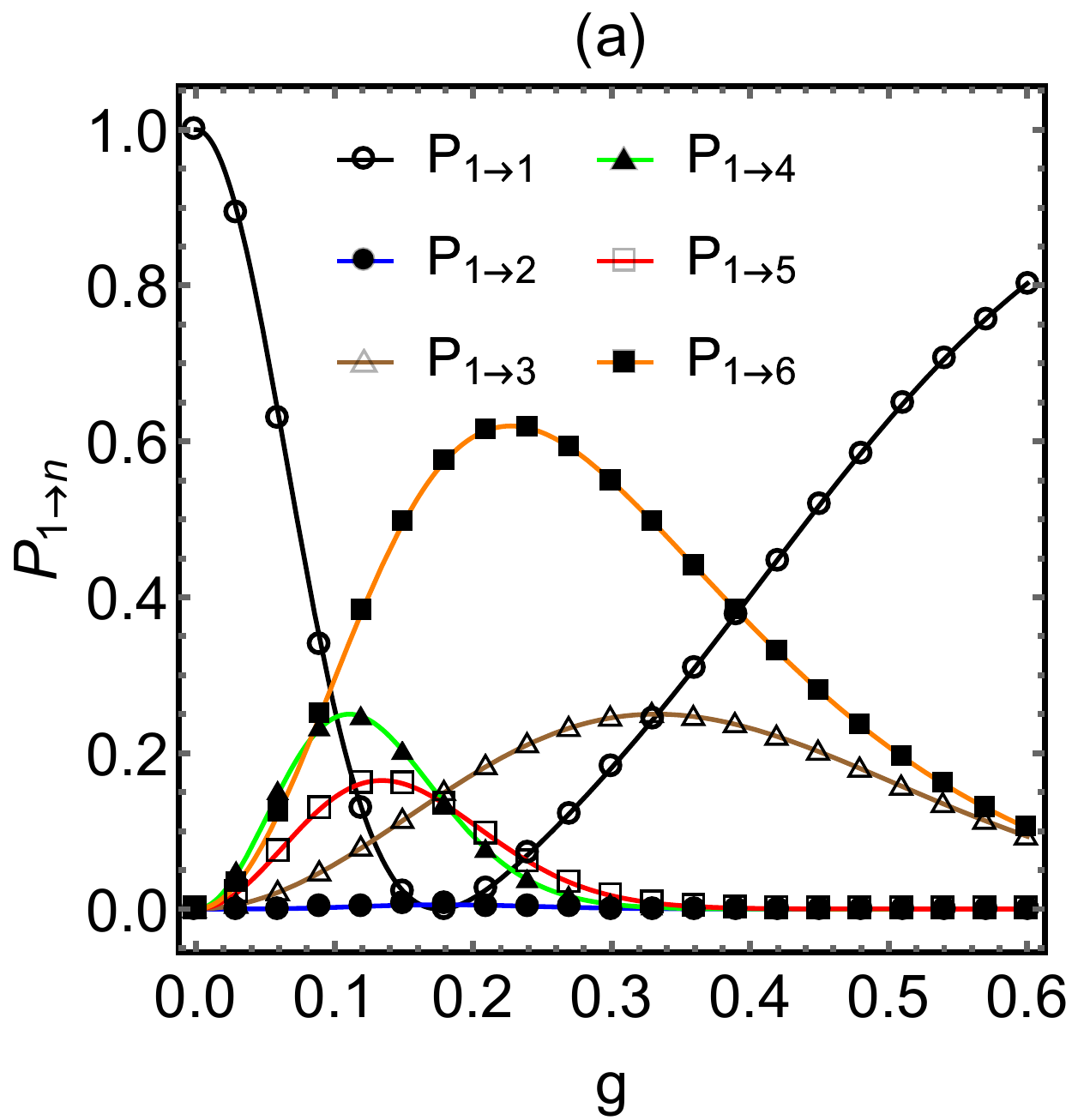}}~ ~ ~
\scalebox{0.5}[0.5]{\includegraphics{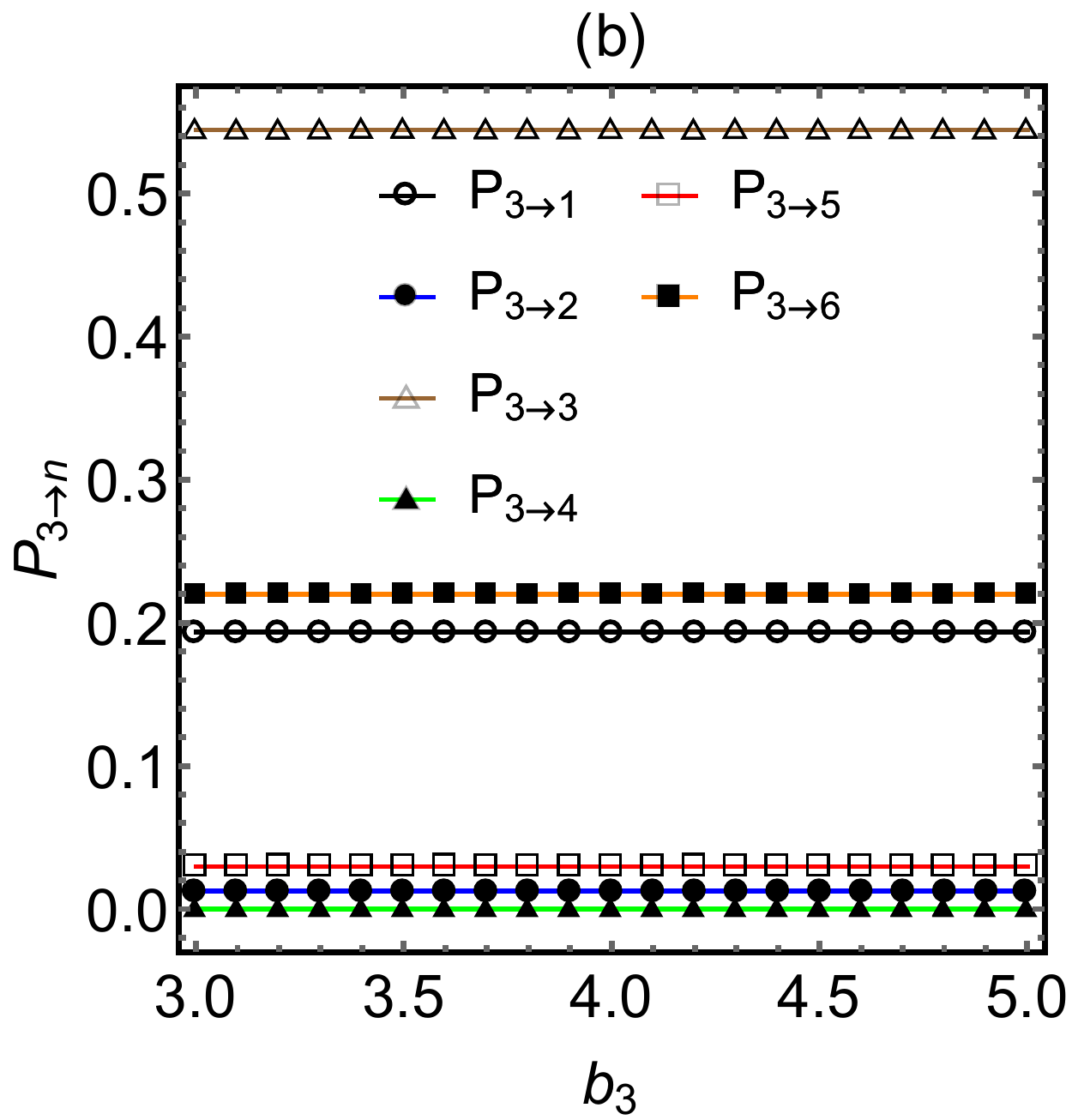}}
\caption{Transition probabilities in a 6-state model. Solid curves are predictions of Eq.~\eqref{P1n_6-state} and discrete points are results of numerical calculations for evolutions from $t=-500$ to $t=500$, with a time step $dt=0.005$. (a) Transition probabilities from level 1 to all diabatic states as functions of coupling $g$. Parameters
are: $e=1$, $\rho=1$, $b=1$, $b_3=4$, $b_4=2$, $b_5=-2.5$, $b_6=-5$; $-\lambda_3=\lambda_4=\lambda_5=\lambda_6=1$;  $g_{13}=g \sqrt{b_3/b - 1 }$, $g_{14}=3g \sqrt{b_4/b - 1 }$, $g_{15}=2g \sqrt{1 - b_5/b}$, $g_{16}=\sqrt 6 g \sqrt{1- b_6/b}$; $e_i$ and $g_{2i}$ are determined by  constraints (\ref{b-con})-(\ref{sign-con}). (b) Transition probabilities from level 3 to all diabatic states as functions of $b_3$. Couplings $g_{13}$ and $g_{23}$ are chosen such that the quantities $g_{13}^2/(b_3- b)$ and $ g_{23}^2/(b_3+b)$ stay constant, $g=0.22$, and all other parameters are the same as in (a).
The agreement between theory and numerics is excellent.
}\label{6-state_numerics}
\end{figure}

The number of different phases is quickly growing with $N$.
Analogous studies for the sector with $N=6$ predict already five phases with different patterns of path interference, as we  show in Fig.~\ref{6-state_diagram}.
We performed a number of numerical tests for a few arbitrarily chosen cases and always found excellent agreement with predictions of the semiclassical ansatz. For example,
Fig.~\ref{6-state_numerics} shows results of numerical tests for transition probabilities 
in the phase with $-\lambda_3=\lambda_4=\lambda_5=\lambda_6=1$. In this case,  transition probabilities from level 1 and from level 3  to all other states read:
\begin{align}\label{P1n_6-state}
&P_{1\rar 1}=(p_3p_4-q_3q_4)^2,\quad P_{1\rar2}=p_4q_3^2, \quad P_{1\rar 3}=p_3q_3,\quad
P_{1\rar 4}=p_4q_4 ,\quad P_{1\rar 5}=p_3p_6q_5, \quad P_{1\rar 6}=p_3q_6,\nn\\
&P_{3\rar 1}= p_3 q_3,\quad   P_{3\rar 2}= p_3p_4q_3,\quad P_{3\rar 3} =p_3^2,
\quad P_{3\rar 4}=0,\quad P_{3\rar 5}= p_6q_3q_5,\quad P_{3\rar 6}=q_3q_6.
\end{align}
In Fig.~\ref{6-state_numerics}(b), we checked numerically that transition probabilities do not change if we vary $b_3$, $g_{13}$ and $g_{23}$ but keep $g_{13}^2/(b_3- b)$ and $g_{23}^2/(b_3+b)$ unchanged, i.e., this figure confirms that deformations that preserve ICs with invariant matrix (\ref{skew}) keep the transition probabilities the same.

\begin{figure}[!htb]
\scalebox{0.5}[0.5]{\includegraphics{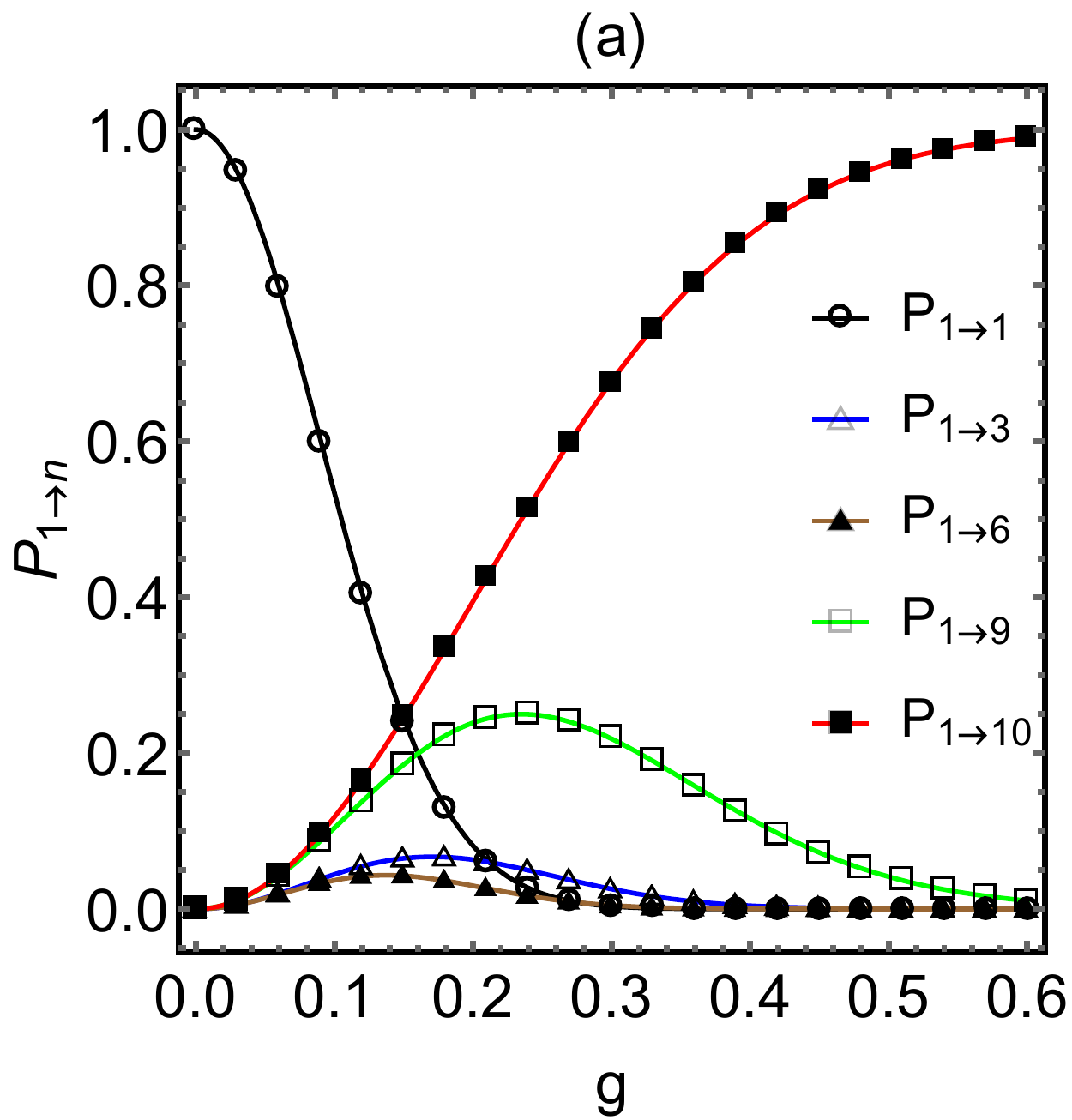}}~ ~ ~
\scalebox{0.515}[0.515]{\includegraphics{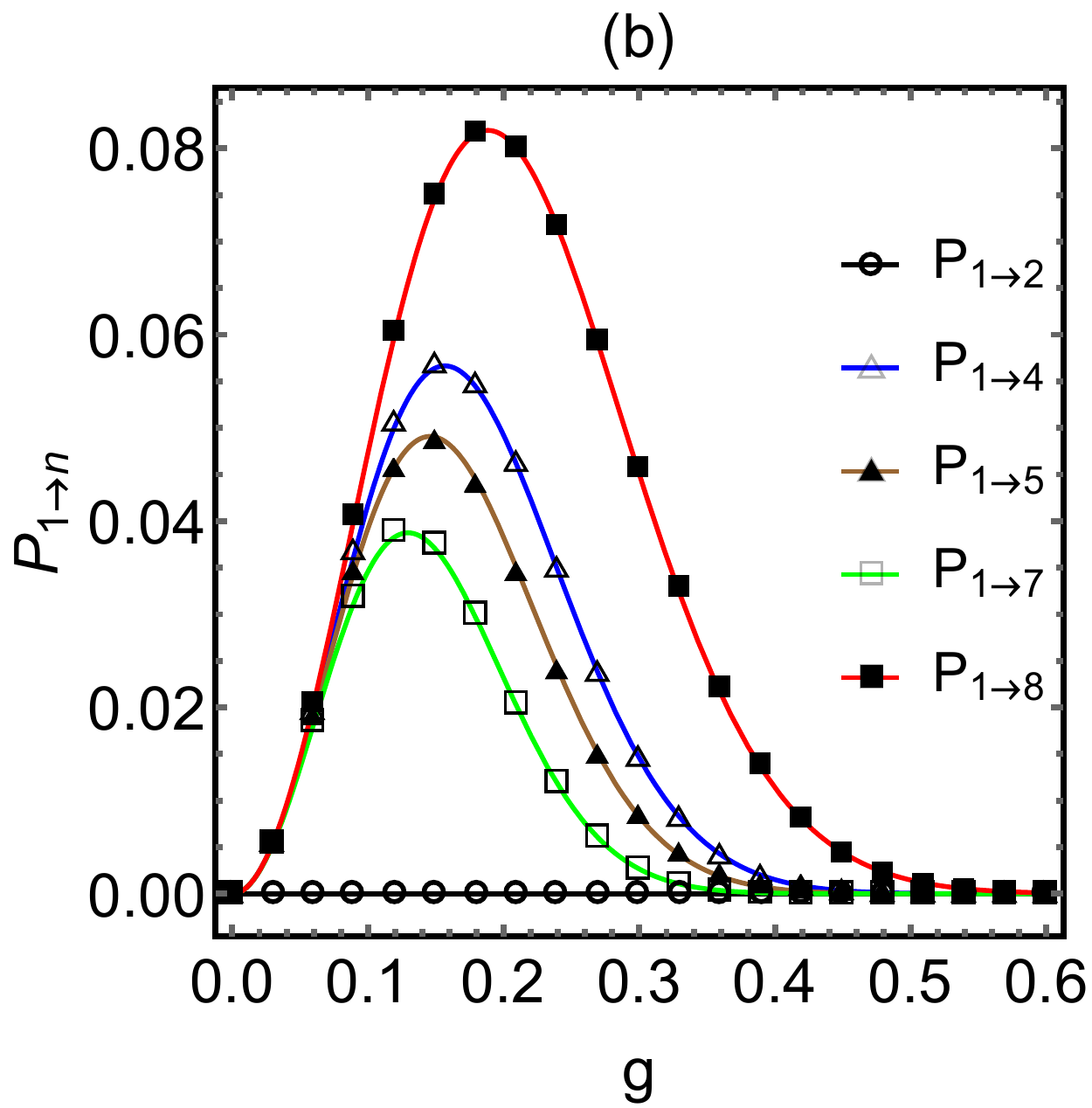}}
\caption{Transition probabilities from level 1 to some diabatic states in a 10-state model. Solid curves are predictions of Eq.~\eqref{P1n_10-state} and discrete points are results of numerical calculations for evolutions from $t=-100$ to $t=100$, with a time step $dt=0.01$.  Parameters are:  $e=1$, $\rho=1$, $b = 1$, $b_3 = 7$, $b_4 = 5$, $b_5 = 4$, $b_6 = 2.5$, $b_7 = 2$, $b_8 = -1.5$, $b_9 = -3$, $b_{10} = -3.5$. All $\lambda_i$'s are 1;   $g_{1i}=g \sqrt{\left|b_i/b - 1\right|}$ for $i=3,4,5,6,7,8$ and $g_{1i}=\sqrt 2 g \sqrt{\left| b_i/b - 1\right|}$ for $i=9,10$; $e_i$ and $g_{2i}$ are determined by constraints \eqref{e-con} and \eqref{g-con-2}.
}\label{10-state_numerics}
\end{figure}

Similarly, we looked at a few cases with  larger $N$. For example,  the 10-state model, whose slopes satisfy $b_3>b_4>b_5>b_6>b_7>b>0>-b>b_8>b_9>b_{10}$ and $\lambda_i=1$ for all $i$, has the following
probabilities of transitions from level 1 to all other states:

\begin{align}\label{P1n_10-state}
&P_{1\rar 1}=p_3 p_4 p_5 p_6p_7p_8p_9p_{10},\quad P_{1\rar2}=0, \quad P_{1\rar 3}=p_8 p_9 p_{10} q_3,\quad
P_{1\rar 4}=p_3 p_8 p_9 p_{10} q_{4},\quad P_{1\rar 5}=p_3 p_4 p_8 p_9 p_{10} q_5, \nn\\
&P_{1\rar 6}=p_3 p_4 p_5 p_8 p_9 p_{10} q_6,\quad P_{1\rar 7}=p_3 p_4 p_5 p_6 p_8 p_9 p_{10} q_7,\quad
P_{1\rar 8}= p_9 p_{10} q_8,\quad P_{1\rar 9} =p_{10} q_9, \quad P_{1\rar 10}=q_{10},
\end{align}
Figure~\ref{10-state_numerics} provides numerical check of this prediction, again confirming the theory perfectly. So, the semiclassical ansatz applies to our model  as to all other MLZ systems that satisfy ICs. We will explain this fact and the emergence of different phases in later section~\ref{mtlz-sec}.


\section{Derivation of model (\ref{hal}) from Integrability Conditions}
\label{derive-sec}
ICs in MLZ theory read \cite{quest}:

(i) All closed paths in the diabatic level diagram should enclose zero areas. Here,  the closed path means that it goes along diabatic levels to produce a closed loop such that switching levels along this path is allowed only at level crossings with nonzero couplings. The area inside such a closed path is the sum of areas of enclosed plaquettes in the diabatic level diagram counting
clockwise and counterclockwise enclosed areas with opposite signs.

(ii) For pairwise level crossings, if the direct coupling between two crossing diabatic levels is zero, there must be an exact energy level crossing near this point in the spectrum of the Hamiltonian for sufficiently small but finite values of nonzero couplings.


Let us denote the time moment of the crossing between levels $i$ and $j$  as $t_{ij}$.
To satisfy condition (i), it suffices to consider the smallest loops that have 4 vertices formed by 4 diabatic levels. Such a loop can be marked as $1\rar i\rar 2\rar j\rar 1$, with $i,j=3,4,\ldots,N$ and $i\ne j$. This  means that 
the loop starts at the crossing of levels 1 and $i$ at $t_{1i}$, goes along level $i$ to the crossing at $t_{2i}$, switches to level $2$ and goes to the crossing at $t_{2j}$, switches to level $j$ and goes to the crossing at $t_{1j}$, and finally returns to level $1$ and goes back to the crossing at $t_{1i}$. The area of such a loop can be conveniently calculated using the shoelace formula \cite{wiki_shoelace}, stating that the area of any $n$-sided polygon in the $xOy$ plane can be expressed through coordinates of its $n$ vertices $(x_1,y_1)$, $(x_2,y_2)$, $\ldots$, $(x_n,y_n)$ as:
\begin{align}\label{}
A=\frac{1}{2}\sum_{k=1}^{n}\det\left(\begin{array}{cc}
x_k&x_{k+1}\\
y_k&y_{k+1}
\end{array}\right), \quad  x_{n+1}=x_1, \quad y_{n+1}=y_1.
\end{align}
This formula  can be safely applied to self-intersecting polygons that encounter in our case.
Time-energy coordinates  of the diabatic level crossings at $t_{1i}$ and at $t_{2i}$ can be written as:
\begin{align}\label{}
&t_{1i}: \, \left(-\frac{e_i}{b_i-b}, -\frac{be_i}{b_i-b} \right),\quad
t_{2i}: \, \left(-\frac{e_i}{b_i+b}, \frac{be_i}{b_i+b} \right).
\end{align}
Thus, from the shoelace formula the area of the loop $1\rar i\rar 2\rar j\rar 1$ reads:
\begin{align}\label{}
A=b \left(\frac{e_i^2}{b^2 - b_i^2} - \frac{e_j^2}{b^2 - b_j^2}\right).
\end{align}
Requiring $A=0$ gives a constraint between $e_i$ and $e_j$:
\begin{align}\label{ICi_eq}
\frac{|e_i|}{|e_j|}=\sqrt{\frac{b^2 - b_i^2}{b^2 - b_j^2}}.
\end{align}
Since the square root should be non-negative we also find an inequality constraint on slopes:
\begin{align}\label{ICi_ineq}
(b^2 - b_i^2)(b^2 - b_j^2)>0.
\end{align}
 Equations~\eqref{ICi_eq} and \eqref{ICi_ineq} work for any choices of $i$ and $j$ with $i,j=3,4,\ldots,N$ and $i\ne j$, so all parameters $e_i$ can be expressed as:
\begin{align}\label{}
e_i=
\lambda_i e\sqrt{\left|\frac{b_i^2}{b^2}-1\right|}, \quad |b_i|>b, \quad i=3,4,\ldots,N,
\label{ei}
\end{align}
 where $\lambda_i\equiv \sgn(e_i)$ can be either $1$ or $-1$.

Consider now IC (ii). It is generally difficult to  prove analytically that some crossing of diabatic levels leads to an exact crossing point of adiabatic energy levels. However, we can write {\it necessary} conditions for this to happen. This is achieved by assuming  formally that all nonzero couplings are small and then requiring that the lowest order perturbative contribution to the gap between the considered two adiabatic levels is zero.

Let us first look at the crossing between levels 1 and 2 at $t_{12}=0$. These two levels would be coupled to each other  at the 2nd order of the perturbation series via interaction with any level  $i=3,4,\ldots,N$. This condition leads to
\begin{align}\label{}
\sum_{i=3}^N \frac{g_{1i}g_{2i}}{E_1(0)-E_i(0)}=0,
\end{align}
where
$$
E_1(t)=bt, \quad E_2(t)=-b t, \quad E_i(t)=b_i t +e_i, \quad i=3,4,\ldots N
$$
are diabatic energies  of the Hamiltonian (\ref{hal}).
Since the crossing is at $t=0$, we have $E_1(0)=0$ and $E_i(0)=e_i$. Using  \eqref{ei}, we find the condition:
\begin{align}\label{constr_t12}
\sum_{i=3}^N \frac{g_{1i}g_{2i}\lambda_i}{\sqrt{\left|\frac{b_i^2}{b^2}-1\right|}}=0.
\end{align}

Let us now consider the crossing of levels $i$ and $j$ with $i,j=3,4,\ldots,N$, $i\ne j$, at $t_{ij}=-(e_i-e_j)/(b_i-b_j)$. The 2nd order perturbative constraint then reads
\begin{align}\label{}
\frac{g_{1i}g_{1j}}{E_{i}(t_{ij})-E_1(t_{ij})}+\frac{g_{2i}g_{2j}}{E_{i}(t_{ij})-E_2(t_{ij})}=0.
\end{align}
Using expressions for $e_i$ and $e_j$, we find that this constraint can be reduced to:
\begin{align}\label{constr_tij}
&\frac{g_{2i}g_{2j}}{g_{1i}g_{1j}}=\sigma_{ij}\lambda_i\lambda_j\sqrt{\frac{(b_i+b)(b_j+b)}{(b_i-b)(b_j-b)}},
\end{align}
where we defined $\sigma_{ij}\equiv\sgn[(b_i-b)(b_j+b)]$. Note that this equation works for all choices of $i$ and $j$ with $i,j=3,4,\ldots,N$ and $i\ne j$. The number of such equations is $(N-3)(N-2)/2$, but they are not all independent. If we multiply Eq.~\eqref{constr_tij} for some $i, j$ by Eq.~\eqref{constr_tij} for $i,k$, and then divide the result by Eq.~\eqref{constr_tij} for $j, k$, we obtain the relation between $g_{1i}$ and $g_{2i}$:
\begin{align}\label{}
\frac{g_{2i}^2}{g_{1i}^2}=\frac{b_i+b}{b_i-b}.
\end{align}
For any real couplings the right hand side has to be non-negative. So, we have $(b_i+b)/(b_i-b)>0$, or $|b_i|>b$. If this is the case, we have
\begin{align}
\label{g2_g1}
g_{2i}=
\tau_i g_{1i}\sqrt\frac{b_i+b}{b_i-b},
\end{align}
where $\tau_i=\pm 1$ is the relative sign between the couplings $g_{1i}$ and $g_{2i}$. Using  Eq.~(\ref{g2_g1}) in \eqref{constr_tij}, we find that
\begin{align}\label{lambda_tau_sigma}
\lambda_i\tau_i \sigma_i=\lambda_j\tau_j\sigma_j \,\,\,\,\, \forall i,j=3,4,\ldots,N,\,\, i\ne j,
\end{align}
where $\sigma\equiv\sgn(b_i)$. Substituting Eq.~\eqref{g2_g1} into Eq.~\eqref{constr_t12} and using \eqref{lambda_tau_sigma}, we find
\begin{align}\label{}
&\sum_{i=3}^N \frac{g_{1i}^2}{b_i-b}=0.
\end{align}

Summarizing all found relations among parameters we obtain the list of constraints (\ref{b-con})-(\ref{sign-con}). 
%
%
Finally, we note that our proof of ICs does not formally apply to the $N=4$ case because we derived Eq.~\eqref{g2_g1} assuming that there are  at least three different levels with indices $i,j,k>2$.
However, the 4-state case can be formally included because it is still solvable and belongs to the class of the 4-state model that was discussed  in \cite{commute,quest} in detail.
The latter  4-state model is more general (within 4-state systems) because it depends on two rather than $N-3=1$ coupling parameters  \cite{quest}.

\section{Proof of quantum integrability}
\label{bowtie-sec}
Let us now show that the solvable model in section~\ref{main-sec} can be generated  from a family of quantum integrable operators satisfying  (\ref{cond1})-(\ref{cond2}).
Consider the Hamiltonian of the generalized bowtie model:
\be
{H}_0(\bm{\tau})= \left( \begin{array}{ccccc}
\tau^1/2 & 0 & \gamma_3 &\gamma_4 & \ldots \\
0 & -\tau^1/2 & \gamma_3 &\gamma_4 & \ldots \\
\gamma_3 &\gamma_3 &\beta_3 \tau^0 &0 &\ldots\\
\gamma_4 &\gamma_4 &0 & \beta_4 \tau^0 &\ddots \\
\vdots & \vdots & \vdots & \ddots & \ddots
\end{array}
\right).
\label{hbt1}
\ee
In the conventional bowtie model, $\tau^0$ is identified with real time while $\tau^1$ is a constant.
A  linear in $\tau^0$ operator that commutes with $\hat{H}_0$ is known  \cite{yuzbashyan-LZ}.  We searched for operator $\hat{H}_1$ that satisfies both conditions (\ref{cond1})-(\ref{cond2}) in the form of the linear combination of the unit matrix and  the commuting with
$\hat{H}_0$ operator. The result is (see also Ref.~\cite{yuzbashyan-an18} for a more detailed discussion of commuting operators of $H_0$)
\be
{H}_1(\bm{\tau})= \left( \begin{array}{ccccc}
\frac{\kappa}{\tau^1}+ \frac{ \tau^0}{2} & -\frac{\kappa}{\tau^1} &-\frac{\gamma_3}{2\beta_3} &-\frac{\gamma_4}{2\beta_4} & \ldots \\
-\frac{\kappa}{\tau^1} & \frac{\kappa}{\tau^1} -\frac{ \tau^0}{2}  &\frac{\gamma_3}{2\beta_3} &\frac{\gamma_4}{2\beta_4} & \ldots  \\
-\frac{\gamma_3}{2\beta_3} &\frac{\gamma_3}{2\beta_3} &\frac{ \tau^1}{4 \beta_3} &0 &\ldots\\
-\frac{\gamma_4}{2\beta_4} &\frac{\gamma_4}{2\beta_4} &0 &\frac{ \tau^1}{4 \beta_4} &\ddots \\
\vdots & \vdots & \vdots & \ddots & \ddots
\end{array}
\right),
\label{hbt2}
\ee
where
\be
\kappa=\sum_{i=3}^N \frac{\gamma_i^2}{\beta_i}.
\label{cond3}
\ee
(Note that $\tau^0$ and $\tau^1$ here have completely different meaning from the previously defined $\tau_i$ with $i\ge 3$, which are just binary sign variables.) In this article we are interested in MLZ systems with only linear time-dependence of Hamiltonians but ${H}_1$ has terms that depend as $\propto 1/\tau^1$ on time. This problem is removed
if we impose an additional constraint
\be
\kappa=0,
\label{cond4}
\ee
which coincides with constraint (\ref{g-con1}).

In addition to (\ref{cond4}), let us now choose the new time contour such that
\be
{\cal P}_t: \tau^0=a t -e, \quad \tau^1 = at+e, \quad a>0,
\label{cont1}
\ee
where $a$ and $e$ are some constants. Along this contour, evolution is described by the Hamiltonian
\be
\hat{H}(t) = a[\hat{H}_0(\bm{\tau}(t)) +\hat{H}_1(\bm{\tau}(t))].
\label{h-comb1}
\ee
The Hamiltonian (\ref{h-comb1}) has the same matrix form as (\ref{hal}), where
\be
b=a^2, \quad b_j=a^2\beta_j + a^2/(4\beta_j),  
\quad  e_j = e\left(a/(4\beta_j)-a\beta_j \right), \nonumber\\
 g_{1j}=a\gamma_j(1-1/(2\beta_j)), \quad g_{2j}=a\gamma_j(1+1/(2\beta_j)), \quad j=3,\ldots, N.
\label{cont11}
\ee
Using these relations and expressing parameters $a$, $\beta_j$, $\gamma_j$ via  new parameters $b$, $b_j$ and $g_{1j}$, we find that remaining parameters $e_j$ and $g_{2j}$ satisfy conditions, respectively,  (\ref{e-con}) and (\ref{g-con-2}). The possibility of different sign choices of couplings follows from the fact that equations (\ref{cont11}) are generally quadratic in terms of old variables. Thus we found that the Hamiltonian (\ref{hal}) with conditions (\ref{b-con})-(\ref{sign-con}) describes evolution along the time contour (\ref{cont1}) with the Hamiltonian (\ref{h-comb1}).

\begin{figure}
\includegraphics[width=8.75 cm]{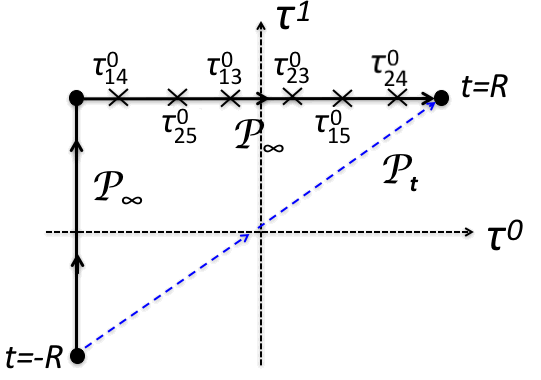}
\centering
\caption{Paths  in the space $(\tau_1,\tau_2)$ for evaluating  transition probabilities
for the  Hamiltonian (\ref{h-comb1}), with $N=5$, $b_3>b_4>0>b_5$, $\lambda_3=\lambda_4=-\lambda_5=1$, and $\rho=1$. On ${\cal P}_t$, $t$ changes from $-R$ to $+R$; all other parameters are fixed. We  deform ${\cal P}_t$ into ${\cal P}_\infty$ without affecting the scattering matrix.
Points $\tau^0_{ij}$ that are marked with crosses indicate nonadiabatic pairwise Landau-Zener transitions between levels $i$ and $j$.}
\label{contour1}
\end{figure}

Having these observations, derivation of transition probabilities in our model can be done completely analogously to the solution of the four-state model in Ref.~\cite{commute}.
For example,  consider the model \eqref{hal} with $N=5$, $b_3>b_4>0>b_5$, $\lambda_3=\lambda_4=-\lambda_5=1$, and $\rho=1$. This is the previously considered Case 4, and its transition probabilities are given in Eq.~\eqref{P_case_4}. We are going to derive the same result using the method of \cite{commute}. First, we transform the physical path ${\cal{ P}}_t$ into ${\cal{ P}}_{\infty}$, as shown in Fig.~\ref{contour1}. We mark the points of diabatic level crossings with crosses. Along ${{\cal P}}_{\infty}$, adiabatic approximation does not hold near six points that are all on the horizontal piece with $\tau_2=aR+e$ and
\begin{align}
\tau^0_{1j}=- \frac{aR+e}{2\beta_j},\quad \tau^0_{2j}=\frac{aR+e}{2\beta_j}, \quad j=3,4,5.
\end{align}
For this choice of parameters, we have $\tau^0_{14}<\tau^0_{13}<0<\tau^0_{23}<\tau^0_{24}$, and $\tau^0_{25}<0<\tau^0_{15}$.
The distances between these points are proportional to $R$, which means that, in the $R\rar \infty$ limit, regions of pairwise nonadiabatic transitions along ${{\cal P}}_{\infty}$ are well apart.

The total evolution matrix $S$ for the path ${\cal P}_{\infty}$ (and hence ${\cal P}_{t}$) factorizes then into an ordered product of  pairwise scattering matrices ${S}^{ab}$, where $a, b$  label  diabatic states experiencing nonadiabatic transitions and diagonal matrices $U^{\alpha,\beta}$ that describe adiabatic evolution between such transition moments $\alpha$ and  $\beta$:
\begin{align}\label{scattering}
{S} = {U}^{aR-e,\tau^0_{24}} {S}^{24} {U}^{\tau^0_{24},
\tau^0_{0,15}}  {S}^{15} {U}^{\tau^0_{15},\tau^0_{23}}
{S}^{23}  {U}^{\tau^0_{23},\tau^0_{13}}
{S}^{13}  {U}^{\tau^0_{13}, \tau^0_{25}}
{S}^{25} {U}^{\tau^0_{25},\tau^0_{14}}{S}^{14}
 {U}^{\tau^0_{14},-aR-e}{U}^{vert},
\end{align}
where  ${U}^{vert}$  stands for the evolution matrix of the vertical piece of $\cal{P}_\infty$, at which $\tau^0={\rm const}$ and no avoided crossing points encounter.
The same arguments as in the four-state example from Ref.~\cite{commute} lead us to conclusion that adiabatic phases, as well as nontrivial Landau-Zener phases, cancel out from the final transition probabilities. Hence, all information about transition probabilities is contained in the product of truncated matrices:

\begin{align}\label{scattering}
{S}^{\rm tr} =  {S}^{24}  {S}^{15}   {S}^{23} {S}^{13}  {S}^{25} {S}^{14} ,
\end{align}
where, in terms of $p_{j}$ and $q_{j}$ defined in Eq.~\eqref{piqi}, we have
\begin{equation}
S^{13}=\left( \begin{array}{ccccc}
\sqrt{p_{3}} & 0 &  i s_{3}\sqrt{q_{3}} & 0 &  0\\
0 & 1 & 0 &  0 &  0\\
i s_{3}\sqrt{q_{3}} & 0  &  \sqrt{p_{3}}  & 0 &0\\
0  & 0 & 0  &  1  &0\\
0 & 0 & 0 & 0   & 1
\end{array}\right),\quad
S^{23}=\left( \begin{array}{ccccc}
1 & 0 &  0 & 0 &  0\\
0 & \sqrt{p_{3}} & i s_{3}\sqrt{q_{3}} &  0 &  0\\
0 & i s_{3}\sqrt{q_{3}} &  \sqrt{p_{3}}  & 0 &0\\
0  & 0 & 0  &  1  &0\\
0 & 0 & 0 & 0   & 1
\end{array}\right),
\label{}
\end{equation}
\begin{equation}
S^{14}=\left( \begin{array}{ccccc}
\sqrt{p_{4}} & 0 & 0&  i s_{4}\sqrt{q_{4}}  &  0\\
0 & 1 & 0 &  0 &  0\\
0  & 0 & 1  &  0  &0\\
i s_{4}\sqrt{q_{4}} & 0& 0  &  \sqrt{p_{4}} &0\\
0 & 0 & 0 & 0   & 1
\end{array}\right),\quad
 S^{24}=\left( \begin{array}{ccccc}
1 & 0 &  0 & 0 &  0\\
0 & \sqrt{p_{4}} &   0 & i s_{4}\sqrt{q_{4}} & 0\\
0  & 0 & 1  &  0  &0\\
0 & i s_{4}\sqrt{q_{4}} & 0 &  \sqrt{p_{4}}  &0\\
0 & 0 & 0 & 0   & 1
\end{array}\right),
\label{}
\end{equation}
\begin{equation}
 S^{15}=\left( \begin{array}{ccccc}
\sqrt{p_{5}} & 0 & 0&  0 & i s_{5}\sqrt{q_{5}}  \\
0 & 1 & 0 &  0 &  0\\
0  & 0 & 1  &  0  &0\\
0 & 0 & 0 & 1   & 0\\
i s_{5}\sqrt{q_{5}} & 0& 0  & 0 &  \sqrt{p_{5}}
\end{array}\right),\quad
S^{25}=\left( \begin{array}{ccccc}
1 & 0 &  0 & 0 &  0\\
0 & \sqrt{p_{5}} &   0 &  0 & i s_{5}\sqrt{q_{5}} \\
0  & 0 & 1  &  0  &0\\
0 & 0 & 0 & 1   & 0\\
0 & i s_{5}\sqrt{q_{5}} & 0   &0&  \sqrt{p_{5}}
\end{array}\right),
\label{}
\end{equation}
and where $s_j=\sgn (g_{1j})$ $(j=3,4,5)$.
The corresponding transition probabilities are given by
\be
P_{i\rar j} = |S_{ji}^{\rm tr}|^2.
\label{ptot1}
\ee
This result coincides with predictions of the semiclassical ansatz. So, the fact that our model belongs to the family of operators satisfying conditions (\ref{cond1})-(\ref{cond2}) can be used to prove validity of
our solution in any sector of the model.

\begin{figure}
\includegraphics[width=8.75 cm]{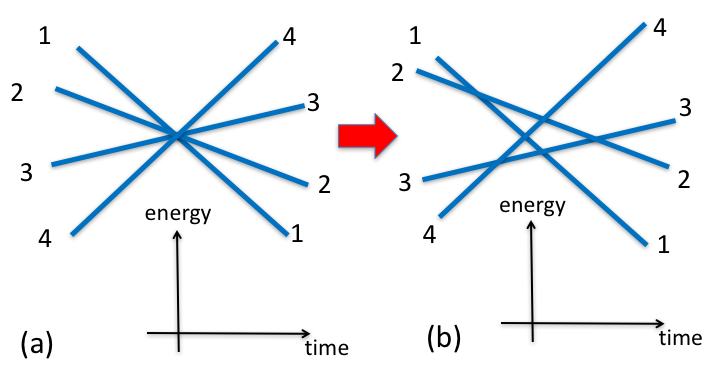}
\centering
\caption{(a) Four energy levels crossing  in one point exactly.  Orders of level indexes before and after the crossing are reverse of each other.  (b) $4\cdot 3/2=6$ crossing points produced by $4$ levels that reverse the order of their indexes when $t$ changes from $-\infty$ to $+\infty$ and cross only in pairwise fashion. }
\label{levels-cross}
\end{figure}

Finally, we are going to show that Eqs.~(\ref{cond1})-(\ref{cond2}) lead to the proof that our N-state model has  $1+(N-2)(N-3)/2$ exact crossing points.
According to (\ref{h-comb1}), the Hamiltonian of our model can be written as the sum of the two bowtie Hamiltonians. Since the Hamiltonians (\ref{hbt1}) and (\ref{hbt2}) commute with each other,  we can write analogous  expression for the eigenvalues of these Hamiltonians:
\be
E^{\alpha}(t) = a[E^{\alpha}_0(\bm{\tau}(t)) + E^{\alpha}_1(\bm{\tau}(t))], \quad \alpha=1,\ldots,N.
\label{ev-bt1}
\ee
Let us order state indexes according to the order of their state energies (eigenvalues of the Hamiltonian) at $t \rar -\infty$, with the highest index corresponding to the lowest energy.  Let us also define the energy level index at arbitrary $t$ by assuming that this {index} does not change if level's energy changes continuously with $t$. Note that near an exact level crossing, the index of each level point remains well defined.
Until levels cross, their indexes remain ordered according to  the order of corresponding state energies. However, if energy levels cross exactly, order of their indexes changes, as shown in Fig.~\ref{levels-cross}(a).

Hamiltonians (\ref{hbt1}) and (\ref{hbt2}) at condition (\ref{cond4}) belong to the bowtie model type \cite{bow-tie}. One can check by inspection that these Hamiltonians have exact crossing points at zero energy and both $\tau^0=0$ and $\tau^1=0$. For example, the Hamiltonian (\ref{hbt1}), as any bowtie Hamiltonian, has a well-known exact crossing point of $N-2$ levels at $\tau^0=0$ \cite{bow-tie}. Another crossing point, of just two levels, appears only when condition (\ref{cond4}) is imposed. At this point, asymmetric combination of first two diabatic states decouples from all other states and becomes an eigenstate with zero energy. Condition (\ref{cond4}) then tunes the energy of one of the remaining states to zero, which leads to the exact  two-level crossing point.

The case of the Hamiltonian  (\ref{hbt2}) at $\kappa=0$
 is similar, except that the role of points at $\tau^0$ and $\tau^1$ interchange. So, if we change $t$ from $-\infty$ to $\infty$, energy levels of both Hamiltonians (\ref{hbt1}) and (\ref{hbt2}) will experience the same
 change of the index order: for each of the Hamiltonians, there will be a pair of levels that  exchange their index orders and, separately, there will be a set of  $N-2$ levels whose mutual index order will reverse too.

 Since energy levels of the Hamiltonian $H(t)$ are given by the linear combination (\ref{ev-bt1}), the order of level indexes for this model will experience the same changes with $t$.
 So, exact level crossings must appear. Their number follows from the fact that all exact crossings happen now in pairwise fashion. Hence, each of the $N-2$ levels that exchange their order
 has to cross all $N-3$ levels of the same set exactly, as we show in Fig.~\ref{levels-cross}(b). There are precisely $(N-2)(N-3)/2$ such crossing points. One more crossing point follows from the pairwise crossing point enforced by condition $\kappa=0$,  leading to the desired number $1+(N-2)(N-3)/2$ of all such crossings. This example shows that existence of crossing  points, and hence validity of IC (ii),  in the model defined in section~\ref{main-sec} are the consequences of conditions (\ref{cond1})-(\ref{cond2}).

\section{Creating new solvable models}
\label{dist-sec}
In this section, we will not discuss the model (\ref{hal}) but rather list three other models that we identified by similar methods. One model
is a trivial family that satisfies conditions (\ref{cond1})-(\ref{cond2}).
The other two were obtained by deforming already known solved systems, as in \cite{quest}. MLZ ICs (i)-(ii) from section~\ref{derive-sec} lead for them to relatively complex nonlinear equations for couplings and level slopes. Nevertheless, we found that in many cases the ansatz that conserves the matrix $\hat{\gamma}$ in (\ref{skew}) does solve these equations. Moreover, necessary conditions on exact adiabatic energy crossings also turned out to be sufficient for such crossings to appear. So, the goal of this section is to demonstrate that many properties of the model (\ref{hal}) that we have already discussed are not unique to this model. They are likely general for a broad class of  solvable MLZ systems.


\subsection{Composite MLZ models and quantum integrability}
\label{composite-subsec}
Let
\be
\hat{H}_{g}(\tau) = \left(
\begin{array}{cc}
\tau & g\\
g & -\tau
\end{array}
\right)
\label{h2}
\ee
be the two-state Landau-Zener Hamiltonian with coupling $g$, and let $\hat{1}$ be a  unit matrix acting in the same two-state space. Consider now  Hamiltonians acting in the space of $M$ independent two level systems:
\be
\label{hams1}
 \hat{H}_k(\tau^0,\ldots \tau^M) &=& a_k^0\hat{H}_{g_0}\left(\sum_{i=0}^M a_i^0 \tau^i \right) \otimes \hat{1} \otimes \ldots \otimes  \hat{1} + \\
\nonumber &+&a_k^1 \hat{1}\otimes \hat{H}_{g_1}\left(\sum_{i=0}^M a_i^1\tau^i \right) \otimes \hat{1} \otimes \ldots \otimes  \hat{1} + \ldots \\
\nonumber &+& a^M_k\hat{1} \otimes\ldots \otimes \hat{1} \otimes \hat{H}_{g_M} \left(\sum_{i=0}^M a_i^M \tau^i \right), \quad k=0,\ldots, M,
\ee
where $a_i^j$, $i,j=0,\ldots, M$ are arbitrary constants.  Apparently, all terms contributing to operators $\hat{H}_k$ commute with each other because they act nontrivially in disjoint subspaces.
Moreover, in each such a subspace, operators $\hat{H}_k$ are different only by a constant pre-factor. So, all operators $\hat{H}_k$ commute with each other, even though coefficients $a_i^j$ can be chosen to make operators $\hat{H}_k$ linearly independent. Condition
$$
\partial \hat{H}_j/\partial \tau^i = \partial \hat{H}_i /\partial \tau^j
$$
is also trivial  to verify. Thus we conclude that models (\ref{hams1}) are quantum integrable in the sense of satisfying conditions (\ref{cond1})-(\ref{cond2}). In principle, they can be solved by the exact WKB approach designed in \cite{commute}. Certainly, in our case this is not needed because different terms contributing to any $\hat{H}_k$ act in different subspaces, so the scattering matrix for each $\hat{H}_k$ factorizes in products of trivial Landau-Zener scattering matrices acting in each two-state subspace, as it was discussed in \cite{multiparticle}.

This example is trivial but useful because it proves that there are families of operators satisfying conditions (\ref{cond1})-(\ref{cond2}) with an arbitrary number of time variables and only linear time-dependence of all Hamiltonians. The latter is the property shared with the model introduced in section~\ref{main-sec}. So, there may be a much larger class of systems with this property.


\subsection{Distorting the driven Tavis-Cummings model}
\label{TC-sec}
First, we consider the driven Tavis-Cummings model \cite{DTCM}. It describes interaction of an arbitrary number of two-level systems (spin-1/2's) with a single bosonic mode, whose frequency depends on time linearly:
\begin{align}
\label{HamiltonianTC}
\hat{H}(t)=-\beta t\hat{a}^\dag \hat{a}+\sum_{i=1}^{N_s}\epsilon_i\hat{\sigma}_i+g\sum_{i=1}^{N_s}(\hat{a}^\dag \hat{\sigma}_i^-+\hat{a} \hat{\sigma}_i^+),
\end{align}
where $N_s$ is the number of spins, $\beta$ is the slope of linear dependence of the bosonic mode frequency, $g$ is the coupling of spins to bosons, $\hat{a}$ is the boson annihilation operator, $\hat{\sigma}_i^{\pm}$ are the $i$th spin's raising and lowering operators, $\epsilon_i$ is the intrinsic level splitting of the $i$th spin, and $\hat{\sigma}_i\equiv(\hat{1}+\hat{\sigma}_z^i)/2$ is the projection operator to spin ``up'' state of the $i$th spin, where $\hat{1}_i$ is a unit matrix acting in the $i$th spin subspace, and $\sigma_z$ is the Pauli $z$-matrix of the $i$th spin.

In this model, the number of bosons plus the number of up-spins is conserved. So if we consider the sector containing the diabatic state with $N_B$ bosons and all spins down, then any diabatic state in this sector can be labelled by the configuration of spins $|\sigma_1,\sigma_2,\ldots,\sigma_{N_s}\ra$, where $\sigma_i$ being $1$ or $0$ corresponds to $i$th spin being up or down. The diabatic energy of the state $|\sigma_1,\sigma_2,\ldots,\sigma_{N_s}\ra$ is, up to a gauge transformation, $b_n t+\sum_{i=1}^{N_s}\epsilon_i \sigma_i$, where the diabatic energies have equidistant slopes:  $b_n=n \beta$, and $n=\sum_{i=1}^{N_s} \sigma_i$ is the number of up-spins. The couplings are non-zero only between two states which are related by flip of a single spin, and the coupling strength is $g_n=g\sqrt{N_B+n}$, where $n$ is the number of up-spins in the
state with lower spin polarization in the pair of coupled states \cite{DTCM}.

\begin{figure}[!htb]
\scalebox{0.52}[0.52]{\includegraphics{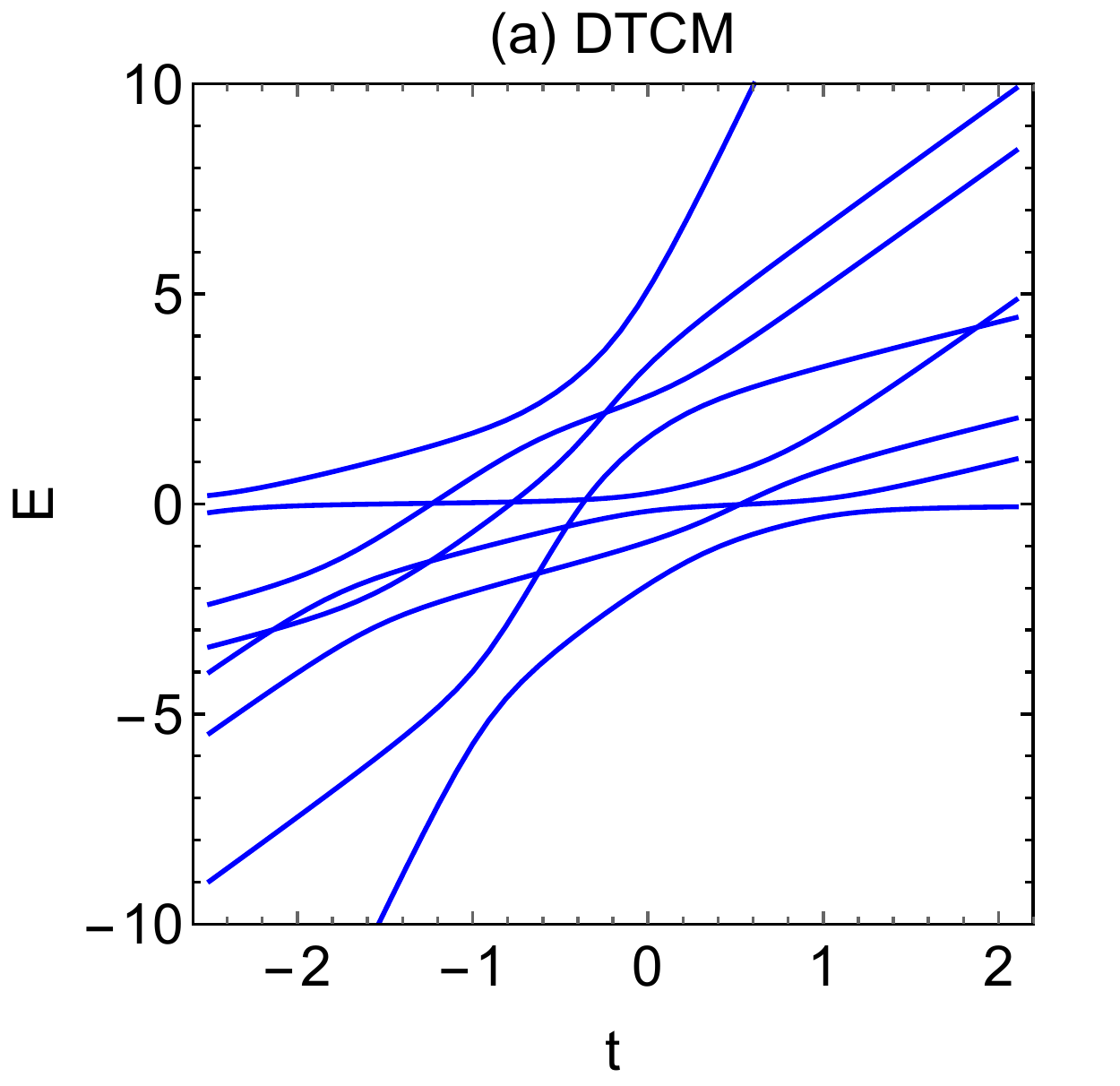}}
\scalebox{0.5}[0.5]{\includegraphics{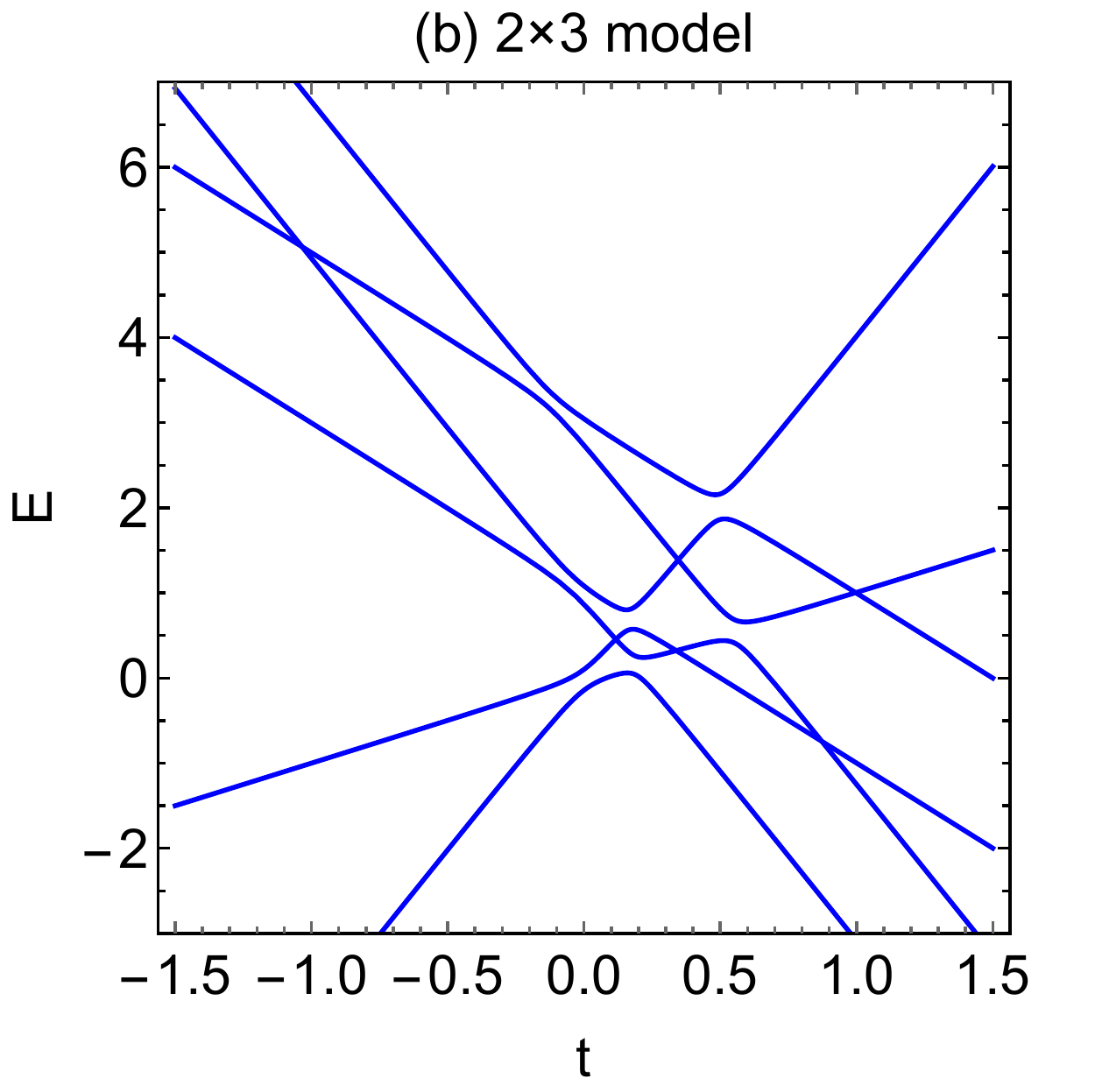}}\\
\caption{Adiabatic energies as functions of time for the distorted version of  (a) the driven Tavis-Cummings model (DTCM) with $N_s=3$ and (b) the $2\times 3$ model. In each figure, the number of exact crossings agrees with the number of zero direct couplings in the Hamiltonian. (a) For $N_s=3$, the number of exact crossings is 10. The slopes are: $b_0=0$, $b_1=1$, $b_2=3$, $b_3=9$ (from Eq.~\eqref{conj_bn}), the constant diagonal elements are given by Eq.~\eqref{conj_en} with $\epsilon_1=2.4$, $\epsilon_1=0$ and $\epsilon_3=-1$, and the couplings are $g_1= g \sqrt{N_B + 1}$, $g_2= g \sqrt{N_B + 2} \sqrt{(b_2 - b_1)/b_1}$ and $g_3= g \sqrt{N_B + 3} \sqrt{(b_3 - b_2)/b_1}$ with $N_B=0$ and $g=0.2$. (b) For the $2\times 3$ model described by the Hamiltonian \eqref{2X3_Hal_Omega}, the number of exact crossings is 6. Parameters are: $b_1 = 4$, $b_2 = 2$, $b_3 = 1$, $e_2 = 1$, $e_3 = 3$, $g_1 = 0.1$, $g_2 = 0.12$, $g_3 =0.15$, and $e_5$ and $e_6$ are given by Eqs.~\eqref{con_e5} and \eqref{con_e6}, both with the negative signs.
}\label{other_models}
\end{figure}

Assuming more general level slopes and couplings, keeping parallel diabatic levels of the model (\ref{HamiltonianTC}) still parallel after the deformations, and then imposing ICs, we find that the driven Tavis-Cummings model can be generalized so that the slopes are no longer equidistant,  i.e., $b_n\ne \beta n$. Let us take the slopes of the states with 0, 1, and 2 up-spins to be $b_0=0$, $b_1=\beta$ and $b_2=(1+\gamma)\beta$, where $\gamma$ is a new parameter. We found that IC (i) is satisfied if the slopes for states with $n\ge 3$ up-spins are chosen as
\begin{align}\label{conj_bn}
&b_{n}=\frac{1+\gamma}{1-\frac{n-2}{n}\gamma}\beta,
\end{align}
and the constant diagonal element of the Hamiltonian for the state $|\sigma_1,\sigma_2,\ldots,\sigma_{N_s}\ra$ is given by
\begin{align}\label{conj_en}
e_{|\sigma_1,\sigma_2,\ldots,\sigma_{N_s}\ra}=
\frac{b_n}{n\beta}\sum_{i=1}^{N_s}\epsilon_i \sigma_i.
\end{align}
Up to $N_s=5$, we derived conditions \eqref{conj_bn} and \eqref{conj_en} analytically. It is likely working equally well for any $N_s$ but we did not pursue the rigorous proof.
We then found that IC (ii) is also satisfied if we modify the couplings so that $\Omega_{ij}$ in (\ref{skew}) is preserved, namely we change $g_n$ so that
\begin{align}\label{g-tc}
g_n\rar \sqrt\frac{b_{n}-b_{n-1}}{\beta}g_n.
\end{align}
Adiabatic energy levels  of such a generalized model for $N_s=3$ are shown in Fig.~\ref{other_models}(a), which has the expected number of exact crossings. So, this model is solvable and its solution is
given by the semiclassical ansatz \cite{DTCM}. We checked numerically for  models with up to $N_s=4$ that this is indeed the case (not shown). The time-independent version of the model (\ref{HamiltonianTC}) has been influential in the theory of the algebraic Bethe ansatz \cite{bethe}. It should be interesting to explore if deformations given by Eqs.~(\ref{conj_bn})-(\ref{g-tc}) are also solvable in this sense but we will not explore this here.

\subsection{The $2\times 3$ model}
The next model is the 6-state model constructed as the direct product of the 2-state Landau-Zener model and the 3-state Demkov-Osherov model. We will call this model the $2\times 3$ model. Its Hamiltonian reads:
\begin{align}\label{}
&\hat H=\hat{H}_{LZ}\otimes \hat I_3+\hat I_2\otimes\hat{H}_{DO},\nn\\
&{H}_{LZ} = \left( \begin{array}{cc}
 \beta_1 t+\epsilon_1 & g_1\\
g_1 & \beta_2 t+\epsilon_2
\end{array}\right),\quad
{H}_{DO} = \left( \begin{array}{ccc}
 \beta_3 t+\epsilon_3 & g_2  &g_3 \\
g_2 & \beta_4 t+\epsilon_4 & 0\\
g_3 & 0 & \beta_4 t+\epsilon_5
\end{array}\right).
\label{}
\end{align}
As a  direct product of two solvable models, this model is  solvable. To distort it, we follow \cite{quest} and assume all nonzero couplings to be independent. We change slopes of  diabatic levels keeping parallel levels still parallel, i.e., we search for the integrable model Hamiltonian in the form
\begin{equation}
 H=\left( \begin{array}{cccccc}
 b_1 t & g_{12} & g_{13} & g_{14} & 0 & 0\\
g_{12} & -b_2 t+e_2 &  0 &  0 & g_{25} & 0\\
g_{13} & 0 &  -b_2 t+e_3  & 0 &0 &g_{36}\\
g_{14} & 0 & 0  & b_3 t  &g_{45} &g_{46}\\
0 & g_{25} & 0 & g_{45} & -b_1 t+e_5  &0\\
0 & 0 & g_{36} & g_{46} & 0  & -b_1 t+e_6
\end{array}\right).
\label{}
\end{equation}
The original model can be recovered by setting $b_3=b_2$, $e_5=e_2$, $e_6=e_3$, $g_{14}=g_{25}=g_{36}=g_1$, $g_{12}=g_{45}=g_2$ and $g_{13}=g_{46}=g_3$. We then use ICs to find constraints on the parameters. From IC (i), we found that  constant diagonal elements should be
\begin{align}\label{}
&e_5=\frac{b_1+b_3}{b_2+b_3}\left(1\pm\sqrt{\frac{(b_1-b_2)(b_1-b_3)}{(b_1+b_2)(b_1+b_3)}}\right)e_2,\label{con_e5}\\
&e_6=\frac{b_1+b_3}{b_2+b_3}\left(1\pm\sqrt{\frac{(b_1-b_2)(b_1-b_3)}{(b_1+b_2)(b_1+b_3)}}\right)e_3.\label{con_e6}
\end{align}
From IC (ii), we then obtained four independent constraints on couplings. We checked that the conjecture of conservation of $\hat{\gamma}$ works, namely, the model with the Hamiltonian
\begin{equation}
\hat H=\left( \begin{array}{cccccc}
 b_1 t & g_{2} & g_{3} & g_{1}\sqrt\frac{b_1-b_3}{b_1-b_2} & 0 & 0\\
g_{2} & -b_2 t+e_2 &  0 &  0 & g_{1} & 0\\
g_{3} & 0 &  -b_2 t+e_3  & 0 &0 &g_{1}\\
g_{1}\sqrt\frac{b_1-b_3}{b_1-b_2} & 0 & 0  & b_3 t  &g_{2}\sqrt\frac{b_1+b_3}{b_1+b_2} &g_{3}\sqrt\frac{b_1+b_3}{b_1+b_2}\\
0 & g_1 & 0 & g_{2}\sqrt\frac{b_1+b_3}{b_1+b_2}  & -b_1 t+e_5  &0\\
0 & 0 & g_{1} & g_{3}\sqrt\frac{b_1+b_3}{b_1+b_2} & 0  & -b_1 t+e_6
\end{array}\right)
\label{2X3_Hal_Omega}
\end{equation}
is solvable. Its adiabatic energy diagram with six exact crossing points is shown in Fig.~\ref{other_models}(b).

\section{Multi-Time  Landau-Zener model}
\label{mtlz-sec}
\subsection{Definition}

Summarizing our findings so far: there are simple empirical rules (i)-(ii), discussed in section~\ref{derive-sec},  that lead to algebraic equations on MLZ model parameters.
 For a given diabatic level crossing pattern, these equations define a model that has an explicit  solution. Once such a solvable model is identified, it is possible to prove validity of its solution rigorously by finding the family of Hamiltonians satisfying conditions
(\ref{cond1})-(\ref{cond2}).

Natural questions then follow. Are previously used ICs optimal for classification of solvable MLZ models, i.e., can we use conditions (\ref{cond1})-(\ref{cond2})  to derive  even simpler equations that determine parameters?
Also, can we prove validity of the previously used ICs more rigorously for a broader class of MLZ systems?
Here, we note that,  alone, Eqs.~(\ref{cond1})-(\ref{cond2}) are too general to resolve these questions.
Any Hamiltonian has a family of commuting operators satisfying Eq.~(\ref{cond1}), and if a Hamiltonian depends on many parameters, it is not surprising that after change of variables some parameter combination  can satisfy Eq.~(\ref{cond2}).

In order to make definition of quantum integrability in MLZ theory more precise, we make an additional restriction. We will call MLZ system (\ref{mlz}) integrable if, first, its Hamiltonian belongs to the family of operators satisfying (\ref{cond1})-(\ref{cond2}) and, second, there is explicitly solvable WKB approximation that allows to find asymptotically exact solution of the system (\ref{cond1})-(\ref{cond2}) in the vicinity of some contour $\cal{P}$ at $|\bm{\tau}|\rar \infty$ that connects real time points $\tau^0=\pm \infty$.

The last restriction is still not straightforward to quantify. It requires understanding of behavior of the system  (\ref{cond1})-(\ref{cond2}) at $|\bm{\tau}|\rar \infty$ for possible choices of time-dependent
Hamiltonians.
However, we expect that the desired property will be found generally in systems with simple time-dependence that leads only to simple pole singularities and, possibly, a low rank irregular point at infinite time.
For example, the model introduced in section~\ref{main-sec} and the family of composite models in section~\ref{composite-subsec}, as well as the four-state
MLZ model that was solved in Ref.~\cite{commute}, depend only linearly on all time variables. Therefore, let us consider the general class of such models satisfying conditions
(\ref{cond1})-(\ref{cond2}) and having only linear dependence on all times:
\begin{eqnarray}
\label{linear-family} H_{j} (\bm{\tau}) = B_{kj} \tau^{k} + A_{j}, \quad k,j=0,\ldots, M,
\end{eqnarray}
where $B_{kj}$ and $A_j$ are real symmetric matrices.  For simplicity, we will assume that matrices $A_j$ have zero diagonal elements and all matrices $B_{kj}$ have nondegenerate eigenvalues.
This is the case of all mentioned models, including our main model defined in section~\ref{main-sec}.
We remind that we reserve lower indexes to mark different matrices $H_j$ and we will assume summation over this index when it is repeated as time-index, e.g., of $\tau^j$.
All equations in (\ref{system1}) belong to the  MLZ type (\ref{mlz}). For this reason, we will refer to models of this class as to {\it Multi-Time Landau-Zener (MTLZ) models}.

\subsection{Integrability conditions in MTLZ problem}
Conditions (\ref{cond1})-(\ref{cond2}), applied to a linear family of Hamiltonians (\ref{linear-family}), lead to constraints on matrices $B_{kj}$ and $A_j$:
\begin{eqnarray}
\label{linear-family-2} B_{kj} = B_{jk}, \;\;\; [B_{jk}, B_{lm}] &=& 0, \\
\label{linear-family-3} [B_{sj}, A_{k}] - [B_{sk}, A_{j}] &=& 0, \\
\label{linear-family-4} [A_{j}, A_{k}] &=& 0, \quad s,k,j,l,m=0,\ldots M.
\end{eqnarray}
This set of constraints resembles equations that were used to classify commuting operators with linear dependence on a single dispersion parameter \cite{yuz-q,yuzbashyan-LZ}.
The difference is only in new first equation in  (\ref{linear-family-2}) and extra constraints of the type~(\ref{linear-family-3}).
We will say that  Eqs.~(\ref{linear-family-2})-(\ref{linear-family-4}) represent  the new {\it integrability conditions for MTLZ problem}.
Let $\Lambda_{kj}^a$, $a=1,\ldots, N$ be eigenvalues of the matrices $B_{kj}$.
We further note that Eq.~(\ref{linear-family-3}) written for off-diagonal components in the diabatic basis set means that if levels $a$ and $b$ are coupled directly then
\begin{eqnarray}
\label{Lambda-to-B} \Lambda_{sj}^{a} - \Lambda_{sj}^{b} = \chi_{s}^{ab} A_{j}^{ab},
\end{eqnarray}
for some $\chi_{s}^{ab}$ (collinearity of two vectors), while the condition $\Lambda_{sj}^{a} = \Lambda_{js}^{a}$, which follows from (\ref{linear-family-2}), implies
\begin{eqnarray}
\label{Lambda-to-B-2} \chi_{s}^{ab} = (\gamma^{ab})^{-1} A_{s}^{ab},
\end{eqnarray}
for some $\gamma^{ab}$, resulting in
\begin{eqnarray}
\label{Lambda-to-B-3} \gamma^{ab} (\Lambda_{kj}^{a} - \Lambda_{kj}^{b}) = A_{k}^{ab} A_{j}^{ab}.
\end{eqnarray}

In what follows, we are going to show that integrable MTLZ models have the required properties of the WKB  approximation at  $|\bm{\tau}|\rar \infty$. Hence, corresponding MLZ models that
they generate  are all explicitly solvable. Moreover, we will show how various previously  identified properties of solvable MLZ systems can be  understood using relation (\ref{Lambda-to-B-3}), which is
a direct consequence of Eqs.~(\ref{linear-family-2})-(\ref{linear-family-4}).

\subsection{Adiabatic regions}
\label{sec:Int-LZ-linear-WKB}

Due to Eq.~(\ref{linear-family-2}) all matrices $B_{jk}$ can be diagonalized in the same orthonormal basis set $(\bar{\bm{e}}_{a} \, | \, a = 1, \ldots, N)$,
hereafter referred to as the diabatic basis set, so that
\begin{eqnarray}
\label{decompose-H-diab} H_{j}(\bm{x}) = \sum_{a} \Lambda_{kj}^{a} \tau^{k} | \bar{\bm{e}}_{a} \rangle \langle \bar{\bm{e}}_{a} | + \sum_{ab}^{a \ne b} A_{j}^{ab} | \bar{\bm{e}}_{a} \rangle \langle \bar{\bm{e}}_{b} |,
\end{eqnarray}
 whereas the off-diagonal elements of $A_{j}$ represent the coupling constants.

The adiabatic energies $E_{j}^{a}(\bm{\tau})$ with the proper accuracy are obtained using the second-order perturbation theory in the coupling constants, resulting in
\begin{eqnarray}
\label{E-adiabatic} E_{j}^{a}(\bm{\tau}) = \Lambda_{kj}^{a} \tau^{k} + \sum_{b, b \ne a} \frac{A_{j}^{ab} A_{j}^{ba}}{(\Lambda_{kj}^{b} - \Lambda_{kj}^{a}) \tau^{k}}.
\end{eqnarray}

 Equation~(\ref{E-adiabatic}) shows that, for nondegenerate eigenvalues, $\Lambda_{kj}^{b} \ne \Lambda_{kj}^{a}$, nonadiabatic correction is generally vanishing in the limit $|\bm{\tau}|\rar \infty$.
Exception is for hyperplanes defined by pairwise degeneracy equations
\be
(\Lambda_{kj}^{b} - \Lambda_{kj}^{a}) \tau^{k}=0, \quad a,b=1,\ldots, N.
\label{hyper1}
\ee
All such hyperplanes contain $\bm{\tau}=0$ point. They divide the multiple-time space into {\it adiabatic regions}, in which approximation (\ref{E-adiabatic}) is valid. It becomes exact at $|\bm {\tau} | \rar \infty$.

Following \cite{commute}, condition (\ref{cond1}) applied to the stationary Schr\"odinger equation gives that  a multidimensional energy surface $E_{j}^{a}(\bm{\tau})$ for level $a$ is a  gradient of some classical action ${\cal S}^{a} (\bm{\tau})$ associated with level $a$:
\be
E_{j}^{a}(\bm{\tau}) = -\partial_{j} {\cal S}^{a}(\bm{\tau}).
\label{en-grad}
\ee
Searching then for a solution of (\ref{cond1})-(\ref{cond2}) in the WKB form, and using that away from hyperplanes the adiabatic basis set asymptotically approaches the diabatic one, we find that up to nonvanishing in $|\bm{\tau}| \rar \infty$ terms the wave function inside arbitrary adiabatic domain $\alpha$ can be written in the form
\begin{eqnarray}
\label{Sol-flat-conn-semiclass} \Psi_{\alpha} (\bm{\tau}) &=& \sum_{a} \Psi_{\alpha}^{a} e^{i({\cal S}^{a}(\bm{\tau}) - {\cal S}^{a}(\bm{\tau}(0))}  | \bar{\bm{e}}_{a}(\bm{\tau}) \rangle, \quad \forall a \ne b,
\end{eqnarray}
with some coefficients $ \Psi_{\alpha}^{a}$ that are fixed by boundary conditions on hyperplanes, and with
\begin{eqnarray}
\label{Sol-flat-conn-semiclass-2}
{\cal S}^{a}(\bm{\tau}) &=& -\frac{1}{2} \Lambda_{jk}^{a} \tau^{j}\tau^{k} -  \sum_{b \ne a} \gamma^{ab} \ln |l^{ab}(\bm{\tau})|,
 \nonumber \\ l^{ab}(\bm{\tau}) &=& \frac{A_{k}^{ab} \tau^{k}}{\sqrt{|\gamma^{ab}|}}.
\end{eqnarray}
The choice of the denominators in the definition of $l^{ab}(\bm{\tau})$ could be arbitrary because the adiabatic action is defined up to an additive constant.  Our choice will simplify some notation in section~\ref{sec:scatter-multi-dim} later.
Importantly, according to Eqs.~(\ref{Lambda-to-B-3}) each hyperplane equation (\ref{hyper1})  corresponds to
\begin{eqnarray}
\label{defined-danger-ab} l^{ab}(\bm{\tau}) = 0
\end{eqnarray}
for the pair of diabatic states $a$ and $b$ at nonzero coupling $A_j^{ab}$ between these states. Unlike (\ref{hyper1}), Eq.~(\ref{defined-danger-ab}) does not depend on the  index of the Hamiltonian $j$, showing that there is actually a single such hyperplane for all Hamiltonians in the family.
Also, since  $ l^{ab}(\bm{\tau})$ is zero on the hyperplane, it  has different signs in different side sectors, so each sector
$\alpha$ has a completely determined sign factors $s_{ab}(\alpha)$, associated with all scattering pairs, which are defined as
\begin{eqnarray}
\label{defined-s-ab-alpha} s_{ab}(\alpha) = {\rm sgn}(l^{ab}(\bm{\tau})), \;\;\; {\rm for} \; \bm{\tau} \in \alpha.
\end{eqnarray}


\begin{figure}
\includegraphics[width=8.75 cm]{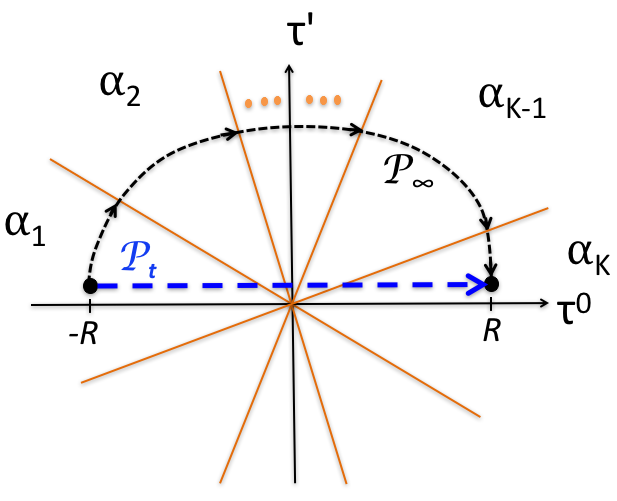}
\centering
\caption{Sectors $\alpha_m$, $m=1, \ldots, K$, separated by hyperplanes that are represented, for convenience, by radial lines in a 2D cross-section $(\tau^0,\tau')$, with $\tau'$ being any linear combination of other than $\tau^0$ time-variables. We can choose the path $\cal{P}_{\infty}$ to lie in this cross-section. Hyperplanes correspond to pairwise degeneracies of diabatic energy levels of the Hamiltonians $\hat{H}_i$, $i=0,\ldots, M$ with corresponding nonzero pairwise level couplings. $K$ is the number of such degeneracies encountered by the path $\cal{P}_{\infty}$ that connects points
$t\equiv \tau^0 \in (-R,R)$ at $R\rar \infty$ and finite constant values of $\tau^j$, $j=1, \ldots, M$.  $\cal{P}_{\infty}$ always remains in the region $|\bm{\tau}| \rar \infty$ with well-justified WKB approximation.
Adiabatic approximation is valid along this contour inside any sector $\alpha_m$ but WKB wavefunction experiences jumps, which are described by the  Landau-Zener formula, when   $\cal{P}_{\infty}$ crosses any of the hyperplanes. The evolution matrix does not depend on a particular choice of the contour  that connects the given initial and final points. So,
the exact WKB solution along the contour $\cal{P}_{\infty}$ reproduces the desired  matrix of evolution along the contour ${\cal{P}}_t$ that has finite constant values of all time variables except $\tau^0 \equiv t$.
}
\label{sectors}
\end{figure}

\subsection{Boundary conditions on hyperplanes}
\label{sec:scatter-multi-dim}

Multidimensional adiabatic approximation holds when $|\bm{\tau}|$ is large enough, say $|\bm{\tau}| > R \rar \infty$. So, it is convenient to choose a path ${\cal P}_{\infty}$, in Fig.~\ref{sectors}, that connects desired points at $\tau_{0}=\pm \infty$, with finite values of other
time variables, while always going through the region $|\bm{\tau}| >R \rar \infty$.

Figure~\ref{sectors} shows, however, that any such a path has to cross a number of dangerous hyperplanes where the adiabatic levels become degenerate, so pure adiabatic approximation is not sufficient to connect asymptotic solutions of a MLZ problem.
The vectors $\bm{\Psi}_{\alpha}$, defined in (\ref{Sol-flat-conn-semiclass}) and associated with different cells, are linearly related. So we can then define connection matrices $ S^{ab}_{\alpha\beta}$ associated with
a hyperplane that represents degeneracy of levels $a$ and $b$ at the border of sectors $\alpha$ and $\beta$:
\begin{eqnarray}
\label{scatter-multidim} \Psi_{\alpha}^{a} = \sum_{b} S^{ab}_{\alpha\beta} \Psi_{\beta}^{b}.
\end{eqnarray}
Fortunately, points at which ${\cal P}_{\infty}$ crosses hyperplanes are separated by distances of at least order $R\rar \infty$, so we can safely disregard presence of other hyperplanes in order to understand behavior of the wavefunction in the vicinity of any one of them. Using the fact that for other levels $c\ne a,b$ the semiclassical action ${\cal{S}}^c(\bm \tau)$ changes continuously across the hyperplane with degeneracy of $a$ and $b$, we can choose the boundary condition
\begin{eqnarray}
\label{scatter-multidim-2} S_{\alpha\beta} = \bar{S}_{\alpha\beta; ab} \otimes \bar{I}_{ab},
\end{eqnarray}
where $\bar{I}_{ab}$ is the $(N-2) \times (N-2)$ unit matrix acting in the space of all levels, except for $a$ and $b$, while $\bar{S}_{\alpha\beta; ab}$ is a $2 \times 2$ matrix acting in the space of levels $a$ and $b$ that can be obtained by considering a standard $2 \times 2$ LZ problem in the vicinity of the scattering hyperplane.

To find $\bar{S}_{\alpha\beta; ab}$, we note that it should not depend on direction of crossing the hyperplane, so we  choose a path with $\tau^k=\tau^k_h$, $k\ne j$ and $\tau_j=\tau^j_h+t$, where $\bm{\tau}_h$ lie on the hyperplane (\ref{hyper1}).
 Restricting dynamics to levels  $a$ and $b$, we find from (\ref{system1}) that corresponding amplitudes $a(t)$ and $b(t)$ change locally as amplitudes in the two-state Landau-Zener model:
 \be
 i\frac{d}{dt}a(t) = (\Lambda_{jj}^at +\varepsilon)a(t) +A_j^{ab} b(t), \quad  i\frac{d}{dt}b(t) = (\Lambda_{jj}^bt +\varepsilon)b(t) +A_j^{ab} a(t),
 \label{two-LZ}
 \ee
 where summation over repeated $j$ is not assumed and $\varepsilon= \Lambda_{kj}^{a} \tau^k_h = \Lambda_{kj}^{b} \tau^k_h= {\rm const}$. Asymptotically, at $t\rar \pm \infty$, there are basis solutions of (\ref{two-LZ}) that depend on time as
 $\psi^{a} =e^{-i [(\Lambda_{jj}/2) t^2 +\varepsilon t + \gamma^{ab} {\rm ln}(|t|) + \phi]}$, where $\phi$ is any constant phase and $ \gamma^{ab}=(A^{ab}_j)^2/[\Lambda_{jj}^a-\Lambda_{jj}^b]$. Comparing this with WKB wavefunctions (\ref{Sol-flat-conn-semiclass}) we find that they coincide along the chosen time contour up to logarithmic phase terms $\propto \sum_{c\ne a,b} \gamma^{ac} {\rm ln}(l^{ac}(\bm{\tau}_h))$. So, up to the phase $\phi_B^{ab}$ that describes this effect of basis change, the matrix  $\bar{S}_{\alpha\beta; ab} $ coincides with the scattering matrix of the Landau-Zener model, which is known explicitly:
\begin{eqnarray}
\label{scatter-multidim-3}  \bar{S}_{\alpha\beta; ab} = S_{LZ,s_{ab}(\alpha)}(\gamma^{ab})e^{i \phi^{ab}_B},
\end{eqnarray}
where
\begin{eqnarray}
\label{scatter-LZ-standard-2} S_{LZ,\pm}^{aa}(\gamma^{ab}) &=& e^{-\pi |\gamma^{ab}|}, \;\;\; S_{LZ,\pm}^{ab}(\gamma^{ab}) = \pm \sqrt{1 - e^{-2\pi |\gamma^{ab}|}}e^{i \varphi_{\pm}(\gamma^{ab})}, \nonumber \\ \varphi_{\pm}(\gamma^{ab}) &=& \pm {\rm sgn}(\gamma^{ab}) \left(\frac{\pi}{4} - {\rm arg}(\Gamma(-i|\gamma^{ab}|))\right),
\end{eqnarray}
and $\Gamma(x)$ being the Euler gamma-function. Note that the sign in the subscript of $S_{\pm}(\gamma^{ab})$ in the r.h.s. is determined by ${\rm sgn} \, (l^{ab}(\bm{\tau}))$ for $\bm{\tau}$ in the sector $\alpha$, according to the definition, given by Eq.~(\ref{defined-s-ab-alpha}).
Importantly, since parameters $\gamma^{ab}$ do not depend on the index $j$, neither does the found boundary condition (\ref{scatter-multidim-3}). This result is the consequence of the fact that conditions (\ref{cond1})-(\ref{cond2}) guarantee that boundary conditions do not depend on the direction of crossing the hyperplane.

We can now construct  solution for a MTLZ problem.
If the regions/sectors $\alpha$ and $\beta$ do not share a border, they can be connected via a sequence $\mu_{1}, \ldots \mu_{k}$ of cells with a border sharing property, so that the connecting matrix can be obtained as
\begin{eqnarray}
\label{scatter-multidim-4} S_{\alpha\beta} = {S}_{\alpha \mu_{k}} S_{\mu_{k} \mu_{k-1}} \ldots S_{\mu_{2} \mu_{1}} S_{\mu_{1} \beta},
\end{eqnarray}
with each factor in the right hand side being defined by Eq.~(\ref{scatter-multidim-2}), combined with Eq.~(\ref{scatter-multidim-3}).

Equation~(\ref{scatter-multidim-4}) can be considered a general solution of the scattering problem for the MTLZ model. It
confirms the previously made conjecture \cite{quest,six-LZ} that the scattering matrices of many solvable MLZ models factorize into products of $N \times N$ matrices, each having exactly one non-trivial $2 \times 2$ block, represented by a $2 \times 2$ scattering matrix, associated with a standard LZ problem.
In other words, we proved that
if a MLZ Hamiltonian can be extended to a nontrivial MTLZ family, solution of this model has the form of the matrix product ansatz (\ref{scatter-multidim-4}).

\subsection{Family of solvable MLZ models}
\label{sec:solvable-MLZ}
\begin{figure}
\includegraphics[width=8.75 cm]{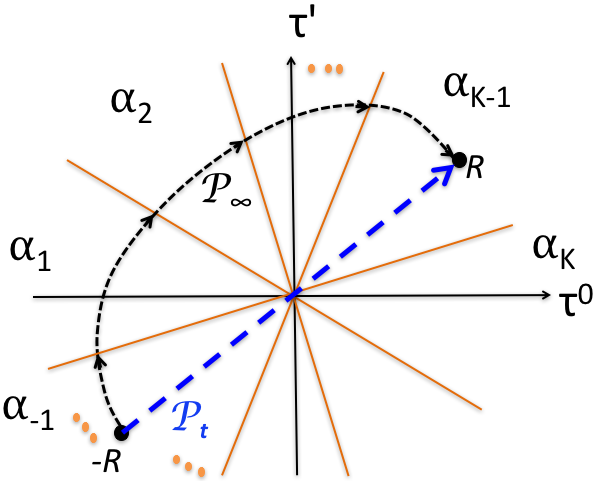}
\centering
\caption{A continuous family of MLZ models is obtained by considering evolution along arbitrary linear time contour ${\cal P}_t$ in the multi-time plane. }
\label{sectors2}
\end{figure}

Let us now address the general prior observation \cite{quest} that solvable MLZ systems tend to belong to families of MLZ models with continuous parameter deformations  that preserve the matrix $\hat{\gamma}$ in (\ref{skew}). Within the MTLZ type of systems this observation has simple explanation. One property of such models is that if we choose a linear time path  via substitution
\begin{eqnarray}
\label{linear-family-pullback} \bm{\tau} (t) = \bm{v}t + \bm{\varepsilon},
\end{eqnarray}
with arbitrary parameter vectors  $\bm{v}$ and $\bm{\varepsilon}$, then Eqs.~(\ref{cond1})-(\ref{cond2}) reduce to a MLZ model (\ref{mlz}), i.e., not only evolution along one of the time variables but also
along an arbitrary linear time contour (\ref{linear-family-pullback})  is of  the MLZ's
 type. The corresponding  Hamiltonian is
 \be
 H(t)= v^iH_i(\bm{\tau}(t)),
 \label{hamt}
 \ee
 Substituting (\ref{linear-family}) in (\ref{hamt}), we find that $H(t)= Bt+A$, where
\begin{eqnarray}
\label{linear-family-pullback-2} A &=& B_{kj} \varepsilon^{k}v^{j} + A_{j} v^{j}, \;\;\; \hat{B} = B_{jk} v^{j}v^{k}, \nonumber \\ b_{a} &=& \Lambda_{jk}^{a} v^{j}v^{k}, \;\;\; e_{a}\equiv A^{aa} = \Lambda_{jk}^{a} \varepsilon^{j}v^{k}, \;\;\; g_{ab} = A_{j}^{ab} v^{j}.
\end{eqnarray}

Figure~\ref{sectors2} illustrates the idea how to find the scattering matrix for the model with the Hamiltonian (\ref{hamt}). Note that the choice of parameters $v_i$ defines the sectors with initial and final time points. Depending on the latter, the detour path ${\cal{P}}_{\infty}$ will cross hyperplanes in different order, so the solutions experience sharp changes when  changes of parameters $v_i$ lead to changes  of
the path endpoint sectors. On the other hand, variations of  $v_i$ that do not lead to change of sectors for endpoints of ${\cal {P}}_t$ will preserve the corresponding scattering matrix $S_{\alpha\beta}$.
Hence,  by varying parameters $v_i$ we can observe sharp changes of behavior of transition probabilities. This explains qualitative features of the phase diagram   that we discussed in sections~\ref{test5-sec} and \ref{testN-sec}, in particular, independence of transition probabilities of parameter $e$ except sharp phase transitions at  point parameter values. Indeed, changes of $e$ keep initial and end points of ${\cal {P}}_t$ within the same sectors. This independence gives additional intuitive explanation for why the semiclassical ansatz is valid: it is because by setting $e\rar \infty$, we make all pairwise crossing points well separated in the diabatic level diagram.

\subsection{Dynamical phase in MLZ problem}
\label{sec:multi-dim-to-MLZ}

There is a  complication to relate matrix $S_{\alpha\beta}$ in (\ref{scatter-multidim-4}) to the scattering matrix of a desired MLZ model (\ref{mlz}) because the latter matrix is usually written in a different basis from our WKB wavefunction. This results in an additional phase, which cancels when transition probabilities are calculated.

Usually, the scattering matrix of a MLZ problem with the Hamiltonian  parametrization (\ref{linear-family-pullback-2}) is written to relate states that behave at $t \to \pm \infty$ as
\begin{eqnarray}
\label{state-adiab-asympt} \psi^{a}(t) &=& \psi_{\pm}^{a} e^{i \hat{\Phi}_{\pm}^{k} (t)}, \;\;\; {\rm for} \; t \to \pm \infty, \nonumber \\ \hat{\Phi}_{\pm}^{a} (t) &=& -\frac{b_{a}t^{2}}{2} - e_{a}t - \sum_{b \ne a} \gamma^{ab} \ln (|\eta^{ab}(t)|), \;\;\; \eta^{ab}(t) = \frac{g_{ab}t}{\sqrt{|\gamma^{ab}|}}.
\end{eqnarray}

Once the asymptotic states are identified, including the phase factors as prescribed by Eq.~(\ref{state-adiab-asympt}), the MLZ-problem scattering matrix $\hat{S}$ can be introduced without any ambiguity, as a connector
\begin{eqnarray}
\label{define-S-MLZ} \psi_{+}^{a} = \sum_{b=1}^{N} \hat{S}_{ab} \psi_{-}^{b}.
\end{eqnarray}

Denoting by $\alpha_{\pm}$ the adiabatic sectors that host $\bm{\tau}(t)$ for $t \to \pm \infty$, respectively  (endpoints of the path ${\cal{P}}_t$ in Fig.~\ref{sectors2}), we can also rewrite asymptotic values of $\psi^{a}(t)$
in terms of the WKB solutions:
\begin{eqnarray}
\label{psi-of-t-asympt}
 \psi^{a}(t) = \Psi^{a}(\bm{x}(t)) = \Psi_{\alpha_{\pm}}^{a} e^{i {\cal S}^{a}(\bm{\tau}(t))}  \;\;\; {\rm for} \; t \to \pm \infty.
\end{eqnarray}

By inspecting the phases for both representations that are given explicitly by Eqs.~(\ref{state-adiab-asympt}) and (\ref{psi-of-t-asympt}) combined with Eq.~(\ref{linear-family-pullback-2}), we observe that the quadratic, linear and logarithmic in $t$ terms fully coincide. The mismatch is a time-independent term that appears in ${\cal S}$ due to the $\bm{\varepsilon}$ term in the expression for $\bm{\tau}(t)$, so that we arrive at
\begin{eqnarray}
\label{asymt-states-relation} \psi_{\pm}^{a} = \Psi_{\alpha_{\pm}}^{a} e^{-i \Phi_{\rm D}^{a} }, \;\;\; \Phi_{\rm D}^{a} = \frac{1}{2} \Lambda_{jk}^{a} \varepsilon^{j} \varepsilon^{k} .
\end{eqnarray}
The quadratic in $\varepsilon_j$ term  does not depend  on level couplings. So, this phase is related to the  semiclassical dynamic phase that is used in formulation of IC (i) in section~\ref{derive-sec}.

Comparing the definitions of the MLZ scattering matrix $\hat{S}$ [Eq.~(\ref{define-S-MLZ})] and the connecting matrix $S_{\alpha\beta}$ associated with a linear multidimensional MLZ problem [Eq.~(\ref{scatter-multidim})], we arrive at
\begin{eqnarray}
\label{S-MLZ-pullback} \hat{S}_{ab} = S^{ab}_{\alpha_{+} \alpha_{-}} e^{i \Phi_{\rm D}^{ab}  }, \;\;\; \Phi_{\rm D}^{ab} = \Phi_{\rm D}^{a} - \Phi_{\rm D}^{b},
\end{eqnarray}
so that Eq.~(\ref{S-MLZ-pullback}) with Eq.~(\ref{scatter-multidim-4}) provide a factorized expression for the scattering matrix of a MLZ problem generated from a MTLZ family. Hence, the dynamical phase $\Phi_{\rm D}^{ab}$, associated with an element $\hat{S}_{ab}$ of the scattering matrix for a MLZ problem, does not generally cancel, and in fact has not only rational, but also not-trivial logarithmic terms. However, it depends only on the initial $b$ and final $a$ diabatic states, and therefore the dynamic phase does not lead to additional complicated interference effects between different scattering pathways. Explicit expressions for the dynamical phase will be presented elsewhere.


\subsection{Demonstration of the first IC}
We are now in a position to demonstrate ICs \cite{quest} that we used in section~\ref{derive-sec}.
By ``demonstrate'' we mean here only that we show that properties (i)-(ii) in section~\ref{derive-sec}  follow directly from MTLZ conditions (\ref{linear-family-2})-(\ref{linear-family-4}).
For IC (i), let us first rewrite the expression for the dynamical  phase $\Phi_{\rm D}^{ab}$ using only parameters of the generated MLZ model (\ref{linear-family-pullback-2}). To that end, we consider two interacting levels, $a$ and $c$ with $g_{ac} \ne 0$, and make use of Eqs.~(\ref{Lambda-to-B-3}) to compute
\begin{eqnarray}
\label{compute-Phi-D} \Phi_{\rm D}^{ac} = \frac{1}{2} (\Lambda_{jk}^{a} - \Lambda_{jk}^{c}) \varepsilon^{j}\varepsilon^{k} = \frac{1}{2 \gamma^{ac}} (A_{j}^{ac} \varepsilon^{j})^{2}.
\end{eqnarray}
and using Eq.~(\ref{linear-family-pullback-2}) we find
\begin{eqnarray}
\label{compute-Phi-D-2} e_{a} - e_{c} = (\Lambda_{jk}^{a} - \Lambda_{jk}^{c}) \varepsilon^{j}v^{k} = \frac{1}{\gamma^{ac}} (A_{j}^{ac} \varepsilon^{j}) (A_{k}^{ac} v^{k}) = \frac{1}{\gamma^{ac}} (A_{j}^{ac} \varepsilon^{j}) g_{ac}.
\end{eqnarray}
Combining Eqs.~(\ref{compute-Phi-D}) and (\ref{compute-Phi-D-2}) we obtain
\begin{eqnarray}
\label{compute-Phi-D-3} \Phi_{\rm D}^{ac} = \frac{\gamma^{ac}}{2 (g_{ac})^{2}} (e_{a} - e_{c})^{2}.
\end{eqnarray}
Computing in a similar way
\begin{eqnarray}
\label{compute-Phi-4} b_{a} - b_{c} = (\Lambda_{jk}^{a} - \Lambda_{jk}^{c}) v^{j}v^{k} = \frac{1}{\gamma^{ac}} (A_{j}^{ac} v^{j})^{2},
\end{eqnarray}
we arrive at
\begin{eqnarray}
\label{compute-Phi-D-5} \Phi_{\rm D}^{ac} = \frac{(e_{a} - e_{c})^{2}}{2 (b_{a} - b_{c})}.
\end{eqnarray}

Generally, diabatic levels $a$ and $c$ may not be coupled directly but they must be connected with a sequence $c, d_{1}, \ldots, d_{k}, a$ of levels with the nearest neighbor interacting property. In this case, similar arguments lead us to an expression for the dynamical phase:
\begin{eqnarray}
\label{compute-Phi-D-final} \Phi_{\rm D}^{ac} = \frac{(e_{a} - e_{d_{k}})^{2}}{2 (b_{a} - b_{d_{k}})} + \frac{(e_{d_{1}} - e_{c})^{2}}{2 (b_{d_{1}} - b_{c})} + \sum_{j=1}^{k-1} \frac{(e_{d_{j+1}} - e_{d_{j}})^{2}}{2 (b_{d_{j+1}} - b_{d_{j}})}.
\end{eqnarray}

Importantly, Eq.~(\ref{asymt-states-relation}) shows that the left hand side of (\ref{compute-Phi-D-final}) does not depend on the choice of the connecting sequence. So, if there is another sequence of pairwise scatterings $c,d'_1,\ldots,d'_k,a$ then difference between dynamic phases that they generate must be zero. This means that if there is any loop in the graph with chain-wise connected diabatic levels having indexes $r_1,\ldots,r_k$ then
\begin{eqnarray}
\label{compute-loop}
{\mathcal A} \equiv \frac{(e_{r_{k}} - e_{r_{1}})^{2}}{2 (b_{r_{k}} - b_{r_{1}})}+\sum_{j=1}^{k-1} \frac{(e_{r_{j+1}} - e_{r_{j}})^{2}}{2 (b_{r_{j+1}} - b_{r_{j}})} =0.
\end{eqnarray}
A simple analysis shows that ${\mathcal A}$ has  geometrical interpretation. Thus, ${\mathcal A}$ is  the area inside a  closed loop on the diabatic level diagram.  Indeed, each term in (\ref{compute-loop}) is the area of a triangle that has two crossing diabatic levels and the energy axis as the boundary.
Each such area is counted with a proper sign, so the sum of all contributions is just the area inside the closed boundary made of crossing levels.
Hence, within the MTLZ class, the first integrability condition (i) can now be considered rigorously proved.


We note also that  Eq.~(\ref{compute-Phi-4})  confirms that in terms of the parameters of the generated MLZ problem  we have
\begin{eqnarray}
\label{compute-gamma} \gamma^{ac} = \frac{|g_{ac}|^{2}}{b_{a} - b_{c}}.
\end{eqnarray}
This explains the observation, made in \cite{quest} and also found in our model (\ref{hal}),  that solvable MLZ models form families that have the same values of  parameter combinations $ |g_{ac}|^{2}/(b_{a} - b_{c})$.

\subsection{Rationalization of the second IC}

Let us remind that the second IC (section~\ref{derive-sec}) states that at sufficiently small but finite values of nonzero couplings, a solvable MLZ model must have an exact energy crossing point, at some $t$, per each pair of diabatic levels that are not coupled directly. Words ``sufficiently small" account for observation that at large values of couplings some of the exact crossing points can merge and annihilate each other \cite{DTCM}.
For MTLZ family with only linear dependence of all operators on time variables, energy rescaling transforms the latter restriction to the condition that couplings can be arbitrary while exact eigenvalue crossing points must appear at sufficiently large time values. So, it is sufficient to prove that there are such exact crossings in the WKB region.

Let $a$ and $b$ be two diabatic levels with $A^{ab}_j=0$ for any $j$. In the WKB region, coupling between corresponding diabatic eigenstates can appear in higher order perturbation series in $1/|\bm{\tau}|$, so the region where an exact crossing can appear must be in the vicinity of the hyperplane defined by Eq.~(\ref{hyper1}). The crucial difference of this hyperplane from the hyperplane that defines the crossing of directly coupled diabatic levels is that the former depends on the index $j$ of the Hamiltonian. Indeed, Eq.~(\ref{Lambda-to-B-3}) that leads to time-index independent Eq.~(\ref{defined-danger-ab}) requires $A^{ab}_j \ne 0$. So, even if (\ref{Lambda-to-B-3}) is valid for some $j$, we have generally
\be
(\Lambda^a_{kn}-\Lambda_{kn}^b) \tau^k \ne 0, \quad {\rm for } \,\, n \ne j.
\label{no-cross1}
\ee
For example, Hamiltonians ${H}_0 (\bm{\tau})$ and ${H}_1(\bm{\tau})$ in Eqs.~(\ref{hbt1})-(\ref{hbt2}) have an exact crossing point of two energy levels of states that evolve from the diabatic states with indexes 1 and 2 at $t\rar -\infty$. For $\hat{H}_0 (\bm{\tau})$, this point appears on the ``hyperplane" $\tau^0=0$ and for $\hat{H}_1 (\bm{\tau})$ this point is already at $\tau^1=0$.

Consider points of the hyperplane defined by Eq.~(\ref{Lambda-to-B-3}) for the Hamiltonian with some index $j$ and $A^{ab}_j=0$.
Assume that there is no exact crossings in the WKB region between these levels, i.e., that higher order corrections lift the degeneracy by introducing small but finite coupling between diabatic states $a$ and $b$.  However small this coupling is,  there is a region then near the hyperplane where the bias between the diabatic levels, i.e. the difference of the diagonal elements of the effective Hamiltonian projected to the $ab$-subspace,   vanishes. Hence, this coupling dominates the effective
Hamiltonian projected on states $a$ and $b$.

Hence, along a path that crosses this hyperplane, there must be the point where eigenstates
of the Hamiltonian are  superpositions of diabatic states: $c_a|a \ra +c_b |b \ra$ with $|c_a|=|c_b|$. On the other hand, since the Hamiltonian $\hat{H}_j$ commutes with $\hat{H}_n$, $n\ne j$, this
eigenstate must also be the eigenstate of $\hat{H}_n$. However, condition (\ref{no-cross1}) means that, in the WKB region of our hyperplane, diabatic level splittings for $n\ne j$ are large. Hence, eigenstates of $\hat{H}_n$ must  coincide with the diabatic states up to vanishingly small $O(1/|\bm{\tau}|)$ corrections. So, superposition of diabatic states with $|c_a|=|c_b|$ cannot be an eigenstate of $\hat{H}_n$. We reached a contradiction, meaning that the eigenvalues that correspond to levels $a$ and $b$ of the operator $\hat{H}_j$ must cross exactly on the hyperplane (\ref{hyper1}) if $A^{ab}_j =0$, at least in the WKB region. This proves IC (ii) for the MTLZ family of models.

Note that our arguments about appearance of the crossing points are not specific to the MTLZ family. What is essential here is the existence of the multi-time WKB region, in which appearance of exact crossing points becomes apparent. Such points carry topological indexes, so they should generally survive inside the family of commuting operators even beyond the WKB regime. This sheds light on the old question about
the origin of exact energy level crossings in integrable quantum models.


\subsection{Path forward: toward classification of solvable models}
We conclude with a brief outline of possible extensions of the approach that we designed in the last section, leaving details to the future publications.

(i) In the case of MTLZ problems, the integrability conditions [Eqs.~(\ref{linear-family-2}), (\ref{linear-family-3}), and (\ref{linear-family-4})], considered as a system of nonlinear equations for a family of an undetermined parameter space dimension can be substantially simplified and, in many cases, solved explicitly, leading to a classification of integrable MTLZ families. The classification starts with an undirected graph $\Gamma$, whose vertices represent the diabatic level of the MLZ problem under consideration; two vertices are connected with an edge if there is a direct coupling between the corresponding diabatic levels. The graph $\Gamma$ should satisfy certain restrictions. We further consider Eq.~(\ref{Lambda-to-B-3}) as a set of compatibility conditions for the forms $A^{ab} = A_{k}^{ab} d\tau^{k}$ that leads to a natural parameterization of the latter in terms of pseudo-orthogonal ${\rm SO}(n, m)$ matrices, associated with simple loops on the graph, where $n+m = {\rm ceiling}(l/2)$, with $l$ being the loop length. The dimensionality of the parameter space is obtained as a part of the solution. Among the graphs that satisfy the aforementioned restrictions are the full bipartite graph $\Gamma = {\rm K}_{2, m}$ and the $n$-dimensional cube graph ${\rm C}^{n}$ with $N = 2 + m$ and $N = 2^{n}$ vertices/levels, respectively. In the ${\rm K}_{2, m}$ case our parameterization leads to a complete solution, resulting in a $2$-dimensional family, i.e., $M=1$, so that the family of the solvable MLZ problems, obtained form the aforementioned $2$-dimensional MTLZ family, using the prescription, presented in subsection~\ref{sec:solvable-MLZ}, is exactly the family of MLZ models, described in section~\ref{main-sec}, and also providing the additional Hamiltonian $H_{1}$, given in section~\ref{bowtie-sec}. In the case of the hypercube graph $\Gamma = {\rm C}^{n}$ an explicit solution of the integrability conditions can be obtained  if one assumes permutation symmetry among the graph vertices, resulting in an $n$-dimensional, i.e., $M = n-1$, MTLZ family, so that the procedure of subsection~\ref{sec:solvable-MLZ}, results in the family of distorted Tavis-Cummings models, described in subsection~\ref{TC-sec}, with $N_{\rm s} = n$ in the limit, when the boson occupation numbers tend to infinity, i.e., when the spins interact with a classical boson/scalar field, whose frequency changes linearly in time.

(ii) To broaden the class of exactly solvable models, as briefly stated earlier in this section, we can allow regular singularities for the Hamiltonians, apart from the infinite time. More formally allowing the parameters $\tau^{j}$ attain complex values and interpret the gauge field ${\cal A}_{j}(\bm{\tau}) = -i H_{j}(\bm{\tau})$, linearly depending on $\tau \in \mathbb{C}^{M+1}$ as a meromorphic gauge field in $\mathbb{C}P^{M+1}$ with an irregular singularity of third order along the infinity $\mathbb{C}P^{M} \subset \mathbb{C}P^{M+1}$, and further allow regular singularities of the gauge field along the $\mathbb{C}P^{M}$, globally  complex-analytically embedded into $\mathbb{C}P^{M+1}$. This introduces additional parameters: the positions of the simple poles along with the matrices that describe the corresponding residues. However, the same parametrization as in the linear case allows the extended system of equations to be treated efficiently. Adding simple poles/regular singularities relaxes the conditions on the underlying graph $\Gamma$, adding, e.g., the complete bipartite graph $\Gamma = {\rm K}_{1, N-1}$ to the list of allowed graphs. Similar to the linear case, the integrability equations for the meromorphic families of the described above class can be explicitly solved for ${\rm K}_{1, N-1}$ and ${\rm K}_{2, N-2}$, resulting in an $(N-1)$- and $2$-dimensional integrable families that allow to solve the Demkov-Osherov (DO) and distorted generalized bowtie models, respectively. The case of ${\rm C}^{N_{\rm s}}$ can be also treated explicitly in the permutation-symmetric case, leading to an exact solution of the distorted Tavis-Cummings model, described in subsection~\ref{TC-sec}.

(iii) It turns out that the most challenging step on the complete characterization of the integrable meromorphic families is identification of the underlying graph $\Gamma$. The issue can be addressed by building a composite model out of some already known integrable ones, following the prescription, described in section~\ref{composite-subsec}, see also \cite{multiparticle,constraints}. The obtained composite model, being also integrable, provides its graph $\Gamma$ that should satisfy the constraints, mentioned above. Therefore the obtained graph can be used as a starting point for solving the integrability equations, whose solutions typically provide a broader class of integrable MLZ, with the aforementioned composite model being a particular member of this class. It is natural to refer to such integrable models as distorted composite models.

(iv) In this manuscript we also considered the dynamical phase in the MLZ scattering for the models based on MTLZ models.  Although not discussed in this manuscript in detail, the logarithmic contributions can be also treated within the same framework, resulting in explicit expressions for the complete dynamical phases that, in particular, contain the logarithmic terms.

(v) The situation with dynamical phases in the meromorphic case is technically more involved, due to the absence of a simple analogue of Eq.~(\ref{Sol-flat-conn-semiclass}) and, especially very explicit Eq.~(\ref{Sol-flat-conn-semiclass-2}) that parameterize the adiabatic expressions for the solutions in the $\bm{\tau}$-independent way. Still, the problem of the dynamical phase identification can be efficiently addressed by considering the spectral curve, associated with the MLZ problem under consideration, that consists of points $(z, \lambda)$, with $z$ and $\lambda$ being the complexified time and an eigenvalue of $H(z)$, respectively. The dynamical phase is then defined by the integral of the so-called dynamical form $\alpha = \lambda dz$ over a proper path in the spectral curve. The described picture of the dynamical phase is obtained from the integrability conditions by considering the spectral manifold, associated with the underlying integrable meromorphic family, that covers the compactified space $\mathbb{C}P^{M+1}$ of the complexified multi-dimensional time $\bm{\tau}$ in a way, similar to how the spectral curve covers the compactified space $\mathbb{C}P^{1}$ of complexified time $z$.

\section{Conclusion}

We compared two different approaches to integrability in the multistate Landau-Zener (MLZ) problem by applying them to find and study a new solvable class. Both, the  empirical rules that used to be called ICs and the approach based on finding   Hamiltonians satisfying conditions (\ref{cond1})-(\ref{cond2}) turned out to be  effective. The latter approach is mathematically justified but, initially, we used it in combination with a fortunate fact that the considered class of models could be generated by the previously known solvable model called the bowtie model. Generally, we do not expect to have such luck in classification of models with other geometries of energy level crossings. So, in the last section we developed an approach that, we believe, leads to very general classification of solvable explicitly time-dependent models, including MLZ systems.

The path to this classification is to combine conditions (\ref{cond1})-(\ref{cond2}) with additional constraints that follow from the requirement that WKB approximation, which emerges at large values of time variables, becomes analytically tractable. This means that there must be a time-path in the WKB region that connects  initial and final physically interesting  points.  Along this path, dynamics should split into pieces with adiabatic evolution separated by distant intervals within which evolution is described by much simpler equations with already known solutions. This restricts us to models with specific dependence of Hamiltonians on time-variables. A good candidate for complete classification  is the class of models with a single low rank irregular point at infinite time  and, possibly, regular singularities at other multiple-time points. Specifically for this article, we considered such a family of models (\ref{cond1})-(\ref{cond2}) with only linear dependence on all time variables, i.e., a single  irregular point at infinity. This  is a natural generalization of the two-state Landau-Zener model (or the parabolic cylinder equation) to multi-state and multi-time dynamics.

We found that this restriction and conditions (\ref{cond1})-(\ref{cond2}) do lead to constraints on model parameters that are sufficient for detailed understanding of the whole multi-time Landau-Zener (MTLZ) family. For example, we proved that corresponding WKB approximation   leads to the explicit solution of associated MLZ models. Moreover, our approach explains a number of previous observations including the matrix product form of the solution,  existence of parameters whose variation does not change transition probability matrices, zero area of the loops in the graph of diabatic levels, and existence of a specific number of exact energy level crossing points.

Certainly, the MTLZ class that we introduced does not exhaust all possibilities to create a tractable WKB approach. For example, the solvable driven Tavis-Cummings model has commuting operators with nonlinear dependence on other time-variables. Moreover, in section~\ref{TC-sec} we showed that this model can be distorted and solved using the same methods that we have studied within the MTLZ class. So, apparently, there is a bigger set of systems (\ref{cond1})-(\ref{cond2}) that contain a MLZ model (\ref{mlz}) but do not reduce to MTLZ. In this regard, the old version of integrability conditions \cite{quest} remains a useful tool to search for new solvable models, although we believe that our approach, introduced  in section~\ref{mtlz-sec}, will eventually outperform the previously used method and lead to a broad classification of explicitly solvable multistate time-dependent quantum problems.  For the future research directions, we also note that the topic of quantum integrability of explicitly time-dependent models has recently experienced progress beyond the MLZ theory \cite{yuzbashyan-an18,gritsev}, and that  there are exact results in MLZ theory beyond the models satisfying all known integrability conditions \cite{cross,constraints}. It should be insightful to understand relations of our method to these alternative developments.

\section*{Acknowledgements}
This work  was supported by the National Science Foundation under Grant No. CHE-1111350 (V.Y.C.). Work at LANL was carried out under the auspices of the National Nuclear Security Administration of the U.S. Department of Energy at Los Alamos National Laboratory under Contract No. DE-AC52-06NA25396. N.A.S. also thanks the support from the LDRD program at LANL.

Authors declare equal contribution to this article.

\end{document}